\documentclass{article}

\usepackage{arxiv}

\usepackage[utf8]{inputenc} 
\usepackage[T1]{fontenc}    
\usepackage{hyperref}       
\usepackage{xurl}           
\usepackage{booktabs}       
\usepackage{amsfonts}       
\usepackage{amsmath}        
\usepackage{amssymb}        
\usepackage{nicefrac}       
\usepackage{microtype}      
\usepackage{lipsum}		
\usepackage{graphicx}
\usepackage[numbers]{natbib}
\usepackage{doi}
\usepackage{tabularx}
\usepackage{makecell}
\usepackage{caption}

\usepackage{booktabs}
\usepackage{pifont}
\newcommand{\cmark}{\ding{51}}
\newcommand{\xmark}{\ding{55}}

\usepackage{algorithm}
\usepackage{algorithmicx}
\usepackage{algpseudocode}

\usepackage{upgreek}
\usepackage{bm}
\usepackage{physics}

\usepackage{placeins}
\usepackage{dblfloatfix}
\usepackage{multirow}
\usepackage{rotating}

\usepackage{authblk} 

\usepackage{listings}
\usepackage{xcolor}
\usepackage{bm}

\definecolor{codegreen}{rgb}{0,0.6,0}
\definecolor{codegray}{rgb}{0.5,0.5,0.5}
\definecolor{codepurple}{rgb}{0.58,0,0.82}
\definecolor{backcolour}{rgb}{0.95,0.95,0.92}

\lstdefinestyle{dimacs}{
    backgroundcolor=\color{backcolour},
    basicstyle=\ttfamily\footnotesize,
    breaklines=true,
    commentstyle=\color{codegreen},
    morecomment=[l]{c},  
    keywordstyle=\color{magenta},
    morekeywords={p, cnf, edge, col},  
    numberstyle=\tiny\color{codegray},
    numbers=left,
    numbersep=5pt,
    showspaces=false,
    showstringspaces=false,
    showtabs=false,
    tabsize=2
}

\title{Comparative study of Potts machine dynamics and performance for Max-$k$-Cut}

\author[]{Bjarke Almer Frederiksen\textsuperscript{*}}
\author[2,3]{Robbe De Prins}
\author[2]{Peter Bienstman}

\affil[1]{Department of Physics, Technical University of Denmark, Lyngby, Denmark}
\affil[2]{Photonics Research Group, Ghent University – imec, Ghent, Belgium}
\affil[3]{Applied Physiscs Research Group, Vrije Universiteit Brussel, Brussels, Belgium}

\renewcommand{\shorttitle}{Comparative study of Potts machine dynamics and performance for Max-$k$-Cut}

\hypersetup{
pdftitle={Comparative study of Potts machine dynamics and performance for Max-$k$-Cut},
pdfsubject={q-bio.NC, q-bio.QM},
pdfauthor={Bjarke Almer Frederiksen, Robbe De Prins, Peter Bienstman},
pdfkeywords={First keyword, Second keyword, More},
colorlinks=true,        
linkcolor=black,        
citecolor=black,        
filecolor=black,        
urlcolor=black          
}

\begin{document}
\pagenumbering{arabic}   
\setcounter{page}{0}     
\newpage
\maketitle
\begingroup\renewcommand\thefootnote{}\footnotetext{* bjarkeaf@proton.me}\endgroup



\begin{abstract}
\addcontentsline{toc}{section}{Abstract}

Combinatorial optimization problems in logistics, finance, energy, and scheduling routinely involve multi-state decision variables. Ising machines (IMs) require binary expansions (e.g., one-hot encoding) to encode such variables, whereas Potts machines (PMs) represent them natively. By doing so, PMs are expected to outperform IMs on multi-state problems. To the best of our knowledge, no systematic study of PM models has yet assessed whether this expectation holds. We therefore benchmark five representative PMs against a reference IM on Max-3-Cut and Max-4-Cut, using 800-vertex GSet graphs and random graphs of up to 50 vertices. Surprisingly, the reference IM still outperforms every PM, and the IM supremacy increases significantly in going from Max-3-Cut to Max-4-Cut. These results provide clear evidence that current PM dynamics underperform relative to binary approaches, even in regimes where they are presumed advantageous. We provide a way forward by quantifying the underperformance of current PMs, as well as by identifying three dynamical properties that correlate strongly with their performance ranking. Our work stresses the need for more systematic assessments of algorithmic performance in order to guide the design of more effective Potts machines.

\end{abstract}

~\newpage

\twocolumn

\section{Introduction} \label{sec:intro}
Combinatorial optimization is concerned with minimizing a cost function over a set of discrete variables. Such problems are key to a wide range of applications, including logistics \cite{weinberg_supply_2023}, finance \cite{herman_quantum_2023}, energy systems \cite{morstyn_annealing_2023}, job scheduling \cite{venturelli_quantum_2015}, and traffic flow \cite{neukart_traffic_2017}. The size of the configuration space of combinatorial optimization problems (COPs) can grow faster than exponentially in the number of objects, making exhaustive search impractical \cite{COPs}. As a result, several algorithms and hardware accelerators have been developed over the years which aim to provide high-quality solutions with minimal consumption of resources such as time and energy \cite{kirkpatrick_optimization_1983,Mohseni2022}.

Ising machines (IMs) \cite{Inagaki2016,Yamamoto2020,Yamamoto2017,Mohseni2022} are hardware accelerators designed to find low-energy states of the \textit{Ising model} of statistical mechanics, defined by the Ising Hamiltonian \cite{ising_beitrag_1925}:
\begin{equation}
    H_\mathrm{Ising}(\left\{\sigma_i\right\}) = -\frac{1}{2}\sum_{i,j}^N J_{ij} \sigma_i \sigma_j - \sum_{i}^N h_i \sigma_i. \label{eq:ising_hamiltonian}
\end{equation}
Here, $\sigma_i \in \{-1, +1\}$ represent binary spins such as the magnetic moment of a particle, which can either be spin-up ($+1$) or spin-down ($-1$). The first term describes the interaction between spins, where $J_{ij}$ is the real-valued coupling coefficient between spins $i$ and $j$, and the second term describes the real-valued external bias field $h_i$ acting on each spin. The goal of an IM is to find a configuration of spins that minimizes the energy of the system, which corresponds to finding a ground state of the Ising Hamiltonian.

While IMs have been widely studied as analog computing platforms for solving many NP-hard COPs \cite{lucas_ising_2014}, many real-world and industry-relevant COPs are naturally defined using multi-level discrete variables. Indeed, many of the Ising formulations catalogued by Lucas \cite{lucas_ising_2014} involve multi-state assignments that map more naturally to Potts spins, including graph coloring \cite{wu_potts_1982}, the traveling salesman problem \cite{peterson_new_1989}, and job scheduling \cite{gislen_complex_1992}. Max-$k$-Cut is an archetypal example of such a problem, closely related to graph coloring. Given a graph, the objective is to assign one of $k$ colors to each vertex such that the number of edges between differently colored vertices is maximized. This can be described as maximizing the following objective function, referred to as the \textit{cut value}:
\begin{equation}
    C(\left\{v_i\right\}) = \frac{1}{2}\sum_{i,j}^N w_{ij} \left(1 - \delta\left(v_i, v_j\right)\right).
\end{equation}
Here, $v_i \in \{1,2,\ldots,k\}$ is the label denoting which set vertex~$i$ belongs to, $w_{ij}$ is the weight of the edge between vertices~$i$ and~$j$, and $\delta(v_i,v_j)$ is the Kronecker delta, which equals 1 if $v_i = v_j$ and 0 otherwise. 

For $k=2$, Max-$k$-Cut reduces to the Max-Cut problem, which maps directly to the Ising model by setting $\sigma_i = 2v_i - 3$, $J_{ij} = -w_{ij}$, and $h_i = 0$, and can readily be addressed by IMs. However, for $k>2$, Max-$k$-Cut cannot be natively represented by binary variables, and requires a binary expansion in order to map the problem to the Ising model \cite{lucas_ising_2014}. Such an expansion incurs significant overhead in terms of embedding size, increasing hardware complexity. Furthermore, the embedding overhead can lead to a less convex energy landscape, resulting in sub-optimal performance. A similar issue has been identified in $k$-SAT problems, where polynomial unconstrained binary optimization (PUBO) outperforms quadratization methods \cite{pedretti_solving_2025,dobrynin_energy_2024,valiante_computational_2021,hizzani_memristor-based_2024} due to the latter providing a less native mapping of the original problem.

To address the limitations in mapping multi-level discrete variable COPs to the Ising model, Potts machines (PMs) have been proposed as extensions of IMs that directly encode multi-state discrete variables \cite{kalinin_simulating_2018,honari-latifpour_combinatorial_2022,roychowdhury_oscillator-based_2022,inoue_coherent_2023,mallick_computational_2022}. PMs are designed to find low-energy states of the Potts model, which comes in two variants: the \textit{standard Potts model} and the \textit{planar Potts model}, defined by the standard Potts Hamiltonian and the planar Potts Hamiltonian, respectively \cite{wu_potts_1982}:
\begin{subequations}
\begin{gather}
  H_\mathrm{Standard\ Potts}(\left\{s_i\right\}) = -\frac{1}{2}\sum_{i,j}^N J_{ij} \delta\left(s_i, s_j\right), \label{eq:std_potts} \\
  H_\mathrm{Planar\ Potts}(\left\{\theta_i\right\}) = -\frac{1}{2}\sum_{i,j}^N J_{ij} \cos{\left(\theta_i - \theta_j\right)}. \label{eq:planar_potts}
\end{gather}
\end{subequations}
For the standard Potts model, $s_i \in \{1, 2, \ldots, q\}$ is the state of spin $i$, while for the planar Potts model, $\theta_i$ is the state of spin $i$, taking one of $q$ equispaced angles in the range $[-\pi, \pi)$. The two Potts models include a coupling term with real-valued coupling coefficients $J_{ij}$. The two Hamiltonians are equivalent for $q\leq 3$, where the angles can be mapped to the discrete states of the standard Potts model. For $q=2$, both variants reduce to the Ising Hamiltonian defined in Eq. \eqref{eq:ising_hamiltonian}, without an external bias field, i.e. $h_i = 0$. The goal of a PM is to find a configuration of spins that minimizes the energy of the Potts model, which corresponds to finding a ground state of a Potts Hamiltonian.

Max-$k$-Cut can be mapped directly to the standard Potts model, simply by setting $q=k$, $s_i = v_i$ and $J_{ij} = -w_{ij}$, such that maximizing the cut value and minimizing the Potts Hamiltonian are equivalent. Max-$k$-Cut is thus the canonical problem for PMs, allowing PMs to solve these problems without embedding overhead. However, for $q>3$, the planar and standard Potts Hamiltonians are not equivalent, so PMs implementing the planar Potts model introduce additional constraints on the problem structure, the implications of which are examined in Section~\ref{sec:decreasing_performance_max3cut_max4cut}.

While the field of analog Ising machines has seen extensive studies of how the computational performance compares across different implementations \cite{bohm_order--magnitude_2021,lamers_using_2024} and studies of the properties that are conducive to high performance \cite{yamamura_geometric_2024}, no such studies have yet been conducted for Potts machines. Many PMs have been proposed thus far. We group these into two categories based on the nature of the spin variables: (i) discrete-spin PMs, where the spins directly take the discrete states of the Potts model, such as in implementations using probabilistic multi-dimensional bits (p-dits) \cite{duffee_extended-variable_2025} and single-photon avalanche diodes (SPADs) \cite{whitehead_cmos-compatible_2023}, or (ii) continuous-spin PMs, also called \textit{analog PMs}, where the spins take continuous values and typically evolve dynamically to approach the discrete states of the Potts model, such as in implementations using physical coupled multi-stable gain-dissipative systems \cite{kalinin_simulating_2018,honari-latifpour_combinatorial_2022,mekawy_enabling_2025,roychowdhury_oscillator-based_2022,inoue_coherent_2023,mallick_computational_2022}.

This work constitutes, to the best of our knowledge, the first comparative study of PMs, both qualitatively and quantitatively. We study only analog continuous-time PMs, which can all be described by coupled ordinary differential equations and simulated within a unified dynamical systems framework. We further focus on adiabatic annealing, in which control parameters are varied slowly enough for the system to track the evolving fixed-point structure. This allows a fair comparison of solution quality across models and connects to the geometric landscape annealing principle identified for coherent Ising machines by Yamamura \emph{et al.} \cite{yamamura_geometric_2024}.
We consider three such PM implementations from the literature:
the nonequilibrium condensate (NEC) model by Kalinin and Berloff \cite{kalinin_simulating_2018,kalinin_networks_2018},
the $q$-photon downconversion ($q$-PDC) model by Honari-Latifpour et al. \cite{honari-latifpour_optical_2020,honari-latifpour_combinatorial_2022},
and the $q$\textsuperscript{th} sub-harmonic injection locking ($q$-SHIL) model by Roychowdhury and Seal \cite{roychowdhury_oscillator-based_2022}. Our study currently does not include the $q$-level phase sensitive amplifier ($q$-PSA) model by Inoue et al. \cite{inoue_coherent_2023}, as it is a discrete-time model that does not fit our continuous-time framework. We also do not include the phase-only PM models by Mallick et al. \cite{mallick_computational_2022} and Crnkić et al. \cite{crnkic_collective_2020}, as their engineered multi-harmonic coupling functions fall outside the standard Kuramoto-type coupling $J_{ij}\sin(\theta_j-\theta_i)$ shared by the models considered here, of which the $q$-SHIL model is the phase-only representative obtained as the fixed-amplitude limit of the $q$-PDC model.

Aside from the three PM models from the literature, we also propose two new PM models: a polynomial PM that unifies common traits among literature models, and a sigmoid PM drawing on insights from analog IMs \cite{bohm_order--magnitude_2021,lucas_ising_2014}. We also include an analog IM model \cite{prins_how_2025} to study the performance of PMs relative to IMs. Through theoretical analysis and numerical simulations, we study the dynamics of the five PM models and identify three dynamical properties that may influence optimization performance: (i) a connected origin-to-solution fixed-point branch, (ii) enforced amplitude homogeneity, and (iii) enforced phase discretization. We then benchmark the models on Max-3-Cut and Max-4-Cut problems from five 800-vertex graphs in the GSet collection, as well as on smaller random graphs with up to 50 vertices, to assess which of these properties are conducive to high performance, with the goal of guiding the design of future Potts machines.





\section{Review of studied models, their governing equations and annealing schemes}

The studied models are summarized in Table \ref{tab:models} and their governing equations are described below. Although the physical mechanisms differ across models, the parameters $\alpha$, $\beta$, and $\gamma$ play analogous functional roles where they appear. Where present, $\alpha$ is the linear amplitude gain, $\beta$ is the inter-spin coupling strength, and $\gamma$ is what we term the \textit{phase discretization strength}, since in every model where it appears the term with prefactor $\gamma$ introduces a potential with $q$-fold rotational symmetry that drives continuous spin phases toward $q$ equally spaced discrete states.

\begin{table*}[t]
  \setlength{\extrarowheight}{3pt}
  \caption{Summary of studied Potts machine (PM) and Ising machine (IM) models.}
  \label{tab:models}
  \centering
  \begin{tabularx}{\textwidth}{lcccXc}
    \toprule
    Model
      & \makecell[c]{Governing\\equations}
      & \makecell[c]{Spin value}
      & \makecell[c]{Amplitude\\nonlinearity}
      & Annealing scheme
      & References \\
    \midrule
    NEC 
      & \eqref{eq:nec}, \eqref{eq:nec_alpha} 
      & $x\in\mathbb{C}$
      & Polynomial
      & Dynamical $\alpha$, adiabatic $\gamma$ increase 
      & \cite{kalinin_networks_2018,kalinin_simulating_2018} \\
    $q$-PDC
      & \eqref{eq:q-pdc}
      & $x\in\mathbb{C}$
      & Polynomial
      & Fixed $\beta$, non-adiabatic (fast) $\gamma$ increase
      & \cite{honari-latifpour_optical_2020,honari-latifpour_combinatorial_2022} \\
    $q$-SHIL
      & \eqref{eq:q-shil}
      & $\theta \in \mathbb{R}$
      & N/A
      & Adiabatic $\gamma$ increase
      & \cite{roychowdhury_oscillator-based_2022} \\
    Polynomial PM
      & \eqref{eq:polynomial_pm}
      & $x \in\mathbb{C}$
      & Polynomial
      & Adiabatic $\beta$ and $\gamma$ increase
      & This work \\
    Sigmoid PM
      & \eqref{eq:sigmoid_pm_r}, \eqref{eq:sigmoid_pm_theta}
      & $x \in\mathbb{C}$
      & Sigmoid
      & Fixed $\alpha$, adiabatic $\beta$ and $\gamma$ increase
      & This work \\
    Reference IM
      & \eqref{eq:sigmoid_im}
      & $x \in \mathbb{R}$
      & Sigmoid
      & Fixed $\alpha$, adiabatic $\beta$ increase
      & \cite{prins_how_2025,lucas_ising_2014,bohm_order--magnitude_2021} \\
    \bottomrule
  \end{tabularx}
  \caption*{The governing equations are provided in the text, while further details on the models and their governing equations, including parameter rescaling and stability analyses, are provided in the Supp. The polynomial PM and sigmoid PM are new models proposed in this work. A reference IM is included for comparison with the PMs. Parameter roles are described in the text.}
\vspace{-2em}
\end{table*}

\subsection{Non-Equilibrium Condensate (NEC) model}
The NEC model \cite{kalinin_simulating_2018,kalinin_networks_2018} is a continuous-time PM based on the dynamics of a Bose-Einstein condensate brought out of equilibrium through a combination of resonant and nonresonant pumping. Kalinin and Berloff describe these dynamics by the complex Ginzburg-Landau equation (CGLE) in the presence of resonant pumpiSng and dynamical gain. The governing equations of the NEC model, in rescaled form, are (Supp.\ \S\ref{app:RescalingNEC}):
\begin{subequations}
\begin{gather}
    \frac{\mathrm{d} x_i}{\mathrm{d} t} = \alpha_i x_i - x_i \lvert x_i\rvert^{n-1} + \gamma (x_i^*)^{q-1} + \sum_{j=1}^{N} J_{i j}\,x_j, \label{eq:nec} \\
    \frac{\mathrm{d} \alpha_i}{\mathrm{d} t} = \varepsilon_\alpha\bigl(\sqrt{r_{\mathrm{target}}} - \sqrt{\lvert x_i\rvert}\bigr).\label{eq:nec_alpha}
\end{gather}
\end{subequations}
The first equation describes the evolution of the complex-valued amplitude $x_i \in \mathbb{C}$ of condensate $i$, where $\alpha_i$ is the local gain, $\gamma$ is the resonant pumping strength at the $q^\mathrm{th}$ resonance, which controls phase discretization (with $q$ corresponding to the number of Potts states, i.e. $q=3$ for Max-3-Cut and $q=4$ for Max-4-Cut), $n$ is the nonlinearity order, $N$ is the number of condensates, and $J_{ij}$ are the coupling coefficients. The four terms represent gain, amplitude saturation, phase discretization, and inter-spin coupling. The second equation describes the evolution of the local gain $\alpha_i$, which is controlled by the gain relaxation rate $\varepsilon_\alpha$ and a target amplitude $r_{\mathrm{target}}$. This rescaled form is obtained by assuming that the single-spin coupling strengths are initialized as the coupling coefficients of the Potts Hamiltonian and that there are no on-site particle interactions (Supp.\ \S\ref{app:RescalingNEC}). The nonlinearity order $n$ is set to 3 in the original model. However, $n\geq q$ is required to ensure bounded amplitudes (Supp.\ \S\ref{app:bounding}). Therefore, we employ $n=3$ for Max-3-Cut ($q=3$) and $n=4$ for Max-4-Cut ($q=4$).

The annealing scheme of the NEC model entails increasing the resonant pumping strength $\gamma$. Meanwhile, the local gains $\alpha_i$ evolve over time as described by Eq. \eqref{eq:nec_alpha}. The local gains are all initialized to $\alpha_i=-\mu_\mathrm{max}$, where $\mu_\mathrm{max}$ is the principal eigenvalue of the coupling matrix $\mathbf{J}$, which is found to be the threshold gain above which the origin becomes unstable (Supp.\ \S\ref{app:StabilityNEC}). The gain relaxation rate $\varepsilon_\alpha$ is set to a small value ($10^{-2}$) for adiabatic operation with respect to Eq. \eqref{eq:nec}. We increase the resonant pumping strength $\gamma$ linearly from zero over the course of the annealing with a rate $\varepsilon_\gamma = \mathrm{d}\gamma/\mathrm{d}t$, which is set to a small value (around $10^{-3}$, see Supp.\ \S\ref{app:SimParams}) to ensure adiabatic annealing.

\subsection{\textit{q}-photon downconversion (\textit{q}-PDC) model}
The $q$-PDC model \cite{honari-latifpour_optical_2020, honari-latifpour_combinatorial_2022} describes a Potts machine based on a network of phase-multistable coupled parametric oscillators. Phase multistability is achieved through the nonlinear process of spontaneous multiphoton downconversion, which is a process where a single pump photon is converted into $q$ photons of lower energy. This mechanism has also been realized in the electronic domain using a parametric frequency divider circuit \cite{mekawy_enabling_2025}. Assuming all spins have roughly an equal number of couplings and equal coupling coefficients, the governing equations of the $q$-PDC model can be rescaled as (Supp.\ \S\ref{app:RescalingQPDC}):
\begin{subequations}
\begin{gather}\label{eq:q-pdc}
    \frac{\mathrm{d} x_i}{\mathrm{~d} t}=-x_i-x_i\left|x_i\right|^{2(q-1)}+\gamma\left(x_i^*\right)^{q-1}+\beta \sum_j^N J_{i j} x_j, \\
    \beta = \frac{1}{\kappa_d},\quad \kappa_d = \kappa_l + \frac{1}{N}\sum_{ij}^{N} \left|J_{ij}\right|,\label{eq:qpdc_beta}
\end{gather}
\end{subequations}
where $x_i \in \mathbb{C}$ is the complex-valued amplitude of spin $i$, $\gamma$ is the small-signal gain, which controls phase discretization, $\beta$ is the reciprocal of the oscillator damping rate $\kappa_d$ and sets the effective coupling strength, $q$ is the number of photons produced in the downconversion process, and $J_{ij}$ are again the coupling coefficients. The damping rate $\kappa_d$ is the sum of the intrinsic loss of the oscillators $\kappa_l$ and the average external losses due to interactions with other spins. We will use the value $\kappa_l = 1$ for all simulations, which is the value used in the original work \cite{honari-latifpour_optical_2020}.

The $q$-PDC model is annealed by linearly increasing $\gamma$, starting from zero, with a rate $\varepsilon_\gamma = \mathrm{d}\gamma/\mathrm{d}t$. The reciprocal damping rate $\beta$ is fully determined by $\kappa_l$ and is thus constant throughout annealing. If $\beta < \mu_\mathrm{max}^{-1}$, where $\mu_\mathrm{max}$ is the principal eigenvalue of the coupling matrix, the origin ($x_i=0, \forall i$) remains a stable equilibrium (attracting) throughout annealing (Supp.\ \S\ref{app:StabilityQPDC}). This means that the system can get trapped in the origin indefinitely. Instead of relying on destabilization of the origin through increasing $\beta$, the oscillators are seeded at a non-zero amplitude and $\gamma$ is increased fast enough to allow the system to reach a stable fixed point occurring through a bifurcation at threshold gain $\gamma_\mathrm{th}=\frac{4}{3^{3/4}} \approx 1.755$ \cite{honari-latifpour_optical_2020} before getting trapped in the origin. As $\gamma$ must be annealed fast enough to avoid trapping, this model does not support adiabatic annealing, except for the case of $\beta > \mu_\mathrm{max}^{-1}$, where the origin is always unstable (repelling). We will employ a seeding amplitude of $r_\mathrm{seed} = 1$, which is the value used in the original work \cite{honari-latifpour_optical_2020}. For the problems studied in this work, an annealing rate $\varepsilon_\gamma$ of around $0.5\gamma_\mathrm{th} \approx 1$ is found to be sufficient to ensure that the system reaches a stable fixed point before getting trapped in the origin.

\subsection{\textit{q}-th sub-harmonic injection locking (\textit{q}-SHIL) model}
The $q$-SHIL model \cite{roychowdhury_oscillator-based_2022} describes a PM implemented as a system of coupled phase oscillators perturbed by an oscillation at the $q^\mathrm{th}$ harmonic of the oscillator frequency. The governing equations of the $q$-SHIL model can be rescaled as follows (Supp.\ \S\ref{app:RescalingQSHIL}). The coupling term takes the same sinusoidal form as in the Kuramoto model \cite{kuramoto_self-entrainment_1975}, augmented with a $q$-th harmonic injection locking term:
\begin{equation}\label{eq:q-shil}
    \frac{\mathrm{d} \theta_i}{\mathrm{~d} t} = -\gamma \sin{q \theta_i} + \sum_{j=1}^{N} J_{ij} \sin{\left(\theta_j - \theta_i\right)},
\end{equation}
Here, $\theta_i \in \mathbb{R}$ is the phase of spin $i$, and $\gamma$ is the injection locking strength, which controls phase discretization. Note that this model corresponds to a simplification of the $q$-PDC model proposed by Honari-Latifpour \emph{et al.} \cite{honari-latifpour_combinatorial_2022} assuming fixed, homogeneous amplitudes.

The $q$-SHIL model is annealed by linearly increasing $\gamma$, starting from zero, with a rate $\varepsilon_\gamma = \mathrm{d}\gamma/\mathrm{d}t$, which we set to $2\times10^{-3}$ for adiabatic annealing. Note that there is no notion of an origin in this model, as it just describes phase dynamics for unit-amplitude spins.

\subsection{Polynomial Potts machine model (Polynomial PM) - this work}
The polynomial PM model is a new model proposed in this work, which serves both as a unifying model for NEC and $q$-PDC, assuming prefactors of the governing equations are homogeneous across all spins (so no local gain as in the original NEC model), and as a Potts extension of the coherent Ising machine (CIM) \cite{bohm_order--magnitude_2021}. The governing equations of the polynomial PM model, in rescaled form, are
\begin{equation}\label{eq:polynomial_pm}
    \frac{\mathrm{d} x_i}{\mathrm{~d} t}=-x_i-x_i\left|x_i\right|^{n-1}+\gamma\left(x_i^*\right)^{q-1}+\beta \sum_j^N J_{i j} x_j.
\end{equation}
Here, $x_i \in \mathbb{C}$ is the complex-valued amplitude of spin $i$, $\beta$ is the coupling strength, $\gamma$ is the phase discretization strength, $n$ is the nonlinearity order, and $q$ is the number of Potts states. This rescaled form is obtained by absorbing arbitrary real-valued prefactors of the polynomial model into $\beta$, $\gamma$, and $n$ (Supp.\ \S\ref{app:RescalingPolynomial}).

We propose annealing the model by increasing $\beta$ and $\gamma$ linearly over the course of the annealing with rates $\varepsilon_\beta = \mathrm{d}\beta/\mathrm{d}t$ and $\varepsilon_\gamma = \mathrm{d}\gamma/\mathrm{d}t = f_\gamma \varepsilon_\beta$, respectively, where $f_\gamma$ is a constant factor. By stability analysis, we find that the origin becomes unstable when $\beta$ crosses the threshold $\beta_\mathrm{th}=\mu_\mathrm{max}^{-1}$, where $\mu_\mathrm{max}$ is the principal eigenvalue of the coupling matrix (Supp.\ \S\ref{app:StabilityPolynomial}). Thus, the spins can be initialized at the origin and, since $\beta$ is initialized at the destabilization threshold (see below), $\beta$ and $\gamma$ can be increased arbitrarily slowly to ensure adiabatic annealing. The initial values for $\beta$ and $\gamma$ are set to $\beta_0 = \beta_\mathrm{th}$ and $\gamma_0 = f_\gamma \beta_\mathrm{th}$, respectively, so that the ratio $\gamma / \beta = f_\gamma$ is maintained from the start of annealing. For our studies, we set $\varepsilon_\beta$ to $2\times 10^{-3}$ and sweep $f_\gamma$ to find the optimal relative annealing rate for each problem instance (Supp.\ \S\ref{app:SimParams}).

\subsection{Sigmoid Potts machine model (Sigmoid PM) - this work} \label{sec:sigmoid}
The sigmoid PM model is a new mathematical model proposed in this work, motivated by the observation that sigmoid nonlinearities enforce amplitude homogeneity and improve performance in analog IMs \cite{bohm_order--magnitude_2021}. Unlike the other models considered here, the sigmoid PM is not directly connected to a specific hardware implementation. For Potts machines with complex-valued spins, simply replacing the polynomial nonlinearity with a complex sigmoid nonlinearity would break the $q$-fold rotational symmetry of the dynamics, because the complex hyperbolic tangent does not commute with phase rotations $x \to e^{i\phi}x$, whereas the polynomial nonlinearity does. Instead, we propose a model that applies a sigmoid nonlinearity to the amplitude of the spins, while keeping the phase dynamics unaltered, enabled by using separate equations for amplitude and phase evolution (Supp.\ \S\ref{app:Sigmoid}). The governing equations of the sigmoid PM model are then:
\begin{subequations}
\begin{align}
  \frac{\mathrm{d} r_i}{\mathrm{d} t}
    &= -\,r_i 
       + \tanh\Bigl(\alpha\,r_i
         + \gamma\,r_i^{\,q-1}\cos(q\theta_i)\notag\\
    &\quad\;+\;\beta\sum_{j=1}^N J_{ij}\,r_j\cos(\theta_j-\theta_i)\Bigr)\,,\label{eq:sigmoid_pm_r}\\
  \frac{\mathrm{d} \theta_i}{\mathrm{d} t}
    &= -\gamma\,r_i^{\,q-2}\sin(q\theta_i)
       + \frac{\beta}{r_i}\sum_{j=1}^N J_{ij}\,r_j\sin(\theta_j-\theta_i)\,.\label{eq:sigmoid_pm_theta}
\end{align}
\end{subequations}
Here, $r_i \in \mathbb{R}$ is the amplitude of spin $i$, $\theta_i \in \mathbb{R}$ is the phase of spin $i$, $\alpha$ is the amplitude gain, $\beta$ is the coupling strength, $\gamma$ is the phase discretization strength, and $q$ is the number of Potts states. 

The sigmoid PM model can be annealed, either by increasing $\alpha$ or $\beta$, or both, along with $\gamma$. We choose to treat $\alpha$ as a constant and increase $\beta$ and $\gamma$ linearly over the course of the annealing with rates $\varepsilon_\beta = \mathrm{d}\beta/\mathrm{d}t$ and $\varepsilon_\gamma = \mathrm{d}\gamma/\mathrm{d}t = f_\gamma \varepsilon_\beta$, respectively, where $f_\gamma$ is a constant factor. Within this annealing scheme, lowering the constant $\alpha$ in practice slows the evolution of the system, which can support adiabatic annealing. The origin becomes unstable when $\beta$ crosses the threshold $\beta_\mathrm{th}=(1-\alpha)\mu_\mathrm{max}^{-1}$ or when $\alpha$ crosses the threshold $\alpha_\mathrm{th}=1-\beta\mu_\mathrm{max}$, where $\mu_\mathrm{max}$ is the principal eigenvalue of the coupling matrix (Supp.\ \S\ref{app:StabilitySigmoid}). Thus, the spins can be initialized at the origin, and $\beta$ and $\gamma$ can be increased arbitrarily slowly to ensure adiabatic annealing. The initial values for $\beta$ and $\gamma$ are set to $\beta_0 = \beta_\mathrm{th}$ and $\gamma_0 = f_\gamma \beta_\mathrm{th}$, respectively. For our studies, we set $\varepsilon_\beta$ to $10^{-3}$ and sweep $f_\gamma$ as well as $\alpha$ for each problem instance (Supp.\ \S\ref{app:SimParams}).

\subsection{Reference Ising machine model (Reference IM)} \label{sec:sigmoid_im}
As a reference Ising machine for comparison with the studied Potts machines, we use a model based on the coherent Ising machine (CIM) with sigmoid nonlinearity proposed by Böhm \emph{et al.} \cite{bohm_order--magnitude_2021}. This model is chosen as a reference IM because it is a well-studied model that has been shown to achieve high performance on Ising optimization problems \cite{bohm_order--magnitude_2021,lamers_using_2024}. The sigmoid nonlinearity approximates the saturation of spin amplitudes commonly observed in experimental analog IMs \cite{prins_how_2025, lamers_using_2024}. The Ising formulation of multi-state problems such as Max-$k$-Cut requires external bias fields in addition to spin-spin couplings. Given an external bias field $h_i$ for each spin $i$, and using the spin sign method, introduced for simulated bifurcation machines by Goto et al. \cite{goto_high-performance_2021} and recently shown to be the most effective method for encoding external fields \cite{prins_how_2025}, the governing equations of the reference IM become
\begin{subequations}
\begin{gather}\label{eq:sigmoid_im}
    \frac{\mathrm{d} x_i}{\mathrm{d} t}
      = -x_i + \tanh\Bigl(\alpha\,x_i
        + \beta I_i\Bigr) \\
        I_i = h_i + \sum_{j=1}^N J_{ij}\,\mathrm{sgn}(x_j)
\end{gather}
\end{subequations}
where $x_i \in \mathbb{R}$ is the spin value of spin $i$, $\alpha$ is the amplitude gain, $\beta$ is the coupling strength, and $I_i$ is the local field experienced by spin $i$. The local field is the sum of the external bias field $h_i$ and coupling contributions from all other spins. 

To represent Max-$k$-Cut problems, we use an Ising graph coloring formulation from the literature \cite{lucas_ising_2014}, which allows us to encode the target problem in the form of the coupling coefficients $J_{ij}$ and the external bias field $h_i$ (Supp.\ \S\ref{app:OneHotMapping}). Following De Prins \emph{et al.} \cite{prins_how_2025}, we additionally rescale the external fields by a constant factor $\zeta$, which corrects for imbalances between the spin couplings and the constraint terms introduced by the one-hot encoding. The factor $\zeta$ is treated as a hyperparameter (Supp.\ \S\ref{app:SimParams}).

The annealing scheme of this model entails increasing the coupling strength $\beta$ linearly over the course of the annealing with a rate $\varepsilon_\beta = \mathrm{d}\beta/\mathrm{d}t$, while keeping $\alpha$ constant. As for the sigmoid PM, lowering the constant $\alpha$ slows the evolution of the system, which can support adiabatic annealing.

\section{Methods} \label{sec:methods}
\subsection{Numerical methods} \label{sec:numerical_methods}

We use the Euler--Maruyama method \cite{maruyama_continuous_1955} to numerically integrate the governing equations of the Potts- and Ising-machine models. This method is a stochastic extension of the Euler method for solving ordinary differential equations and is suitable for simulating continuous-time dynamical systems with noise. See Supp.\ \S\ref{app:euler-maruyama} for a more detailed description of the method and its implementation.

For all simulations, the noise factor of the stochastic process is set to $f_\xi=10^{-4}$ to model a scenario with low experimental noise. The time step size is chosen to ensure numerical stability and accuracy, in the range of $10^{-3}$ to $10^{-2}$, depending on the model. See Supp.\ \S\ref{app:SimParams} for details on the choice of time step size for each model, and \S\ref{app:convergence_tests} for convergence tests informing these parameter choices.

\subsection{Problem selection and encoding}\label{sec:problem_selection}
We use two benchmark problem sets in this work:

\textbf{GSet benchmark problems.} We select 5 graphs from the GSet collection \cite{gset}, all with 800 vertices and 19176 edges (6\% edge frequency). The GSet collection contains undirected graphs with edge weights of +1 or –1. We select 5 graphs with purely +1 edge weights (G1--G5), as we implement unweighted Max-$k$-Cut. These graphs are interpreted as both Max-3-Cut and Max-4-Cut problems, referred to as the GSet Max-3-Cut and GSet Max-4-Cut benchmark problems, respectively. The chosen graphs and their best known Max-3-Cut and Max-4-Cut values, obtained via a multiple operator heuristic (MOH) \cite{ma_multiple_2017}, are listed in Supp.\ Table~\ref{tab:gset_graphs}.

\textbf{ER50 benchmark problems.} We additionally evaluate the models on 60 Max-3-Cut problems on Erd\H{o}s--R\'{e}nyi random graphs generated according to the $G(n,p)$ model with edge probability $p=0.5$. These graphs have sizes ranging from 5 to 50 vertices (10 instances per size) and all +1 edge weights. This problem set is referred to as ER50 due to the graph model and the 50\% edge probability. The optimum Max-3-Cut values for these problems are found using the publicly available \texttt{max\_k\_cut} package \cite{max_k_cut}.

The Max-$k$-Cut problems are encoded as Potts Hamiltonians by having one Potts spin with $q=k$ states represent each vertex and setting coupling coefficients $J_{ij}=J_{ji}=-1$ for edges with weight +1 between vertices $i$ and $j$, and 0 otherwise. For the Ising machine, each vertex is represented by $k$ Ising spins using one-hot encoding, yielding $Nk$ spins in total, with external fields rescaled as described in section~\ref{sec:sigmoid_im} (Supp.\ \S\ref{app:OneHotMapping}).

\subsection{Performance metrics} \label{sec:performance_metrics}
Success rate, the fraction of runs that achieve the optimum cut value, is often used as a performance metric for combinatorial optimization problem solvers:
\begin{equation}
\mathrm{Success\ rate} = \frac{n_\mathrm{success}}{n_\mathrm{runs}},
\end{equation}
where $n_\mathrm{success}$ is the number of runs that achieve the optimum cut value, and $n_\mathrm{runs}$ is the total number of runs.

For smaller problems, models achieve significant success rates, making success rate the preferred metric. We therefore report success rates for the ER50 benchmark problems, which have 5 to 50 vertices.

However, for larger benchmark problems, success rates are typically zero or near-zero, such that success rate cannot resolve performance differences between models. For the GSet benchmark problems, we therefore report the optimality gap, defined as the difference between the optimum cut value of the problem and the maximum cut value found by the model. It can be normalized by the optimum cut value for a relative optimality gap that can be compared across different problems:
\begin{equation}
\label{eq:opt_gap}
\mathrm{Relative\ optimality\ gap}
= \frac{C_\mathrm{opt} - C_\mathrm{model}}{C_\mathrm{opt}},
\end{equation}
where $C_\mathrm{model}$ is the maximum cut value found by the model, and $C_\mathrm{opt}$ is the optimum cut value of the problem, either the global optimum or the best known solution to the problem.

Time-to-solution, defined as the time taken by a model to reach a solution with a relative optimality gap below a given threshold, is another commonly used performance metric. However, this study focuses on solution quality under adiabatic annealing schemes, where annealing rates are deliberately set to small values to ensure the system tracks the evolving fixed-point structure, so differences in wall-clock time reflect the chosen annealing scheme rather than any intrinsic advantage of the underlying dynamics. Time-to-solution is therefore not an informative axis for distinguishing models whose annealing rates are tuned for quality rather than speed.

\subsection{Benchmarking procedure} \label{sec:benchmarking_procedure}
Each model \(m\) is evaluated on each problem graph \(g\) by sweeping its hyperparameters \(\boldsymbol{\Theta}\). For each triplet \((m,g,\boldsymbol{\Theta})\), we run \(n_{\rm runs}\) independent simulations of \(n_{\rm steps}\) time steps, record the maximum cut value per run, and compute the relative optimality gap (Eq.~\eqref{eq:opt_gap}).

We record \(n_{\rm runs} = 100\) independent runs per configuration to characterize the run-to-run variability of the reported metrics. The number of time steps \(n_{\rm steps}\) depends on the model and is chosen to ensure that the system has enough time to converge. See Supp.\ \S\ref{app:SimParams} for details on the choice of \(n_{\rm steps}\) for each model and \S\ref{app:convergence_tests} for convergence tests informing these parameter choices. 


\section{Qualitative study of model dynamics}

In this section, we qualitatively study the dynamics of the five Potts machine (PM) models presented in the previous section. The goal of this qualitative study is to identify dynamical properties that characterize the evolution of the system. We will later show that these properties indeed correlate with the performance of the models in solving Max-$k$-Cut.

\subsection{Selected dynamical properties for analysis}

We focus on the following three dynamical properties:

\textbf{Connected origin-to-solution branch} requires that the trivial fixed point at the origin is initially stable and bifurcates into a nontrivial solution branch that evolves continuously from the origin. 

\textbf{Enforced amplitude homogeneity} requires a mechanism that drives all spin amplitudes to the same value over time.


\textbf{Enforced phase discretization} requires a mechanism that drives the continuous spin phases to converge to the discrete set of phases corresponding to the Potts model states over time.



\subsection{Evaluation of dynamical properties}

To evaluate the dynamical properties of the Potts machine models, we analyze their governing equations under the annealing schemes specified in Table~\ref{tab:models}. We highlight key findings here and refer to the Supp.\ \S\ref{app:dynamical_properties} for the details of the evaluation. Table~\ref{tab:dynamic_properties} summarizes the evaluation for each model.

\begin{figure*}[t!]
    \centering
    \includegraphics[scale=.8]{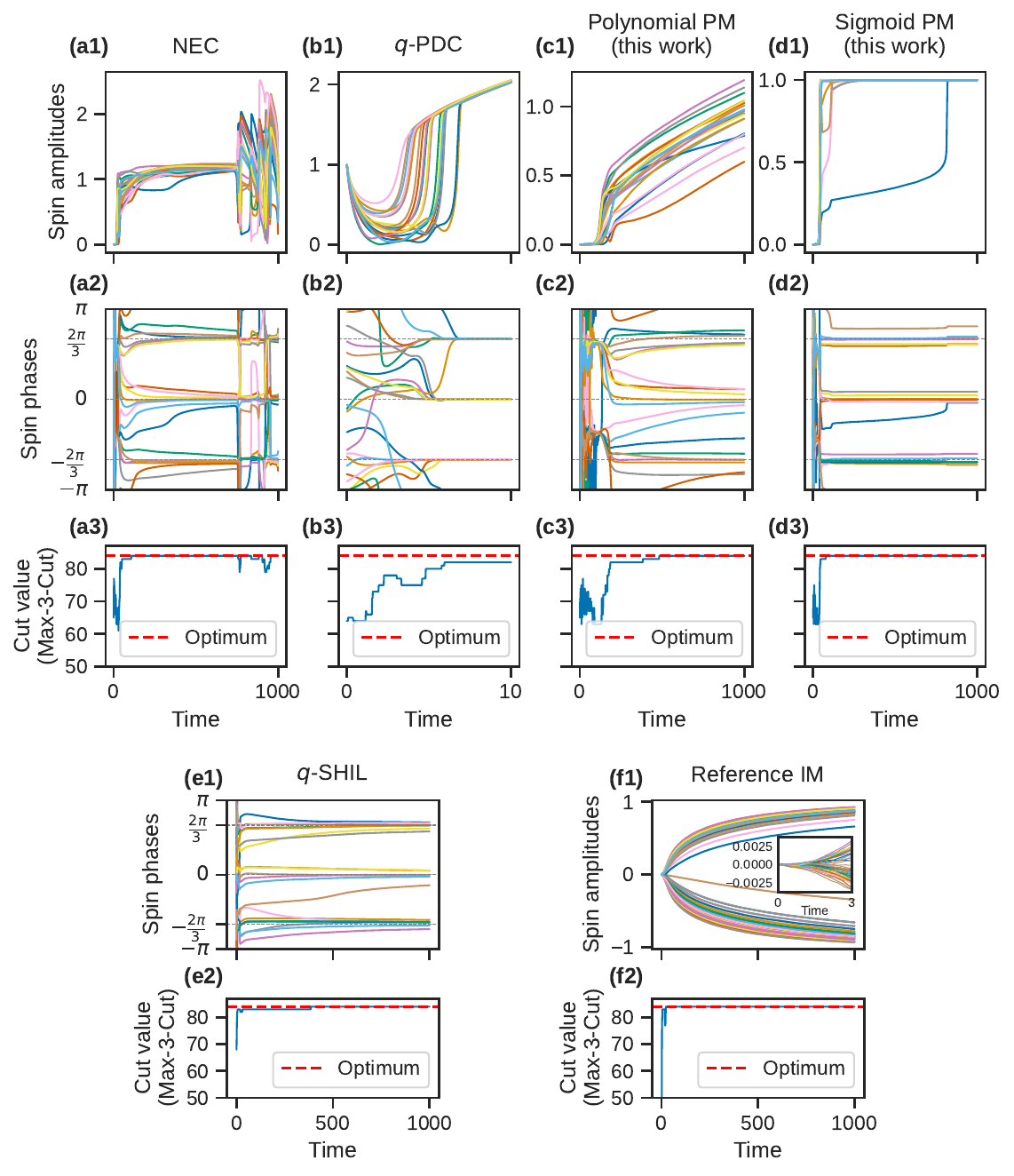}
    \caption{Time traces for each studied model on a 20-vertex Max-3-Cut problem from the ER50 set. We visualize the evolution of the amplitudes \textbf{(a1 - d1, f1)}, phases \textbf{(a2 - d2, e1)}, and the cut value \textbf{(a3 - d3, e2 - f2)} of the system. For the amplitude and phase plots, curves of different spins are shown in different colors, while the gray dashed lines indicate the discrete Potts phase values $\theta = 2\pi k / q$. For cut value plots, the red dashed line indicates the optimum cut value of the problem. For the Reference IM model, an inset shows its time trace from $t=0$ to $t=3$ to better visualize the initial dynamics. Default parameters listed in the Supp.\ \S\ref{app:SimParams} are used for each model.}
    \label{fig:dynamics}
\vspace{3em}
\end{figure*}

\begin{table*}[t]
\centering
\setlength{\extrarowheight}{3pt}
\caption{Summary of dynamical properties for each model.}
\label{tab:dynamic_properties}
\setlength{\tabcolsep}{4pt} 
\begin{tabular}{lccccc}
  \toprule
  & NEC & $q$-PDC & $q$-SHIL & Polynomial PM & Sigmoid PM  \\
  \midrule
  Connected origin-to-solution branch & \cmark & \xmark & NA & \cmark & \cmark  \\
  Enforced amplitude homogeneity & \cmark & \xmark & \cmark & \xmark & \cmark  \\
  Enforced phase discretization & \cmark & \cmark & \cmark & (\cmark) & (\cmark)  \\
  \bottomrule
\end{tabular}
\setlength{\tabcolsep}{6pt} 
\caption*{The evaluation of these properties is detailed in the Supp.\ \S\ref{app:dynamical_properties}. A checkmark, \cmark, indicates that the model satisfies the property, while a cross, \xmark, indicates that it does not. The parenthesized checkmark, (\cmark), indicates that the property can be satisfied under an alternative annealing scheme. NA indicates that the property is not applicable to the model.}
\vspace{-2em}
\end{table*}

Figure~\ref{fig:dynamics} shows the time traces from numerical simulation of the five PM models and the IM model on a sample Max-3-Cut problem, illustrating the dynamical properties in the evolution of the system. The numerical simulations are described in section~\ref{sec:numerical_methods} with default parameters specified in the Supp.\ \S\ref{app:SimParams}.

\textbf{The NEC model} possesses a connected origin-to-solution branch through gain-controlled bifurcation, enforces amplitude homogeneity via dynamic gain adjustment, and achieves enforced phase discretization through increasing coupling strength. These properties are visible in Figure~\ref{fig:dynamics}, where the amplitudes in panel~(a1) grow from the origin and, while not perfectly uniform, cluster more tightly than in the polynomial PM, and the phases in panel~(a2) converge to the discrete Potts values. The later portions of panels~(a1) and~(a2) also exhibit irregular oscillatory episodes in which settled configurations are disrupted, a behavior we discuss further in section~\ref{sec:nec_oscillatory_dynamics}.

\textbf{The $q$-PDC model} lacks a connected origin-to-solution branch because the origin remains stable throughout annealing, and it does not enforce amplitude homogeneity. Amplitude seeding and rapid increase of $\gamma$ nevertheless allow the spins to escape the origin, yielding enforced phase discretization. Panels~(b1) and~(b2) reflect these findings, with spins starting at a seeded nonzero amplitude and initially being attracted toward the origin until the phases approach a low-energy configuration, after which the spin phases converge to the discrete Potts values. Any apparent amplitude clustering in panel~(b1) at the end of the simulation reflects the particular trajectory rather than an enforcement mechanism. We also note the much shorter time axis in panels~(b1--b3), reflecting the fast, non-adiabatic annealing required by this model.

\textbf{The $q$-SHIL model} has intrinsically fixed and homogeneous amplitudes and enforces phase discretization. The connected-branch property is not applicable because the model has no origin fixed point. Accordingly, only phase dynamics are shown for this model, with the phases in panel~(e1) converging to the discrete Potts values.

\textbf{The polynomial PM model} exhibits a connected origin-to-solution branch but does not enforce amplitude homogeneity. Panel~(c1) shows the amplitudes growing from the origin but settling at visibly different values. Phase discretization depends on the annealing scheme. With nonlinear variation of $\gamma$ versus $\beta$, e.g. $\gamma = f_\gamma \beta^2$, the property can be satisfied. However, with linear variation of $\gamma$ versus $\beta$, as used in this work, the property is not satisfied. In panel~(c2), some spin phases have already settled at persistent nonzero offsets from the discrete Potts values, while others with larger offsets are still approaching their asymptotic offsets.

\textbf{The sigmoid PM model} shows a connected origin-to-solution branch and enforces amplitude homogeneity. Both properties are visible in panel~(d1), where the amplitudes grow from the origin and converge tightly to a common value, in contrast to the spread visible in panel~(c1) for the polynomial PM. However, as for the polynomial PM model, phase discretization depends on the annealing scheme. With nonlinear variation of $\gamma$ versus $\beta$ the property can be satisfied, but with linear variation as used in this work it is not, as the persistent phase offsets at the end of the simulation in panel~(d2) confirm.



\section{Benchmarking model performance}

In this section, we perform a quantitative study of the performance of the Potts- and Ising machine models in solving combinatorial optimization problems. We evaluate on two benchmark sets: (1) the GSet benchmark, consisting of 5 graphs with 800 vertices interpreted as either Max-3-Cut or Max-4-Cut problems (10 total problems), and (2) the ER50 benchmark, consisting of 60 random Max-3-Cut problem graphs with 5 to 50 vertices. The problem selection and encoding were discussed in section \ref{sec:problem_selection}.  

For the large GSet problems, we use relative optimality gap as the performance metric, since success rates are near-zero for these difficult problems. For the smaller ER50 problems, we use success rate, since models achieve significant success rates on these graphs. For each benchmark set, hyperparameters are optimized to minimize mean optimality gap (GSet) or maximize mean success rate (ER50) over a sweep of parameter combinations.

\begin{figure*}[!ht]
  \centering
  \includegraphics[scale=.75]{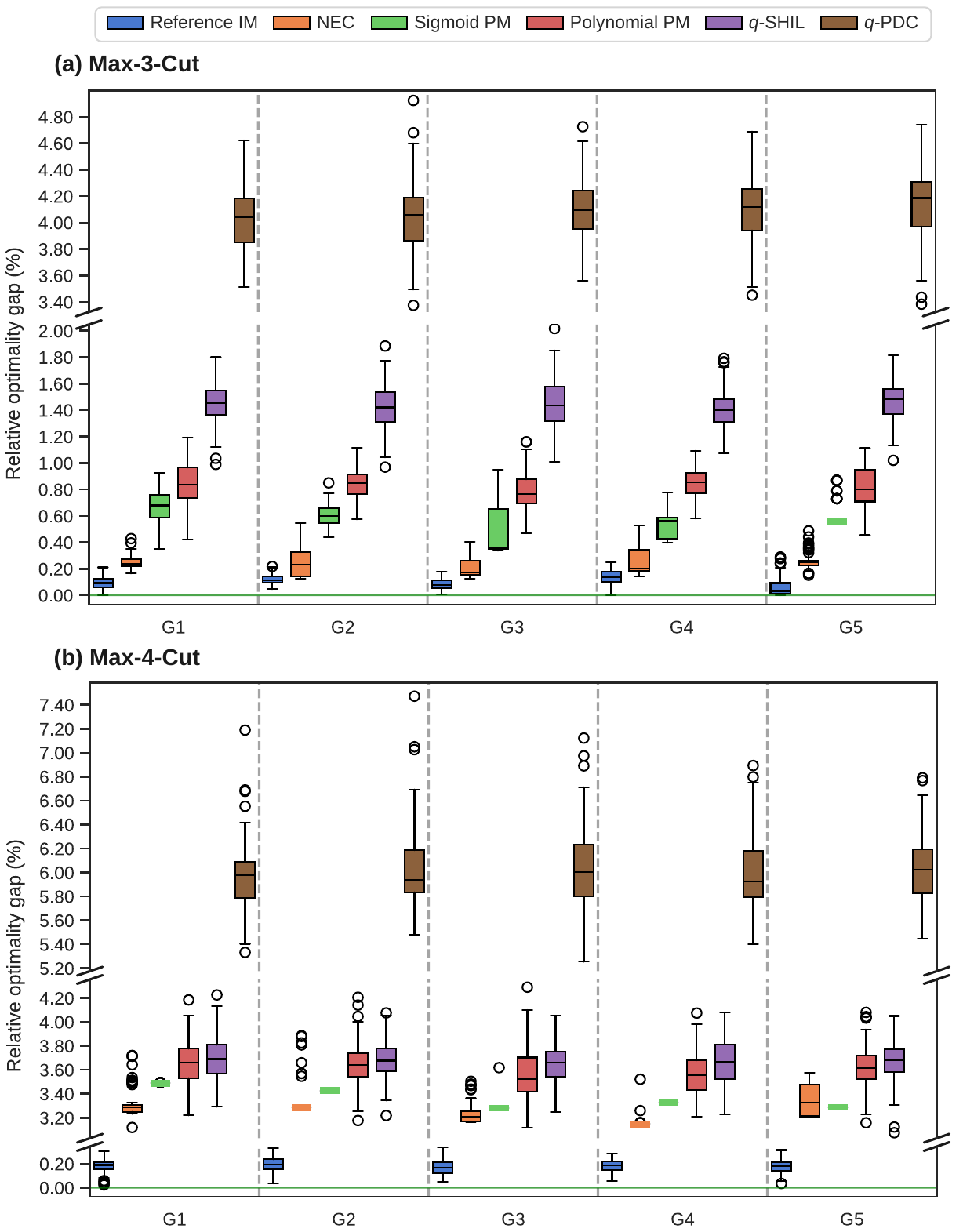}
  \caption{Relative optimality gap distributions for each model on \textbf{(a)} Max-3-Cut problems and \textbf{(b)} Max-4-Cut problems across GSet benchmark graphs G1-G5. Each model was evaluated using hyperparameters optimized for minimum mean optimality gap. Box plots show the distribution of relative optimality gaps (\%) across 100 independent runs per problem-model combination. Lower values indicate better performance, with 0\% representing the optimal solution (green horizontal line). The reference Ising machine consistently achieves the smallest optimality gaps, followed by the NEC Potts machine model. See Supp.\ \S\ref{app:benchmark_results} for detailed results, including hyperparameter tuning heatmaps for each model and graph, and \S\ref{app:SimParams} for hyperparameter sweep ranges.}
  \label{fig:gset_rel_gap}
\end{figure*}

Figure \ref{fig:gset_rel_gap} shows the relative optimality gap distributions for the GSet Max-3-Cut and Max-4-Cut problems. Supp.\ \S\ref{app:benchmark_results} provides more detailed results, including heatmaps of the hyperparameter tuning sweeps for each model and benchmark graph. For more details on the simulation settings and hyperparameter optimization sweeps, refer to the Supp.\ \S\ref{app:SimParams}.

The GSet results show that the reference IM model consistently achieves the smallest optimality gaps across all benchmark graphs, with median gaps around 0.1\% for Max-3-Cut and around 0.2\% for Max-4-Cut problems. The NEC model, which exhibits all three identified dynamical properties, follows closely behind the reference IM in performance for Max-3-Cut, achieving median gaps below 0.2\%, but the median gap increases to around 3.3\% for Max-4-Cut. The polynomial PM and sigmoid PM models show competitive performance compared to other PMs with median gaps around 0.7\% for Max-3-Cut and around 3.5\% for Max-4-Cut. The $q$-SHIL performs slightly worse than the polynomial and sigmoid PMs, with median gaps around 1.2\% for Max-3-Cut and around 4.5\% for Max-4-Cut. The $q$-PDC model trails with higher gaps of around 4\% for Max-3-Cut and around 6\% for Max-4-Cut. 

Note that in going from Max-3-Cut to Max-4-Cut, all Potts machine models show notably degraded performance. Interestingly, the reference IM maintains stable performance across both problem types. We attribute this degradation to a mismatch between the planar Potts Hamiltonian implemented by the PM dynamics and the standard Potts Hamiltonian that correctly represents Max-$k$-Cut. The two Hamiltonians are equivalent for $k \leq 3$, but for $k > 3$ the planar model imposes an angular separation constraint that is not part of the target problem. This is discussed in detail in section~\ref{sec:decreasing_performance_max3cut_max4cut}.

\begin{figure*}[!ht]
  \centering
  \includegraphics[scale=.75]{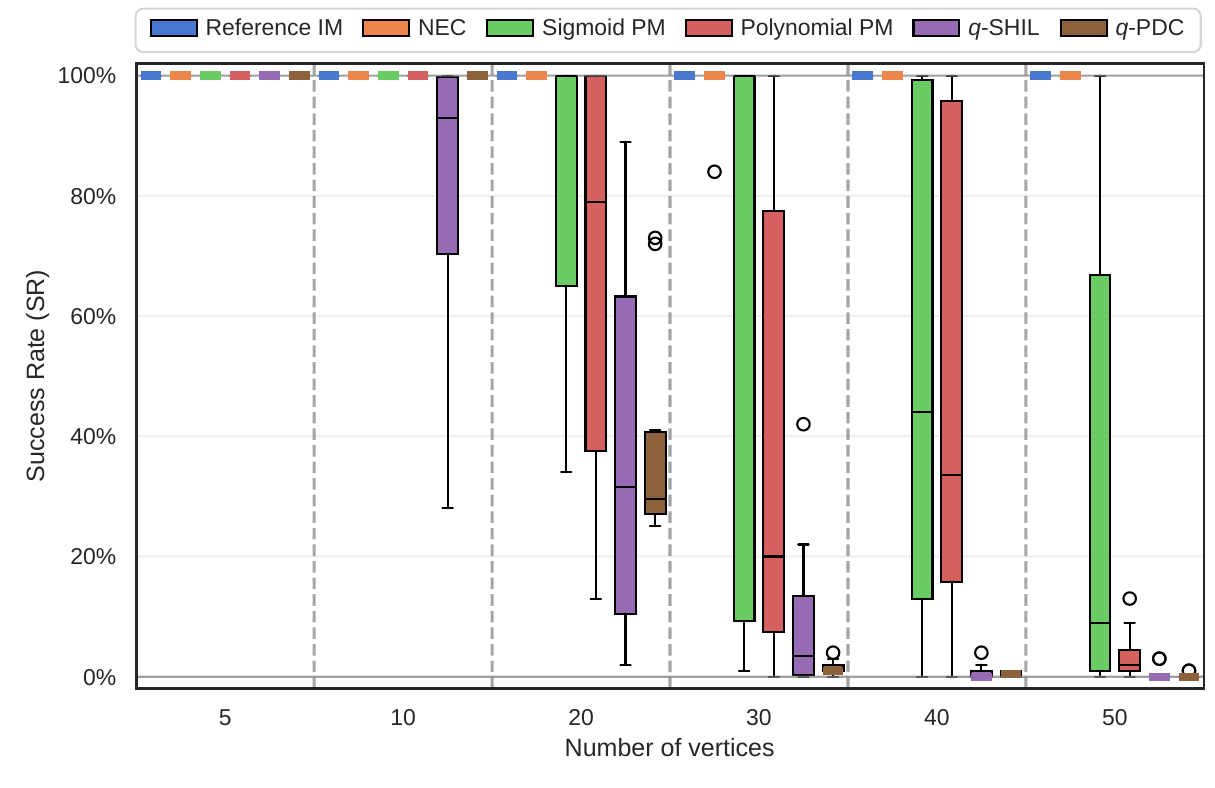}
  \caption{Success rate distributions for each model on ER50 Max-3-Cut benchmark problems (5--50 vertices, 10 instances per size). Each model was evaluated using hyperparameters optimized for maximum success rate, with 100 independent runs per graph-model combination. Box plots show the distribution of success rates across graphs of each size. The reference IM and NEC model achieve consistently high success rates across all graph sizes, while the $q$-PDC and $q$-SHIL models show lower performance, particularly for larger graphs. See Supp.\ \S\ref{app:SimParams} for hyperparameter sweep ranges.}
  \label{fig:g05_success_rate}
\end{figure*}

Figure~\ref{fig:g05_success_rate} shows the success rate distributions for the ER50 Max-3-Cut problems. The results confirm the performance ranking observed on the GSet benchmarks: the reference IM achieves near-perfect success rates across all graph sizes, followed by the NEC model. The polynomial and sigmoid PMs show intermediate performance, while the $q$-SHIL and $q$-PDC models exhibit lower success rates, in particular for larger problems.

Convergence tests presented in the Supp.\ \S\ref{app:convergence_tests} show that the reported performance metrics have saturated with respect to simulation time, time step, and annealing rate for all models.

\FloatBarrier

\clearpage

\section{Discussion}\label{sec:discussion}

\subsection{Relevance of identified dynamical properties}

The performance of a physical solver, such as a Potts machine (PM) or an Ising machine (IM), is determined by its system dynamics, which are governed by its underlying equations and operational parameters, such as the annealing schedule. In the following, we argue that the three identified dynamical properties are associated with higher performance on Max-$k$-Cut problems.
a
Our benchmarking results on both GSet and ER50 problems show that models satisfying more of these properties tend to perform better. The NEC model, which satisfies all three properties (see Figure~\ref{fig:dynamics}), consistently outperforms the other PMs. The polynomial and sigmoid PMs, each satisfying two of the three properties, also demonstrate competitive performance. Conversely, the $q$-PDC model, which satisfies only one property, performs poorly. The $q$-SHIL model fulfills the two applicable properties but lacks the connected origin-to-solution branch, which may be particularly important for performance, as will be discussed in section~\ref{sec:impact_of_origin_connection}. These findings suggest that the identified dynamical properties are conducive to the performance of PMs in solving Max-$k$-Cut problems.

The identified properties are not only relevant for the studied PM models, but could also be applied when evaluating new PM proposals or designing new architectures. Beyond the models studied here, these results suggest that dynamical analysis of the governing equations should complement physical and hardware feasibility considerations when designing and evaluating Potts machine architectures.

\subsection{Impact of origin-to-solution branch} \label{sec:impact_of_origin_connection}

The superior performance of the sigmoid PM over the $q$-SHIL model, despite both satisfying two dynamical properties, highlights that a connected origin-to-solution branch may be particularly important for performance. This property allows the system to traverse a low-amplitude regime where the interplay between stochastic noise and the deterministic drift term of the governing equations is pronounced, facilitating a more effective exploration of the energy landscape. In contrast, the $q$-SHIL model, which lacks amplitude dynamics and therefore lacks this property, operates in a fixed high-amplitude regime where the dynamics are dominated by the drift term, hindering landscape exploration. This contrast is visible in Figure~\ref{fig:dynamics}, where the sigmoid PM amplitudes in panel~(d1) rise from the origin and saturate, while the $q$-SHIL model in panel~(e1) evolves entirely at fixed amplitude without any low-amplitude transient. This suggests that the performance of fixed-amplitude models like $q$-SHIL might be improved by incorporating a noise annealing schedule, analogous to simulated annealing \cite{kirkpatrick_optimization_1983}, to emulate the exploratory benefits provided by an origin-connected solution branch.

\subsection{Oscillatory dynamics in the NEC model} \label{sec:nec_oscillatory_dynamics}

Panels~(a1) and~(a2) of Figure~\ref{fig:dynamics} show that the NEC model exhibits irregular oscillatory episodes at later times, in which settled spin configurations are disrupted and the system temporarily explores alternative phase assignments. These episodes appear to involve frustrated spin pairs, that is, antiferromagnetically coupled spins occupying the same Potts phase. The behavior is more pronounced at low target amplitudes, where the cubic saturation term in Eq.~\eqref{eq:nec} provides weaker restoring forces. A full characterization of this phenomenon and its relationship to the performance of the NEC model is left for future work.

\subsection{Underperformance of Potts machine models compared to Ising machine model}

A key finding of this study is the general underperformance of the studied PMs compared to the benchmark IM in terms of solution quality. This result is surprising, as PMs are specifically formulated to solve multi-state problems such as Max-$k$-Cut. The IM, conversely, requires a problem mapping that introduces significant overhead in the number of spins and couplings to represent the multi-state variables \cite{lucas_ising_2014}. This overhead introduces energy barriers between valid one-hot states and is generally expected to increase the ruggedness of the energy landscape and degrade performance \cite{prins_convex_2025}. The empirical rescaling of the external fields applied to our reference IM \cite{prins_how_2025} has been shown to partially mitigate these barriers \cite{prins_convex_2025}. Despite this overhead, the IM model consistently outperforms the PM models across both benchmark sets, whether measured by optimality gap (GSet) or success rate (ER50). However, this comparison is based on solution quality, and the ranking could be different when considering other metrics such as time-to-solution, which we have not studied here.

The underperformance of the PM models could be due to several factors. Significant research efforts have been dedicated to identifying high-performance IMs, more so than for PMs, which may partly explain the performance gap observed here. The IM model used in this work \cite{prins_how_2025} represents one of the most effective IMs in the literature. In contrast, the study of PMs remains in its early stages, so the most promising architectures and implementations may yet be undiscovered. By establishing a framework for characterizing PM dynamics and introducing new PM models for evaluation, this work provides a foundation for guiding such efforts toward more competitive PM architectures.

\subsection{Decreasing performance of Potts machine models going from Max-3-Cut to Max-4-Cut} \label{sec:decreasing_performance_max3cut_max4cut}

Our benchmarks show a significant performance degradation for all PM models when transitioning from Max-3-Cut to Max-4-Cut problems, a trend not observed for the reference IM model. This is especially surprising, as the representational overhead for an IM increases with $k$, while it remains constant for a PM. The explanation may lie in a fundamental mismatch between the objective function optimized by the studied PMs and the \textit{standard Potts model} that correctly represents Max-$k$-Cut. The governing equations of the studied PMs permit a Lyapunov function (i.e. energy landscape) with a coupling term proportional to $-\sum J_{ij}\cos(\theta_i - \theta_j)$, which depends on the angular separation between spin phases $\theta_i$ and $\theta_j$. This is, in fact, the Hamiltonian of the \textit{planar Potts model} \cite{wu_potts_1982}, which in general differs from the standard Potts model Hamiltonian, $-\sum J_{ij}\delta(s_i,s_j)$, which depends only on the discrete Potts states $s_i$ and $s_j$.


The Max-$k$-Cut cut value and the standard Potts Hamiltonian are defined in terms of the number of edges between pairs of spins in different states. The planar Potts Hamiltonian, on the other hand, is defined in terms of the angular separation between pairs of spins. The angular separation will be the same for all pairs of unique spin states for $k\leq 3$, but not for $k\geq 4$, where the angular separation between pairs of spin states depends on which pair is considered, i.e. whether they are adjacent, next-nearest neighbors, and so on. Therefore, for Max-$k$-Cut problems with $k\geq 4$, the considered PM models try not just to maximize the number of edges between pairs of spins in different states, but also to maximize the angular separation between pairs of connected spins. This constraint is not present in the reference IM model, given its one-hot encoding of the problems \cite{lucas_ising_2014,prins_how_2025}. The Ising formulation consists of two soft penalty terms that reduce to the standard Potts Hamiltonian when the one-hot constraints are satisfied (Supp.\ \S\ref{app:OneHotMapping}). This mismatch is likely the reason for the observed performance degradation from Max-3-Cut to Max-4-Cut.

Future PM designs should therefore aim for governing equations that do not impose unnecessary constraints on the target problem, such as the angular separation constraint identified here.

\section{Conclusion}
In this work, we have presented the first systematic comparison between analog Potts machines (PMs) for solving Max-3-Cut and Max-4-Cut problems and compared their performance to a reference Ising machine (IM) model. We studied three PM models from the literature: nonequilibrium condensate (NEC), $q$-photon downconversion ($q$-PDC), and $q$-th sub-harmonic injection locking ($q$-SHIL). In addition, we introduced two new PM models, the polynomial PM and the sigmoid PM. An analog IM with sigmoid nonlinearity and spin‐sign encoding was included as a reference. Through a combination of dynamical‐systems analysis and numerical benchmarking, we identified three key dynamical properties that are conducive to performance: a connected origin-to-solution branch, enforced amplitude homogeneity, and enforced phase discretization.

Our benchmarks on Max-3-Cut and Max-4-Cut instances reveal that, despite their native representation of multi‐state variables, all PMs exhibited larger mean and minimum optimality gaps than the reference IM, with the NEC model achieving the best performance among PMs. Both the polynomial and sigmoid PMs, which satisfy around two of the three dynamical properties, achieved competitive but still inferior solution quality relative to the NEC model, which satisfies all three properties. The $q$-PDC and $q$-SHIL models, crucially lacking a connected origin-to-solution branch, showed the poorest performance. Notably, all PMs suffered a pronounced performance drop when transitioning from Max-3-Cut to Max-4-Cut, whereas the IM did not show this degradation. We attribute this decline to an angular separation constraint arising from the planar Potts Hamiltonian implemented by the PM dynamics, which does not match the standard Potts model for Max-4-Cut and in general for Max-$k$-Cut with $k > 3$.

Two main insights emerge. First, solvers exhibiting more of the identified dynamical properties achieve higher solution quality, and these properties can therefore guide the design of future Potts machines. Second, the considered PM formulations require fundamental revision to implement the standard Potts Hamiltonian for state counts larger than three, as the planar variant introduces a systematic performance degradation for Max-$k$-Cut with $k > 3$ that cannot be addressed through parameter tuning alone.

More broadly, these findings motivate a dynamics-informed approach to PM design, in which analysis of the governing equations complements physical and hardware feasibility considerations in guiding architectural choices and parameter selection.



\section{Data availability}
The code used for the numerical simulations and the benchmarking procedure as well as the resulting data are available from the authors upon reasonable request. The GSet benchmark graphs are publicly available \cite{gset} and the best known cut values for the Max-3-Cut and Max-4-Cut problems are reported in the literature \cite{ma_multiple_2017}.

\section{Author contributions}
B.A.F. performed the simulations and wrote the manuscript. R.D.P. and P.B. supervised the project and provided critical feedback on the manuscript. All authors reviewed and approved the final manuscript.

\clearpage

\bibliographystyle{unsrtnat-custom}
\bibliography{references}  
\addcontentsline{toc}{section}{References}    

\newpage

\appendix
\onecolumn
\setcounter{equation}{0}
\setcounter{figure}{0}
\setcounter{table}{0}
\setcounter{page}{1}
\renewcommand{\thesection}{S\arabic{section}}
\renewcommand{\theequation}{\thesection.\arabic{equation}}
\renewcommand{\theHequation}{\theequation}
\renewcommand{\thefigure}{\thesection.\arabic{figure}}
\renewcommand{\thetable}{\thesection.\arabic{table}}
\renewcommand{\thepage}{S\arabic{page}}
\cleardoublepage
\phantomsection
\part*{Supplementary Materials}
\addcontentsline{toc}{part}{Supplementary Materials}
\label{sec:supplementary}

\section{Rescaling and deriving governing equations} \label{app:rescaling-rewriting}

\subsection{Non-Equilibrium Condensates (NEC) model} \label{app:RescalingNEC}
For the non-equilibrium condensate model, assuming coupling strengths are initialized to the target coupling coefficients ($\Upsilon_{ij} = J_{ij}$) and setting the exponent of the magnitude $|x_i|$ to $n-1$, the single-spin dynamics
equation reads \cite{kalinin_networks_2018}
\begin{equation}\frac{\dd x_{i}}{\dd t} = \alpha_{i}x_{i} - \sigma_{c}\left| x_{i} \right|^{n-1}x_{i} + h_{i}\left( x_{i}^{*} \right)^{q - 1} + \sum_{j}^{N}{J_{ij}x_{j}}\ \end{equation}
Note that in the original formulation, the exponent of $|x_i|$ is 2, corresponding to $n=3$. However, we shall see in section \ref{app:bounding} that a higher $n$ is required to ensure amplitude bounding when $q>3$, such as for Max-4-Cut problems, where $q=4$. Scaling the complex amplitude by \(x_{i}' = \sigma_{c}^{s}x_{i}\),
where \(s\) is currently unspecified, yields
\begin{equation}\sigma_{c}^{- s}\frac{\dd x_{i}'}{\dd t} = \sigma_{c}^{- s}\alpha_{i}x_{i}' - \sigma_{c}^{1 - sn}\left| x_{i}' \right|^{n-1}x_{i}' + \sigma_{c}^{(1 - q)s}h_{i}\left( {x_{i}'}^{*} \right)^{q - 1} + \sigma_{c}^{- s}\sum_{j}^{N}{J_{ij}x_{j}'}\end{equation}
\begin{equation}\Rightarrow \frac{\dd x_{i}'}{\dd t} = \alpha_{i}x_{i}' - \sigma_{c}^{1 - s(n-1)}\left| x_{i}' \right|^{n-1}x_{i}' + \sigma_{c}^{(2 - q)s}h_{i}\left( {x_{i}'}^{*} \right)^{q - 1} + \sum_{j}^{N}{J_{ij}x_{j}'}\end{equation}
It is evident that setting \(s = 1/(n-1)\) gives
\begin{equation}\frac{\dd x_{i}'}{\dd t} = \alpha_{i}x_{i}' - \left| x_{i}' \right|^{n-1}x_{i}' + \sigma_{c}^{\frac{2 - q}{n-1}}h_{i}\left( {x_{i}'}^{*} \right)^{q - 1} + \sum_{j}^{N}{J_{ij}x_{j}'}\end{equation}
With the effective parameter
\(\gamma = \sigma_{c}^{\frac{2 - q}{n-1}}h_{i}\), the equation now reads
\begin{equation}\label{eq:nec_rescale}
    \frac{\dd x_{i}'}{\dd t} = \alpha_{i}x_{i}' - \left| x_{i}' \right|^{n-1}x_{i}' + \gamma\left( {x_{i}'}^{*} \right)^{q - 1} + \sum_{j}^{N}{J_{ij}x_{j}'}
\end{equation}

\subsection{$q$-Photon Down-Conversion ($q$-PDC) model} \label{app:RescalingQPDC}
In the \(q\)-PDC model, the differential equation governing the
time-evolution of spin \(i\) reads \cite{honari-latifpour_combinatorial_2022}
\begin{equation}\frac{\dd x_{i}}{\dd t} = - \gamma_{i}x_{i} - g_{s}x_{i}\left| x_{i} \right|^{2(q - 1)} + g\left( x_{i}^{*} \right)^{q - 1} + \sum_{j}^{N}{J_{{ij}}x_{j}}\end{equation}
Assuming all spins share a uniform damping factor
\(\gamma_{i} = \gamma_{d}\), this corresponds to the polynomial model
with
\begin{equation}n = 2q - 1,\ \ c_{1} = \gamma_{d},\ \ c_{2} = g_{s},\ \ c_{3} = g,\ \ c_{4} = 1\end{equation}
The equation can therefore be rescaled as (see eq. \ref{eq:poly_pm_rescale})
\begin{equation} \label{eq:q-pdc_rescale}
\frac{\dd x_{i}'}{\dd t'} = - x_{i}' - x_{i}'\left| x_{i}' \right|^{2(q - 1)} + \gamma\left( {x_{i}^{*}}' \right)^{q - 1} + \beta\sum_{j}^{N}{J_{{ij}}x_{j}'},\end{equation}
\begin{equation}t' = \gamma_{d}t,\ \ x_{i}' = \left(\frac{g_{s}}{\gamma_{d}}\right)^{\frac{1}{n-1}}x_{i},\ \ \gamma = \frac{g}{\gamma_{d}}\left( \frac{g_{s}}{\gamma_{d}} \right)^{\frac{2 - q}{n - 1}},\ \ \beta = \frac{1}{\gamma_{d}}\end{equation}

\subsection{$q^\mathrm{th}$-harmonic Subharmonic Injection Locking ($q$-SHIL) model}\label{app:RescalingQSHIL}
In the \(q^{\text{th}}\)-harmonic subharmonic injection locking model,
the phase dynamics for oscillator \(i\) are described by the differential
equation
\begin{equation}\frac{\dd \theta_{i}}{\dd t} = -K_{s}\sin\left( q\theta_{i} \right) + K\sum_{j}^{N}{J_{{ij}}\sin{(\theta_{j} - \theta_{i})}}\end{equation}
Assuming \(K \neq 0\), rescaling the time by \(t' = Kt\) results in
\begin{equation}\frac{\dd \theta_{i}}{\dd t'} = -\frac{K_{s}}{K}\sin\left( q\theta_{i} \right) + \sum_{j}^{N}{J_{{ij}}\sin{(\theta_{j} - \theta_{i})}}\end{equation}
By introducing the effective parameter \(\gamma = K_{s}/K\), the
equation reads
\begin{equation}\frac{\dd \theta_{i}}{\dd t'} = -\gamma\sin\left( q\theta_{i} \right) + \sum_{j}^{N}{J_{{ij}}\sin{(\theta_{j} - \theta_{i})}}\end{equation}

\subsection{Polynomial Potts machine model}\label{app:RescalingPolynomial}
In the polynomial model, the differential equation governing the time-evolution of spin \(i\) reads
\begin{equation}
\frac{\dd x_{i}}{\dd t} = - c_{1}x_{i} - c_{2}x_{i}\left| x_{i} \right|^{n - 1} + c_{3}\left( x_{i}^{*} \right)^{q - 1} + c_{4}\sum_{j}^{N}{J_{{ij}}x_{j}},\ \ x_{i}\mathbb{\in C,\ \ }c_{1},c_{2},c_{3},c_{4}\mathbb{\in R} \setminus \{0\}\end{equation}
Scaling the time by \(t' = c_{1}t\) results in
\begin{equation}
\frac{\dd x_{i}}{\dd t'} = - x_{i} - \frac{c_{2}}{c_{1}}x_{i}\left| x_{i} \right|^{n - 1} + \frac{c_{3}}{c_{1}}\left( x_{i}^{*} \right)^{q - 1} + \frac{c_{4}}{c_{1}}\sum_{j}^{N}{J_{{ij}}x_{j}}\end{equation}
Scaling the spin complex amplitude \(x_{i}' = sx_{i}\), where $s$ is currently unspecified, gives
\begin{equation}
s^{- 1}\frac{\dd x_{i}'}{\dd t'} = - s^{- 1}x_{i}' - \frac{c_{2}}{c_{1}}s^{- n}x_{i}'\left| x_{i}' \right|^{n - 1} + \frac{c_{3}}{c_{1}}s^{- (q - 1)}\left( {x_{i}^{*}}' \right)^{q - 1} + \frac{c_{4}}{c_{1}}s^{- 1}\sum_{j}^{N}{J_{{ij}}x_{j}'}\end{equation}
\begin{equation}
\Rightarrow \frac{\dd x_{i}'}{\dd t'} = - x_{i}' - \frac{c_{2}}{c_{1}}s^{1 - n}x_{i}'\left| x_{i}' \right|^{n - 1} + \frac{c_{3}}{c_{1}}s^{2 - q}\left( {x_{i}^{*}}' \right)^{q - 1} + \frac{c_{4}}{c_{1}}\sum_{j}^{N}{J_{{ij}}x_{j}'}\end{equation}
It is now evident that setting \(s = \left( \frac{c_{2}}{c_{1}} \right)^{\frac{1}{n - 1}}\) such that \(s^{1 - n} = \left( \frac{c_{2}}{c_{1}} \right)^{- 1}\) yields
\begin{equation}
\frac{\dd x_{i}'}{\dd t'} = - x_{i}' - x_{i}'\left| x_{i}' \right|^{n - 1} + \frac{c_{3}}{c_{1}}\left( \frac{c_{2}}{c_{1}} \right)^{\frac{2 - q}{n - 1}}\left( {x_{i}^{*}}' \right)^{q - 1} + \frac{c_{4}}{c_{1}}\sum_{j}^{N}{J_{{ij}}x_{j}'}\end{equation}
Now, the following two effective parameters are introduced
\begin{equation}
\gamma = \frac{c_{3}}{c_{1}}\left( \frac{c_{2}}{c_{1}} \right)^{\frac{2 - q}{n - 1}},\ \ \beta = \frac{c_{4}}{c_{1}},\end{equation}
such that \(\gamma\) is the prefactor on the phase-discretizing term of
the equation, whereas \(\beta\) is the prefactor on the coupling term. The equation then reads
\begin{equation} \label{eq:poly_pm_rescale}
    \frac{\dd x_{i}'}{\dd t'} = - x_{i}' - x_{i}'\left| x_{i}' \right|^{n - 1} + \gamma\left( {x_{i}^{*}}' \right)^{q - 1} + \beta\sum_{j}^{N}{J_{{ij}}x_{j}'}
\end{equation}
In models that anneal by decreasing \(c_{1}\), this annealing corresponds to increasing $\beta$ and having
\(\gamma\) follow \(\beta\) via the expression
\begin{equation}\label{eq:gamma_rel}
    \gamma(\beta) = \frac{c_{3}}{c_{1}}\left( \frac{c_{2}}{c_{1}} \right)^{\frac{2 - q}{n - 1}} = \frac{\beta c_{3}}{c_{4}}\left( \frac{\beta c_{2}}{c_{4}} \right)^{\frac{2 - q}{n - 1}} = \frac{c_{3}}{c_{4}}\left( \frac{c_{2}}{c_{4}} \right)^{\frac{2 - q}{n - 1}}\beta^{1 + \frac{2 - q}{n - 1}} = f_{\gamma}\beta^{1 + \frac{2 - q}{n - 1}}
\end{equation}
Here, $f_\gamma$ is the $\gamma$-factor given by
\begin{equation}f_{\gamma} = \frac{c_{3}}{c_{4}}\left( \frac{c_{2}}{c_{4}} \right)^{\frac{2 - q}{n - 1}}\end{equation}

\subsection{Sigmoid Potts machine model}\label{app:PolynomialBreakingUp}\label{app:Sigmoid}

To derive the sigmoid Potts machine model presented in the main text section \ref{sec:sigmoid}, we start from the rescaled polynomial model equation given in eq. \ref{eq:poly_pm_rescale}, rewrite it in terms of the separate amplitude and phase equations, and then apply the sigmoid function to the amplitude equation only. 

For the rescaled polynomial model, the differential equation for spin
\(i\) is
\begin{equation}\frac{\dd x_i}{\dd t} = - x_{i} - x_{i}\left| x_{i} \right|^{n - 1} + \gamma\left( x_{i}^{*} \right)^{q - 1} + \beta\sum_{j}^{N}{J_{ij}x_{j}}\end{equation}
Expressing the complex amplitude by \(x_{i} = r_{i}e^{i\theta_{i}}\),
where \(r_{i}\) is its amplitude and \(\theta_{i}\) its phase, the left-hand
side of the equation is expanded using the product rule:
\begin{equation}\frac{\dd x_i}{\dd t} = e^{i\theta_{i}}\frac{\dd}{\dd t}r_{i} + r_{i}\frac{\dd}{\dd t}e^{i\theta_{i}} = e^{i\theta_{i}}\frac{\dd r_{i}}{\dd t} + ir_{i}e^{i\theta_{i}}\frac{\dd\theta_{i}}{\dd t} = e^{i\theta_{i}}\left( \frac{\dd r_{i}}{\dd t} + ir_{i}\frac{\dd\theta_{i}}{\dd t} \right)\end{equation}
When expressing \(x_{i}\) in terms of \(r_{i}\) and \(\theta_{i}\), the
right-hand side of the spin dynamics equation yields
\begin{equation}\frac{\dd x_i}{\dd t} = - r_{i}e^{i\theta_{i}} - e^{i\theta_{i}}\left| r_{i} \right|^{n} + \gamma r_{i}^{q - 1}e^{- i(q - 1)\theta_{i}} + \beta\sum_{j}^{N}{J_{ij}r_{j}e^{i\theta_{j}}}\end{equation}
Inserting the expanded left-hand side gives
\begin{equation}\frac{\dd x_i}{\dd t} = e^{i\theta_{i}}\left( \frac{\dd r_{i}}{\dd t} + ir_{i}\frac{\dd\theta_{i}}{\dd t} \right) = - r_{i}e^{i\theta_{i}} - e^{i\theta_{i}}\left| r_{i} \right|^{n} + \gamma r_{i}^{q - 1}e^{- i(q - 1)\theta_{i}} + \beta\sum_{j}^{N}{J_{ij}r_{j}e^{i\theta_{j}}}\end{equation}
\begin{equation}\Rightarrow \frac{\dd r_{i}}{\dd t} + ir_{i}\frac{\dd\theta_{i}}{\dd t} = - r_{i} - \left| r_{i} \right|^{n} + \gamma r_{i}^{q - 1}e^{- iq\theta_{i}} + \beta\sum_{j}^{N}{J_{ij}r_{j}e^{i(\theta_{j} - \theta_{i})}}\end{equation}
\begin{equation}= - r_{i} - \left| r_{i} \right|^{n} + \gamma r_{i}^{q - 1}\left\lbrack \cos{(q\theta_i)} - i\sin{(q\theta_i)} \right\rbrack + \beta\sum_{j}^{N}{J_{ij}r_{j}\left\lbrack \cos\left( \theta_{j} - \theta_{i} \right) + i\sin{(\theta_{j} - \theta_{i})} \right\rbrack}\end{equation}
From this, we can separate the real and imaginary parts:
\begin{equation}\label{eq:polynomial_amplitude_eq_rescaled}
    \frac{\dd r_{i}}{\dd t} = - r_{i} - \left| r_{i} \right|^{n} + \gamma r_{i}^{q - 1}\cos(q\theta_i) + \beta\sum_{j}^{N}{J_{ij}r_{j}\cos\left( \theta_{j} - \theta_{i} \right)},
\end{equation}
and
\begin{equation}ir_{i}\frac{\dd\theta_{i}}{\dd t} = -i\gamma r_{i}^{q - 1}\sin{(q\theta_i)} + i\beta\sum_{j}^{N}{J_{ij}r_{j}\sin{(\theta_{j} - \theta_{i})}}\end{equation}
\begin{equation}\label{eq:poly_phase_rescaled}
    \Rightarrow \frac{\dd\theta_{i}}{\dd t} = -\gamma r_{i}^{q - 2}\sin{(q\theta_i)} + \frac{\beta}{r_{i}}\sum_{j}^{N}{J_{ij}r_{j}\sin{(\theta_{j} - \theta_{i})}}
\end{equation}
Note that to divide by \(r_{i}\) in the phase equation, \(r_{i} \neq 0\) must hold, meaning that for spins with zero amplitude, the phase is undefined.

By separating the amplitude and phase parts of the governing equation like this, the sigmoid saturation of the amplitudes can be introduced without affecting phase dynamics. Similarly to how the reference Ising machine differs from the polynomial Ising machine \cite{bohm_order--magnitude_2021}, the polynomial amplitude equation is altered by placing the coupling term, the phase-discretizing term, and a linear term with prefactor $\alpha$ inside a hyperbolic tangent nonlinearity:
\begin{equation} \label{eq:sigmoid_derivation}
    \frac{\dd r_{i}}{\dd t} = - r_{i} + \tanh\left( \alpha r_{i} + \gamma r_{i}^{q - 1}\cos\left( q\theta_{i} \right) + \beta\sum_{j}^{N}{J_{{ij}}r_{j}\cos\left( \theta_{j} - \theta_{i} \right)} \right)
\end{equation}

\subsection{One-hot encoding of Max-\texorpdfstring{$k$}{k}-Cut and the standard Potts Hamiltonian}\label{app:OneHotMapping}

The reference IM encodes Max-$k$-Cut using the graph coloring formulation of Lucas \cite{lucas_ising_2014}, with the external-field rescaling introduced by Prins \emph{et al.} \cite{prins_how_2025}. Here we show that, when the one-hot constraints are satisfied, the resulting cost function reduces to the standard Potts Hamiltonian.

Each vertex $u$ of the graph is represented by $k$ binary variables $x_{u,1}, \ldots, x_{u,k} \in \{0,1\}$, where $x_{u,c} = 1$ indicates that vertex $u$ is assigned color $c$. The one-hot constraint requires that exactly one variable per vertex is active,
\begin{equation}\label{eq:one_hot_constraint}
    \sum_{c=1}^{k} x_{u,c} = 1, \quad \forall\, u.
\end{equation}
The cost function to be minimized consists of a penalty term enforcing the one-hot constraint and an objective term penalizing monochromatic edges \cite{lucas_ising_2014,prins_how_2025},
\begin{equation}\label{eq:qubo_maxkcut}
    H = A \sum_{u} \left(1 - \sum_{c=1}^{k} x_{u,c}\right)^{\!2}
      + B \sum_{(u,v) \in E} \sum_{c=1}^{k} x_{u,c}\, x_{v,c},
\end{equation}
where $A > 0$ and $B > 0$ control the relative strengths of the two terms. Both terms are soft penalties, meaning that neither the one-hot condition nor the edge penalty is enforced structurally in the dynamics of the machine.

When the one-hot constraints are satisfied, the penalty term vanishes and exactly one variable per vertex equals 1. The product $x_{u,c}\,x_{v,c}$ then equals 1 if and only if both vertices share color $c$, so that
\begin{equation}\label{eq:kronecker_identity}
    \sum_{c=1}^{k} x_{u,c}\, x_{v,c} = \delta(s_u,\, s_v),
\end{equation}
where $s_u$ and $s_v$ denote the colors of vertices $u$ and $v$ and $\delta$ is the Kronecker delta. Substituting into Eq.~\eqref{eq:qubo_maxkcut} yields
\begin{equation}\label{eq:qubo_reduced}
    H\big|_{\text{one-hot}}
      = B \sum_{(u,v) \in E} \delta(s_u,\, s_v),
\end{equation}
which matches the standard Potts Hamiltonian (Eq.~\eqref{eq:std_potts} in the main text) with effective coupling $J_{uv} = -B$ for $(u,v) \in E$. Minimizing this expression is equivalent to maximizing the Max-$k$-Cut cut value. The cost function treats all pairs of distinct colors equally, with no dependence on angular separation, which is why the reference IM does not exhibit the $k \geq 4$ performance degradation observed for the PM models.

The binary variables are mapped to Ising spins $\sigma_{u,c} \in \{-1, +1\}$ via $x_{u,c} = (\sigma_{u,c} + 1)/2$, which introduces coupling coefficients and external bias fields into the Ising Hamiltonian. Following Prins \emph{et al.} \cite{prins_how_2025}, the external fields are rescaled by a factor $\zeta$ to correct for imbalances between the coupling and constraint terms that arise with continuous spin amplitudes, as described in the main text.
\section{Stability and fixed-point analysis} \label{app:stability}
The stability of a given fixed point of a system of equations can be
determined from the eigenvalues of the Jacobian matrix \(\mathcal{U}\)
at the fixed point. A fixed point \(\mathbf{x^{*}}\) is defined as a point
where the time-derivative of all variables is zero:
\begin{equation}\left. \ \frac{\dd \mathbf{x}}{\dd t} \right|_{\mathbf{x} = \mathbf{x^{*}}} = \mathbf{0}\end{equation}
The Jacobian matrix is the matrix of all first-order
derivatives of the system of equations
\begin{equation}\mathcal{U}\mathbf{=}\left\lbrack \begin{matrix}
\frac{\partial}{\partial x_{1}}\frac{\dd\mathbf{x}}{\dd t} & \frac{\partial}{\partial x_{2}}\frac{\dd\mathbf{x}}{\dd t} & 
\cdots & \frac{\partial}{\partial x_{N}}\frac{\dd\mathbf{x}}{\dd t}
\end{matrix} \right\rbrack =\begin{bmatrix}
\frac{\partial}{\partial x_{1}}\frac{\dd x_{1}}{\dd t} & \mathbf{\cdots} & \frac{\partial}{\partial x_{N}}\frac{\dd x_{1}}{\dd t} \\
\mathbf{\vdots} & \mathbf{\ddots} & \mathbf{\vdots} \\
\frac{\partial}{\partial x_{1}}\frac{\dd x_{N}}{\dd t} & \mathbf{\cdots} & \frac{\partial}{\partial x_{N}}\frac{\dd x_{N}}{\dd t}
\end{bmatrix}\mathbf{,}\end{equation}
with elements
\begin{equation}\mathcal{U}_{ij} = \frac{\partial}{\partial x_{j}}\frac{\dd x_{i}}{\dd t}\end{equation}
When evaluated at a fixed point, a negative eigenvalue indicates that the fixed point is attracting in the direction of the eigenvector corresponding to the eigenvalue. Vice versa, a positive eigenvalue indicates that the fixed point is repelling in the direction of the eigenvector corresponding to the eigenvalue. Therefore, when all eigenvalues are negative, the fixed point is attracting in all directions and is said to be stable. When the principal eigenvalue of the Jacobian matrix evaluated at a fixed point changes from negative to positive, that fixed point becomes unstable.

\subsection{Stability of the origin fixed point}\label{app:StabilityOrigin}
Here, the stability of the origin fixed point
\(\mathbf{x^{*} = 0}\) is analyzed for the different Potts machine
models presented in the main text, except for the $q$-SHIL model, which does not have amplitudes, and thus no origin. For all other Potts machine models, the origin is a fixed point, as \(\left.\ \dd\mathbf{x}/\dd t \right|_{\mathbf{x} = \mathbf{0}} = \mathbf{0}\). 

\subsubsection[NEC, q-PDC, and polynomial Potts machine models]{NEC, \(q\)-PDC, and polynomial Potts machine models}\label{app:StabilityPolynomial} \label{app:StabilityNEC} \label{app:StabilityQPDC}
The NEC, \(q\)-PDC, and polynomial Potts machine governing equations can be written in the form of equation \ref{eq:poly_pm_rescale} with a linear term prefactor $\alpha_i$:
\begin{equation}\frac{\dd x_{i}}{\dd t} = \alpha_i x_{i} - x_{i}\left| x_{i} \right|^{n - 1} + \gamma\left( x_{i}^{*} \right)^{q - 1} + \beta\sum_{l}^{N}{J_{il}x_{l}},\end{equation}
where for the NEC model, $\alpha_i$ is given by the dynamic gain equation \eqref{eq:nec_alpha} and $\beta = 1$. For the \(q\)-PDC and polynomial Potts machine models, $\alpha_i = -1 \forall i$. In the following, the case where all spin-dependent gains are set to the same value \(\alpha_{i} = \alpha\forall i\) is considered. Assuming $n > 1$ and $q > 2$, linearizing the system of equations around the origin ($\abs{x_i} \ll 1$) by neglecting the terms of higher than first order in \(x_{i}\) gives
\begin{equation}\label{eq:linearized_general}
    \frac{\dd \mathbf{x}}{\dd t} = \alpha \mathbf{x} + \beta\mathbf{Jx} = \left( \alpha \mathbf{I} + \beta\mathbf{J} \right)\mathbf{x} = \left. \ \mathcal{U} \right|_{\mathbf{x} = \mathbf{0}} \mathbf{x},
\end{equation}
where \(\mathbf{I}\) is the \(N \times N\) identity matrix, $\mathbf{J}$ is the coupling matrix with elements \(J_{ij}\), and \(\left. \ \mathcal{U} \right|_{\mathbf{x} = \mathbf{0}} \) is the Jacobian matrix evaluated at the origin fixed point. Eigenvalues of the Jacobian matrix at the origin fixed point are therefore
\begin{equation}\lambda_{k} = \alpha + \beta\mu_{k},\end{equation}
where \(\mu_{k}\) is the \(k^\text{th}\) eigenvalue of the coupling matrix. The principal eigenvalue of the coupling matrix is denoted \(\mu_{\max}\), such that given \(\beta > 0\), the principal eigenvalue of the Jacobian matrix is
\begin{equation}\lambda_{\max} = \alpha + \beta\mu_{\max},\end{equation}
While $\lambda_{\max} < 0$, the origin is stable, as all other eigenvalues are smaller than the principal eigenvalue and thus also negative. The principal eigenvalue changes sign from negative to positive at the threshold given by
\begin{equation} \label{eq:GeneralAlphaBetaThreshold}
    0 = \alpha_\mathrm{th} + \beta\mu_{\max} \Rightarrow \alpha_\mathrm{th} = - \beta\mu_{\max}\ \text{ or }\ \beta_\mathrm{th} = -\frac{\alpha}{\mu_{\max}},
\end{equation}
depending on which of the two parameters, $\alpha$ or $\beta$, is varied during annealing. As the value of \(\lambda_{\max}\) increases with \(\alpha\) and \(\beta\), the origin is therefore stable for \(\alpha < \alpha_\mathrm{th}\) (or \(\beta < \beta_\mathrm{th}\)) and unstable for \(\alpha > \alpha_\mathrm{th}\) (or \(\beta > \beta_\mathrm{th}\)). When the gain-modulating equation of the NEC model is introduced, the origin might not remain stable indefinitely as $\alpha$ will increase automatically while $r_\mathrm{th} > 0$. Yet the origin is still attracting in all directions for \(\alpha < \alpha_\mathrm{th}\) and repellent in at least one
direction for \(\alpha > \alpha_\mathrm{th}\).

\subsubsection{Sigmoid Potts machine model}\label{app:StabilitySigmoid}
The sigmoid model is governed by separate equations for the spin
amplitude and spin phase:
\begin{equation}\frac{\dd r_{i}}{\dd t} = - r_{i} + \tanh\left( \alpha r_{i} + \gamma r_{i}^{q - 1}\cos\left( q\theta_{i} \right) + \beta\sum_{j}^{N}{J_{ij}r_{j}\cos\left( \theta_{j} - \theta_{i} \right)} \right)\end{equation}
\begin{equation}\frac{\dd \theta_{i}}{\dd t} = -\gamma r_{i}^{q - 2}\sin{(q\theta_{i})} + \frac{\beta}{r_{i}}\sum_{j}^{N}{J_{ij}r_{j}\sin{(\theta_{j} - \theta_{i})}}\end{equation}
To simplify notation the following quantities are introduced:
\begin{equation}R_{i} = \alpha r_{i} + \gamma r_{i}^{q - 1}\cos\left( q\theta_{i} \right) + \beta\sum_{j}^{N}{J_{ij}r_{j}\cos\left( \theta_{j} - \theta_{i} \right)}\end{equation}
\begin{equation}S_{i} = \frac{\dd r_{i}}{\dd t} + r_{i} = \tanh{(R_{i})}\end{equation}
\begin{equation}T_{i} = r_{i}\frac{\dd \theta_{i}}{\dd t} = -\gamma r_{i}^{q - 1}\sin{(q\theta_{i})} + \beta\sum_{j}^{N}{J_{ij}r_{j}\sin{(\theta_{j} - \theta_{i})}}\end{equation}
It is noted that at the origin \(\mathbf{x} = \mathbf{r} = \mathbf{0}\),
all these quantities are zero:
\begin{equation}\left. \ R_{i} \right|_{\mathbf{x} = \mathbf{0}} = 0,\ \ \left. \ S_{i} \right|_{\mathbf{x} = \mathbf{0}} = \left. \ \tanh R_{i} \right|_{\mathbf{x} = \mathbf{0}} = 0,\ \ \left. \ T_{i} \right|_{\mathbf{x} = \mathbf{0}} = 0\end{equation}
As shown in section \ref{app:PolynomialBreakingUp}, the
amplitude and phase equations can be combined into an equation for the
complex amplitude by
\begin{equation}\frac{\dd x_i}{\dd t} = e^{i\theta_{i}}\left( \frac{\dd r_{i}}{\dd t} + ir_{i}\frac{\dd \theta_{i}}{\dd t} \right) = e^{i\theta_{i}}\left( S_{i} - r_{i} + iT_{i} \right) = - x_{i} + e^{i\theta_{i}}\left( S_{i} + iT_{i} \right)\end{equation}
Element \(ij\) of the Jacobian matrix is therefore
\begin{equation}\mathcal{U}_{ij} = \frac{\partial}{\partial x_{j}}\frac{\dd x}{\dd t} = - \delta_{ij} + \frac{\partial}{\partial x_{j}}e^{i\theta_{i}}S_{i} + \frac{\partial}{\partial x_{j}}ie^{i\theta_{i}}T_{i}\end{equation}
The term with \(S_{i}\) can be expanded with the product rule:
\begin{equation}\frac{\partial}{\partial x_{j}}e^{i\theta_{i}}S_{i} = S_{i}\frac{\partial}{\partial x_{j}}e^{i\theta_{i}} + e^{i\theta_{i}}\frac{\partial}{\partial x_{j}}S_{i}\end{equation}
At the origin, only the second of these terms is non-zero due to
\(S_{i}\) being zero at the origin:
\begin{equation}\left. \ \frac{\partial}{\partial x_{j}}e^{i\theta_{i}}S_{i} \right|_{\mathbf{x} = \mathbf{0}} = \left. \ S_{i}\frac{\partial}{\partial x_{j}}e^{i\theta_{i}} \right|_{\mathbf{x} = \mathbf{0}} + \left. \ e^{i\theta_{i}}\frac{\partial}{\partial x_{j}}S_{i} \right|_{\mathbf{x} = \mathbf{0}} = \left. \ e^{i\theta_{i}}\frac{\partial}{\partial x_{j}}S_{i} \right|_{\mathbf{x} = \mathbf{0}}\end{equation}
The partial derivative of \(S_{i} = \tanh R_{i}\) can be expanded using
the chain rule:
\begin{equation}\frac{\partial}{\partial x_{j}}S_{i} = \left( \frac{\partial}{\partial R_{i}}S_{i} \right)\left( \frac{\partial}{\partial x_{j}}R_{i} \right)\end{equation}
The first factor simply takes unit value at the origin:
\begin{equation}\frac{\partial}{\partial R_{i}}S_{i} = \textrm{sech}^{2}R_{i},\ \ \left. \ \frac{\partial}{\partial R_{i}}S_{i} \right|_{\mathbf{x} = \mathbf{0}} = 1\end{equation}
From this follows that
\begin{equation}\left. \ \frac{\partial}{\partial x_{j}}S_{i} \right|_{\mathbf{x} = \mathbf{0}} = \left. \ \frac{\partial}{\partial x_{j}}R_{i} \right|_{\mathbf{x} = \mathbf{0}}\end{equation}
and
\begin{equation}\left. \ \frac{\partial}{\partial x_{j}}e^{i\theta_{i}}S_{i} \right|_{\mathbf{x} = \mathbf{0}} = \left. \ e^{i\theta_{i}}\frac{\partial}{\partial x_{j}}S_{i} \right|_{\mathbf{x} = \mathbf{0}} = \left. \ e^{i\theta_{i}}\frac{\partial}{\partial x_{j}}R_{i} \right|_{\mathbf{x} = \mathbf{0}}\end{equation}
Now, it is noted that
\begin{equation}\left. \ \frac{\partial}{\partial x_{j}}e^{i\theta_{i}}R_{i} \right|_{\mathbf{x} = \mathbf{0}} = \left. \ R_{i}\frac{\partial}{\partial x_{j}}e^{i\theta_{i}} \right|_{\mathbf{x} = \mathbf{0}} + \left. \ e^{i\theta_{i}}\frac{\partial}{\partial x_{j}}R_{i} \right|_{\mathbf{x} = \mathbf{0}} = \left. \ e^{i\theta_{i}}\frac{\partial}{\partial x_{j}}R_{i} \right|_{\mathbf{x} = \mathbf{0}},\end{equation}
such that the above is equivalent to
\begin{equation}\left. \ \frac{\partial}{\partial x_{j}}e^{i\theta_{i}}S_{i} \right|_{\mathbf{x} = \mathbf{0}} = \left. \ \frac{\partial}{\partial x_{j}}e^{i\theta_{i}}R_{i} \right|_{\mathbf{x} = \mathbf{0}}\end{equation}
Turning our attention back to the Jacobian matrix, its elements at the
origin are given by
\begin{align}
\begin{split}
\mathcal{U}_{ij} \bigg|_{\mathbf{x} = \mathbf{0}} &= - \delta_{ij} 
+ \left. \frac{\partial}{\partial x_{j}} \left(e^{i\theta_{i}} S_{i}\right) \right|_{\mathbf{x} = \mathbf{0}} 
+ \left. \frac{\partial}{\partial x_{j}} \left(i e^{i\theta_{i}} T_{i}\right) \right|_{\mathbf{x} = \mathbf{0}} \\
&= - \delta_{ij} 
+ \left. \frac{\partial}{\partial x_{j}} \left(e^{i\theta_{i}} R_{i}\right) \right|_{\mathbf{x} = \mathbf{0}} 
+ \left. \frac{\partial}{\partial x_{j}} \left(i e^{i\theta_{i}} T_{i}\right) \right|_{\mathbf{x} = \mathbf{0}}
\end{split}
\end{align}
Combining the last two terms yields
\begin{equation}\left. \ \mathcal{U}_{ij} \right|_{\mathbf{x} = \mathbf{0}} = - \delta_{ij} + \left. \ \frac{\partial}{\partial x_{j}}e^{i\theta_{i}}(R_{i} + iT_{i}) \right|_{\mathbf{x} = \mathbf{0}}\end{equation}
Most of the work in this derivation thus far was about showing that
\(R_{i}\) can substitute \(S_{i}\) in the above expression. This
was done as \(e^{i\theta_{i}}(R_{i} + iT_{i})\) is easily differentiable
with respect to \(x_{j}\), whereas \(e^{i\theta_{i}}(S_{i} + iT_{i})\)
is not.
To see this, first the sum \(R_{i} + iT_{i}\) is calculated:
\begin{equation}R_{i} + iT_{i} = \alpha r_{i} + \gamma r_{i}^{q - 1}e^{- iq\theta_{i}} + \beta\sum_{j}^{N}{J_{ij}r_{j}e^{i\left( \theta_{j} - \theta_{i} \right)}}\end{equation}
Factoring in \(e^{i\theta_{i}}\) yields an expression in terms of
complex amplitudes instead of amplitudes and phases:
\begin{equation}e^{i\theta_{i}}\left( R_{i} + iT_{i} \right) = \alpha x_{i} + \gamma r_{i}^{q - 1}e^{- i(q - 1)\theta_{i}} + \beta\sum_{j}^{N}{J_{ij}r_{j}e^{i\theta_{j}}} = \alpha x_{i} + \gamma\left( x_{i}^{*} \right)^{q - 1} + \beta\sum_{j}^{N}{J_{ij}x_{j}}\end{equation}
Taking the partial derivative is now straightforward:
\begin{equation}\frac{\partial}{\partial x_{j}}e^{i\theta_{i}}\left( R_{i} + iT_{i} \right) = \alpha\delta_{ij} + \beta J_{ij}\end{equation}
The Jacobian matrix elements at the origin are therefore
\begin{equation}\left. \ \mathcal{U}_{ij} \right|_{\mathbf{x} = \mathbf{0}} = (\alpha - 1)\delta_{ij} + \beta J_{ij},\end{equation}
corresponding to Jacobian matrix
\begin{equation}\left. \ \mathcal{U} \right|_{\mathbf{x} = \mathbf{0}} = (\alpha - 1)\mathbf{I} + \beta\mathbf{J}\end{equation}
with eigenvalues
\begin{equation}\lambda_{k} = \alpha - 1 + \beta\mu_{k}\end{equation}
Assuming \(\beta > 0\), the principal eigenvalue is
\begin{equation}\lambda_{\max} = \alpha - 1 + \beta\mu_{\max}\end{equation}
The destabilization threshold where \(\lambda_{\max} = 0\) can then either
be expressed in terms of \(\alpha\) or \(\beta\):
\begin{equation}\label{eq:sigmoid_beta_threshold}
    \alpha_\mathrm{th} = 1 - \beta\mu_{\max},\ \ \beta_\mathrm{th} = \frac{1 - \alpha}{\mu_{\max}}
\end{equation}

\subsection{Evolution of fixed points following origin destabilization} \label{app:FixedPointEvolution}

\subsubsection{NEC, q-PDC, and polynomial Potts machine models} \label{app:FixedPointEvolutionPolynomial}

For the NEC, $q$-PDC, and polynomial Potts machine models, the governing equations linearized around the origin (Eq. \ref{eq:linearized_general}) take the form
\begin{equation}
    \frac{\dd \mathbf{x}}{\dd t} = \left. \ \mathcal{U} \right|_{\mathbf{x} = \mathbf{0}}  \mathbf{x},
\end{equation}
where $\left. \ \mathcal{U} \right|_{\mathbf{x} = \mathbf{0}} $ is the Jacobian matrix evaluated at the origin, given by
\begin{equation}
    \left. \ \mathcal{U} \right|_{\mathbf{x} = \mathbf{0}}  = \alpha \mathbf{I} + \beta \mathbf{J}.
\end{equation}
The first eigenvalue of the Jacobian matrix that crosses zero and destabilizes the origin is the principal eigenvalue $\lambda_{\max} = \alpha + \beta \mu_{\max}$, with corresponding eigenvector $\mathbf{v}_{\max}$. Therefore, when the origin becomes unstable, the system will evolve along the direction of $\mathbf{v}_{\max}$, such that
\begin{equation}
    \mathbf{x}(t) = c(t) \mathbf{v}_{\max},
\end{equation}
where $c(t) \in \mathbb{C}$ is a time-dependent complex-valued coefficient. As the coupling matrix is real-valued and symmetric, the eigenvector $\mathbf{v}_{\max}$ can be chosen to be real-valued. Therefore, all spin complex amplitudes will initially evolve away from the origin either parallel (if $v_{\max,i} > 0$) or anti-parallel (if $v_{\max,i} < 0$) to the complex coefficient $c(t)$, such that the phases are $\theta_i \in \left\{\arg(c), \arg(c) + \pi\right\} \forall i$. We note that $\lambda_{\max} = \alpha + \beta\mu_{\max}$ and write $v_i = v_{\max,i}$ for brevity. To describe the evolution of $c(t)$ after the origin destabilization, the higher-order terms must be included in the governing equations:
\begin{equation}
    \frac{\dd x_{i}}{\dd t} = \frac{\dd c}{\dd t} v_{i} =
    \lambda_{\max} c v_{i} - c v_{i} \abs{c}^{n - 1} \abs{v_{i}}^{n - 1} + \gamma (c^{*})^{q - 1} v_{i}^{q - 1}
\end{equation}
If we choose the principal eigenvector to be normalized such that $\lVert\mathbf{v}\rVert_2 = 1$, we can easily extract the time-evolution of the global complex coefficient $c(t)$:
\begin{align}
    \frac{\dd c}{\dd t} & = \frac{\dd c}{\dd t} \sum_{i}^{N}{v_{i}^2} =
     \sum_{i}^{N}{v_{i}\frac{\dd c}{\dd t} v_i} = \sum_{i}^{N}{v_{i}\frac{\dd x_{i}}{\dd t}} \\
    &= \lambda_{\max} c - C_n c \abs{c}^{n - 1} + \gamma K_q (c^{*})^{q - 1}
\end{align}
where
\begin{equation}
    C_n = \sum_{i}^{N}{\abs{v_{i}}^{n + 1}},\ \ K_q = \sum_{i}^{N}{v_{i}^{q}}.
\end{equation}
Here, $C_n$ and $K_q$ are constants that depend on the principal eigenvector and the nonlinearity order and number of Potts states, respectively. At a steady state, $\dd c / \dd t = 0$, and the equation for $c$ becomes
\begin{equation}
    \left(\lambda_{\max} - C_n \abs{c}^{n - 1}\right) c = -\gamma K_q (c^{*})^{q - 1}.
\end{equation}
Multiplying both sides by $c^{q - 1}$ gives
\begin{equation}
    \left(\lambda_{\max} - C_n \abs{c}^{n - 1}\right) c^q = -\gamma K_q \abs{c}^{2(q - 1)}.
\end{equation}
The right-hand side is real-valued as $\gamma K_q \in \mathbb{R}$, and the left-hand side must therefore also be real-valued, meaning that $c^q$ is real-valued and of the opposite sign to $K_q$, assuming $\gamma > 0$. Therefore, the phase of $c$ must be one of the $q$ discrete values given by
\begin{equation}
    \arg(c) = \frac{2\pi k + \arg(-K_q)}{q} = \frac{2\pi k + \delta}{q},\quad k = 0, \ldots, q - 1,\quad
    \delta = \begin{cases}
        \pi, & K_q > 0,\\
        0, & K_q < 0.
    \end{cases}
\end{equation}
These correspond to the $q$ discrete phases of the planar Potts model in $[0, 2\pi]$, offset by $\pi/q$ if $K_q > 0$.

The amplitude $\abs{c}$ can be found by taking the absolute value of the steady-state equation and dividing by $\abs{c}>0$:
\begin{equation}
    \abs{\lambda_{\max} - C_n \abs{c}^{n - 1}} = \gamma \abs{K_q} \abs{c}^{q - 2}.
\end{equation}
In section \ref{app:bounding}, it is shown that $n\geq q$ is required to ensure amplitude bounding for all coupling matrices. Assuming $n \geq q$ and $q>2$, the $\abs{c}^{q-2}$ term will dominate close to the origin where $\abs{c} \ll 1$, and leading order terms give the fixed-point amplitude coefficient as
\begin{equation}
    \abs{c} \approx \left(\frac{\lambda_{\max}}{\gamma \abs{K_q}}\right)^{\frac{1}{q - 2}},\ \ \abs{c} \ll 1.
\end{equation}
Thus, when $\alpha$ or $\beta$ is increased beyond the origin destabilization threshold, the fixed points will initially evolve from the origin according to
\begin{equation}
    \mathbf{x} \approx \left(\frac{\lambda_{\max}}{\gamma \abs{K_q}}\right)^{\frac{1}{q - 2}} e^{i\frac{2\pi k + \delta}{q}} \mathbf{v}_{\max},
\end{equation}
where
\begin{gather}
    \lambda_{\max} = \alpha + \beta\mu_{\max} \ll 1, \\
    k = 0, \ldots, q - 1, \\
    K_q = \sum_{i}^{N}{v_{i}^{q}}, \\
    \delta = \begin{cases}
        \pi, & K_q > 0,\\
        0, & K_q < 0,
    \end{cases}
\end{gather}
and $\mathbf{v}_{\max}$ is the principal eigenvector of the coupling matrix $\mathbf{J}$ with eigenvalue $\mu_{\max}$.  

The above shows that after the origin is destabilized, the system bifurcates and all spins initially evolve parallel or anti-parallel to one of these $q$ discrete phases, depending on the sign of the corresponding element in the principal eigenvector. The amplitude grows as a power law in $\lambda_{\max}$ with exponent $1/(q-2)$ close to the origin. As $\lambda_{\max}$ is increased further from zero, the amplitude will deviate from this power law as higher-order terms become significant.

\subsubsection{Sigmoid Potts machine model} \label{app:FixedPointEvolutionSigmoid}
As found in section \ref{app:StabilitySigmoid}, for the sigmoid Potts machine model, the Jacobian matrix at the origin is given by
\begin{equation}
    \left. \ \mathcal{U} \right|_{\mathbf{x} = \mathbf{0}}  = (\alpha - 1)\mathbf{I} + \beta \mathbf{J},
\end{equation}
with principal eigenvalue $\lambda_{\max} = \alpha - 1 + \beta \mu_{\max}$ and corresponding eigenvector $\mathbf{v}_{\max}$, which is also the principal eigenvector of the coupling matrix $\mathbf{J}$ with eigenvalue $\mu_{\max}$. Linearizing the governing equations around the origin right after the principal eigenvalue crosses zero (i.e. $\lambda_{\max} \ll 1$) therefore gives
\begin{equation}
    \frac{\dd \mathbf{x}}{\dd t} = \left. \ \mathcal{U} \right|_{\mathbf{x} = \mathbf{0}}  \mathbf{x},
\end{equation}
and the system will initially evolve along the direction of $\mathbf{v}_{\max}$, such that
\begin{equation}
    \mathbf{x}(t) = c(t) \mathbf{v}_{\max},
\end{equation}
where $c(t) \in \mathbb{C}$ is a time-dependent complex coefficient. As the coupling matrix is real-valued and symmetric, the eigenvector $\mathbf{v}_{\max}$ can be chosen to be real-valued and with $\lVert \mathbf{v}_{\max} \rVert_2 = 1$. The real-valued amplitudes of the spins can therefore be expressed as
\begin{equation}
    \mathbf{r}(t) = \rho(t) \mathbf{a},\quad \rho(t) = \abs{c(t)},\quad a_i = \abs{v_{\max,i}},
\end{equation}
where $\rho(t) \in \mathbb{R}_{\geq 0}$ is a global amplitude coefficient. The phases of the spins will initially be either $\theta_i = \arg(c)$ if $v_{\max,i} > 0$ or $\theta_i = \arg(c) + \pi$ if $v_{\max,i} < 0$.

The amplitude and phase equations of the sigmoid PM model were defined in section \ref{app:Sigmoid} as
\begin{equation}
    \frac{\dd r_{i}}{\dd t} = - r_{i} + \tanh\left( \alpha r_{i} + \gamma r_{i}^{q - 1}\cos\left( q\theta_{i} \right) + \beta\sum_{j}^{N}{J_{{ij}}r_{j}\cos\left( \theta_{j} - \theta_{i} \right)} \right)
\end{equation}
\begin{equation}\label{eq:poly_phase_stability}
    \frac{\dd\theta_{i}}{\dd t} = -\gamma r_{i}^{q - 2}\sin{(q\theta_i)} + \frac{\beta}{r_{i}}\sum_{j}^{N}{J_{ij}r_{j}\sin{(\theta_{j} - \theta_{i})}}
\end{equation}
As $\theta_i \in \left\{\arg(c), \arg(c) + \pi\right\}$ initially, the phase differences $\theta_j - \theta_i$ will be either 0 or $\pi$, such that $\sin(\theta_j - \theta_i) = 0 \forall i,j$. Fixed points of the phase equations, where $\dd \theta_i / \dd t = 0$, must therefore initially satisfy
\begin{equation}
    0 = -\gamma r_{i}^{q - 2}\sin{(q\theta_i)} \Rightarrow \sin{(q\theta_i)} = 0,
\end{equation}
which has the solutions
\begin{equation}
    \theta_{i} = \frac{m\pi}{q},\ \ m \in \mathbb{Z}.
\end{equation}
As $\theta_i \in \left\{\arg(c), \arg(c) + \pi\right\}$, the phase of $c$ must therefore be one of the $q$ discrete values given by
\begin{equation}
    \arg(c) = \frac{2k\pi}{q},\quad k = 0, \ldots, q - 1,
\end{equation}
corresponding to the $q$ discrete phases of the planar Potts model in $[0, 2\pi]$.

The argument of the hyperbolic tangent in the amplitude equation can be expressed as
\begin{align}
    z_i &= \alpha r_{i} + \gamma r_{i}^{q - 1}\cos\left( q\theta_{i} \right) + \beta\sum_{j}^{N}{J_{{ij}}r_{j}\cos\left( \theta_{j} - \theta_{i} \right)}\\
    &= (\lambda_{\max} + 1) \rho a_i + \gamma s_i \rho^{q - 1} a_i^{q - 1},
\end{align}
where $s_i = \cos(q\theta_i) \in \{+1, -1\}$, and where it is used that $\cos(\theta_j - \theta_i) = \mathrm{sign}(v_{i} v_{j}) = \sigma_i \sigma_j$ with $\sigma_i = \mathrm{sign}(v_{i})$, such that
\begin{equation}
    \sum_{j}^{N}{J_{{ij}}r_{j}\cos\left( \theta_{j} - \theta_{i} \right)} = \rho\sum_{j}^{N}{J_{{ij}}a_{j}}\sigma_i \sigma_j = \rho \sigma_i \sum_{j}^{N}{J_{{ij}}(\sigma_j a_{j})} = \rho \sigma_i \mu_{\max} v_{i} = \mu_{\max} \rho a_i.
\end{equation}
Note that $v_i = v_{\max,i} = \sigma_i a_i$. For $z_i \ll 1$, the hyperbolic tangent can be approximated to the third order as $\tanh(z_i) = z_i - z_i^3/3 + \mathcal{O}(z_i^5)$. Assuming $q>2$ and $\rho \ll 1$, keeping only the leading order term of the cubed argument gives
\begin{equation}
    z_i^3 = (\lambda_{\max} + 1)^3 \rho^3 a_i^3 + \mathcal{O}(\rho^{q+1}).
\end{equation}
Therefore, the amplitude equation becomes
\begin{align}
    \frac{\dd r_{i}}{\dd t} = \frac{\dd \rho}{\dd t} a_i &= - r_{i} + z_i - \frac{1}{3} z_i^3 + \mathcal{O}(z_i^5)\\
    &= - \rho a_i + (\lambda_{\max} + 1) \rho a_i + \gamma s_i \rho^{q - 1} a_i^{q - 1} - \frac{1}{3} (\lambda_{\max} + 1)^3 \rho^3 a_i^3 + \mathcal{O}(\rho^5, \rho^{q + 1})\\
\end{align}
Similarly to how we extracted the time-evolution of $c(t)$ in section \ref{app:FixedPointEvolutionPolynomial}, we now multiply by $a_i$ and sum over all spins to extract the time-evolution of the global amplitude coefficient $\rho(t)$, given that $\sum_i a_i^2 = 1$:
\begin{align}
    \frac{\dd \rho}{\dd t} &= \sum_{i}^{N}{a_i \frac{\dd \rho}{\dd t} a_i} = \sum_{i}^{N}{a_i \frac{\dd r_{i}}{\dd t}} \\
    &= \lambda_{\max} \rho + \gamma S_q \rho^{q - 1} - \frac{1}{3} (\lambda_{\max} + 1)^3 C_3 \rho^3 + \mathcal{O}(\rho^5, \rho^{q + 1})
\end{align}
where
\begin{equation}
    S_q = \sum_{i}^{N}{s_i a_i^{q}} = \sum_{i}^{N}{\cos(q\theta_i) a_i^{q}},\quad C_3 = \sum_{i}^{N}{a_i^{4}}.
\end{equation}
At a steady state, $\dd \rho / \dd t = 0$, and the equation for $\rho$, after dividing by $\rho > 0$, becomes
\begin{equation}
    0 = \lambda_{\max} + \gamma S_q \rho^{q - 2} - \frac{1}{3} (\lambda_{\max} + 1)^3 C_3 \rho^2 + \mathcal{O}(\rho^4, \rho^{q}).
\end{equation}
The smallest power of $\rho$ will dominate close to the origin where $\rho \ll 1$. Depending on the number of Potts states $q$, this gives the leading order terms as
\begin{equation}\label{eq:rho_equation}
    \rho \approx \begin{cases}
        \frac{\lambda_{\max}}{\gamma \abs{S_q}}, & q = 3,\\[10pt]
        \sqrt{\frac{\lambda_{\max}}{\abs{\frac{1}{3}(\lambda_{\max} + 1)^3 C_3 - \gamma S_q}}}, & q = 4,\\[10pt]
        \sqrt{\frac{3\lambda_{\max}}{(\lambda_{\max} + 1)^3 C_3}} \approx \sqrt{\frac{3\lambda_{\max}}{C_3}}, & q \geq 5,\\[10pt]
    \end{cases}
\end{equation}
Note that for $q = 3$, the amplitude $\rho$ grows linearly with $\lambda_{\max}$ close to the origin, whereas for $q \geq 4$, $\rho$ grows as $\sqrt{\lambda_{\max}}$.

In summary, when $\alpha$ or $\beta$ is increased beyond the origin destabilization threshold, the fixed points will initially evolve from the origin according to
\begin{equation}
    \mathbf{x} \approx \rho e^{i\arg{c}} \mathbf{v}_{\max},
\end{equation}
where $\rho \ll 1$ is given by Eq.~\eqref{eq:rho_equation},
\begin{gather}
    \arg(c) = \frac{2k\pi}{q},\quad k = 0, \ldots, q - 1, \\
    \lambda_{\max} = \alpha - 1 + \beta\mu_{\max} \ll 1,
\end{gather}
and $\mathbf{v}_{\max} \in \mathbb{R}^N$ is the principal eigenvector of the coupling matrix $\mathbf{J}$ with eigenvalue $\mu_{\max}$.  

Thus, after the origin is destabilized, the system bifurcates and all spins initially evolve parallel or anti-parallel to one of $q$ discrete phases, depending on the sign of the corresponding element in the principal eigenvector. For $q = 3$, the amplitude grows linearly with $\lambda_{\max}$ close to the origin. For $q \geq 4$, the amplitude grows as $\sqrt{\lambda_{\max}}$. As $\lambda_{\max}$ is increased further from zero, the amplitude will deviate from this behavior as higher-order terms become significant.

\subsection{Amplitudes of stable fixed points} \label{app:AmplitudeHomogeneity}
To determine whether the different Potts machine models enforce amplitude homogeneity, we analyze the amplitudes of stable non-origin fixed points. We focus on fixed points where all spin amplitudes are non-zero ($r_i > 0\ \forall i$) and phases are discretized to the planar Potts states ($\theta_i = \theta_i^{(k)} = 2\pi k_i / q$ for some $k_i \in \{0, \ldots, q-1\}$). This simplification allows us to isolate the amplitude dynamics from the phase dynamics.

For the polynomial model, the amplitude equation derived in supplementary materials \ref{app:PolynomialBreakingUp} is
\begin{equation}\label{eq:polynomial_amplitude_eq}
    \frac{\dd r_{i}}{\dd t} = \alpha r_{i} - r_{i}^{n} + \gamma r_{i}^{q - 1}\cos\left( q\theta_{i} \right) + \beta\sum_{j}^{N}{J_{ij}r_{j}\cos\left( \theta_{j} - \theta_{i} \right)}.
\end{equation}
This applies directly to the NEC model (with $\beta = 1$ and dynamic $\alpha_i$), the $q$-PDC model (with $\alpha = -1$), and serves as the basis for the sigmoid model. The $q$-SHIL model has fixed amplitudes by definition, so amplitude homogeneity is trivially enforced.

\subsubsection{Fixed point amplitudes with discretized phases}
Assuming phases are discretized to Potts states, $\cos(q\theta_i) = 1$ for all $i$. We define the effective coupling matrix
\begin{equation}
    \tilde{J}_{ij} = J_{ij}\cos(\Delta_{ij}),\quad \Delta_{ij} = \theta_j^{(k)} - \theta_i^{(k)} = \frac{2\pi}{q}(k_j - k_i),
\end{equation}
which accounts for phase differences between spins. The amplitude equation becomes
\begin{equation}
    \frac{\dd r_{i}}{\dd t} = \alpha r_{i} - r_{i}^{n} + \gamma r_{i}^{q - 1} + \beta\sum_{j}^{N}{\tilde{J}_{ij}r_{j}}.
\end{equation}
At a fixed point, $\dd r_i / \dd t = 0$, giving
\begin{equation}\label{eq:amplitude_fixed_point}
    r_{i}^{n} - \gamma r_{i}^{q - 1} - \alpha r_{i} = \beta\sum_{j}^{N}{\tilde{J}_{ij}r_{j}}.
\end{equation}

\subsubsection{Existence of homogeneous amplitude fixed points}
A homogeneous amplitude fixed point has $r_i = r^* > 0$ for all $i$. Substituting into Eq.~\eqref{eq:amplitude_fixed_point} and dividing by $r^* > 0$ gives
\begin{equation}
    (r^*)^{n-1} - \gamma (r^*)^{q - 2} - \alpha = \beta \sum_{j}^{N}{\tilde{J}_{ij}} = \beta \tilde{\rho}_i,
\end{equation}
where $\tilde{\rho}_i = \sum_j \tilde{J}_{ij}$ is the $i$-th row sum of the effective coupling matrix. For a homogeneous fixed point to exist for all spins simultaneously, either:
\begin{enumerate}
    \item All row sums are equal: $\tilde{\rho}_i = \tilde{\rho}\ \forall i$, or
    \item The coupling is weak: $\beta \to 0$.
\end{enumerate}
For general coupling matrices and phase configurations, the row sums $\tilde{\rho}_i$ differ across spins, so homogeneous amplitude fixed points generically do not exist. This is the fundamental reason why models without additional mechanisms cannot enforce amplitude homogeneity.

\subsubsection{Stability of homogeneous amplitude fixed points}
To analyze stability, we linearize the amplitude equations around a homogeneous fixed point $r_i = r^* + \delta r_i$ where $|\delta r_i| \ll r^*$. The Jacobian matrix of the amplitude equations with respect to the amplitudes, evaluated at the homogeneous fixed point, has elements
\begin{equation}
    \mathcal{W}_{ij} = \left.\frac{\partial}{\partial r_j}\frac{\dd r_i}{\dd t}\right|_{r_k = r^*} = \left(\alpha - n(r^*)^{n-1} + (q-1)\gamma (r^*)^{q-2}\right)\delta_{ij} + \beta \tilde{J}_{ij}.
\end{equation}
This can be written in matrix form as
\begin{equation}
    \mathcal{W} = \lambda_{\mathrm{self}} \mathbf{I} + \beta \tilde{\mathbf{J}},
\end{equation}
where
\begin{equation}
    \lambda_{\mathrm{self}} = \alpha - n(r^*)^{n-1} + (q-1)\gamma (r^*)^{q-2}
\end{equation}
is the self-interaction term arising from the nonlinear amplitude dynamics. The eigenvalues of $\mathcal{W}$ are
\begin{equation}
    \nu_k = \lambda_{\mathrm{self}} + \beta \tilde{\mu}_k,
\end{equation}
where $\tilde{\mu}_k$ are the eigenvalues of the effective coupling matrix $\tilde{\mathbf{J}}$.

For stability, all eigenvalues must be negative: $\nu_k < 0\ \forall k$. This requires
\begin{equation}
    \lambda_{\mathrm{self}} + \beta \tilde{\mu}_{\max} < 0,
\end{equation}
where $\tilde{\mu}_{\max}$ is the largest eigenvalue of $\tilde{\mathbf{J}}$. Rearranging gives the stability condition
\begin{equation}\label{eq:amplitude_stability_condition}
    n(r^*)^{n-1} - (q-1)\gamma (r^*)^{q-2} > \alpha + \beta \tilde{\mu}_{\max}.
\end{equation}

\subsubsection{Application to specific models}

\paragraph{Polynomial and $q$-PDC models:}
For these models, the right-hand side of Eq.~\eqref{eq:amplitude_fixed_point} is spin-dependent through the row sums $\tilde{\rho}_i$, which generically vary across spins. Therefore, homogeneous amplitude fixed points do not exist for general coupling matrices, and amplitude homogeneity is not enforced. Even if a homogeneous fixed point existed (e.g., for specially structured coupling matrices), whether condition \eqref{eq:amplitude_stability_condition} is satisfied depends on the specific values of $r^*$, $\gamma$, and the coupling spectrum, with no general guarantee of stability.

\paragraph{$q$-SHIL model:}
In this model, all amplitudes are fixed to unity by definition ($r_i = 1\ \forall i$). Amplitude homogeneity is therefore trivially enforced without requiring any stability analysis.

\paragraph{NEC model:}
The NEC model augments the amplitude dynamics with spin-dependent gains $\alpha_i$ that evolve according to
\begin{equation}
    \frac{\dd \alpha_i}{\dd t} = \varepsilon_\alpha\bigl(\sqrt{r_{\mathrm{target}}} - \sqrt{r_i}\bigr),
\end{equation}
where $\varepsilon_\alpha > 0$ is the gain relaxation rate and $r_{\mathrm{target}} > 0$ is the target amplitude. At steady state, $\dd \alpha_i / \dd t = 0$ requires $r_i = r_{\mathrm{target}}$ for all $i$, so the only steady state has homogeneous amplitudes.

We now linearize the coupled system around $(r_i, \alpha_i) = (r_{\mathrm{target}}, \alpha^*)$ where $\alpha^*$ satisfies the amplitude fixed point equation. Writing $r_i = r_{\mathrm{target}} + \delta r_i$ and $\alpha_i = \alpha^* + \delta \alpha_i$, the linearized dynamics become
\begin{align}\label{eq:nec_linearized}
    \frac{\dd \delta r_i}{\dd t} &= \lambda_{\mathrm{self}}\,\delta r_i + r_{\mathrm{target}}\,\delta \alpha_i + \sum_{j}{\tilde{J}_{ij}\,\delta r_j}, \\
    \frac{\dd \delta \alpha_i}{\dd t} &= -\frac{\varepsilon_\alpha}{2\sqrt{r_{\mathrm{target}}}}\,\delta r_i,
\end{align}
where we used $\beta = 1$ for the NEC model. The gain dynamics introduce negative feedback: if $\delta r_i > 0$, then $\dd \delta \alpha_i / \dd t < 0$, which causes $\alpha_i$ to decrease. A decreasing $\alpha_i$ reduces $\dd r_i / \dd t$ (since the amplitude equation has a positive $\alpha_i r_i$ term), eventually making it negative and driving $r_i$ back toward $r_{\mathrm{target}}$. The converse holds when $\delta r_i < 0$. This feedback drives all amplitudes toward $r_{\mathrm{target}}$ regardless of the coupling structure.

\paragraph{Sigmoid model:}
The sigmoid model replaces the polynomial amplitude equation with
\begin{equation}
    \frac{\dd r_{i}}{\dd t} = - r_{i} + \tanh\left( \alpha r_{i} + \gamma r_{i}^{q - 1}\cos\left( q\theta_{i} \right) + \beta\sum_{j}^{N}{J_{ij}r_{j}\cos\left( \theta_{j} - \theta_{i} \right)} \right).
\end{equation}
At steady state, $r_i = \tanh(R_i)$ where $R_i$ is the argument. As the parameters $\alpha$ and $\beta$ increase during annealing, $|R_i| \to \infty$ for all spins. When $\tanh(R_i) \to 1$, we have $\dd r_i / \dd t = -r_i + \tanh(R_i) \approx 1 - r_i$, which is positive for $r_i < 1$ and negative for $r_i > 1$, such that the amplitudes must converge to $r_i = 1$. Thus all amplitudes converge to unity regardless of coupling structure.

\subsection{Phases of stable fixed points}
To determine whether the different Potts machine models enforce phase discretization, we analyze the phases of stable non-origin fixed points. We focus on fixed points where all spin amplitudes are non-zero ($r_i > 0\ \forall i$), ensuring that phases are well-defined. This does not exclude the existence of other fixed points where some spins may have zero amplitude. The stable fixed-point phases are determined using the phase equations derived by expressing the complex governing equations in polar form $x_{i} = r_{i}e^{i\theta_{i}}$.

For the polynomial model, the derivation of the phase equations is outlined in section \ref{app:PolynomialBreakingUp}. This derivation applies directly to the $q$-PDC model and to the NEC model when fixing $\beta$ to 1. The sigmoid model uses the same phase equations by definition. The $q$-SHIL model phase equations are by definition the same as for the polynomial model with $\beta=1$ and all amplitudes fixed to unity. The phase equations derived for the polynomial model,
\begin{equation}
    \frac{\dd\theta_{i}}{\dd t} = -\gamma r_{i}^{q - 2}\sin{(q\theta_i)} + \frac{\beta}{r_{i}}\sum_{j}^{N}{J_{ij}r_{j}\sin{(\theta_{j} - \theta_{i})}},
\end{equation}
apply to all Potts machine models discussed in this work. For the NEC model, $\beta = 1$. For the $q$-SHIL model, $\beta = 1$ and all amplitudes are fixed to unity ($r_i = 1\ \forall i$). Fixed points of the phase equations satisfy $\dd \theta_i / \dd t = 0$.

\subsubsection[Stable fixed point phases without coupling (beta = 0)]{Stable fixed point phases without coupling (\texorpdfstring{$\beta = 0$}{beta = 0})}
When there is no coupling between spins ($\beta = 0$), the fixed point equation reduces to
\begin{equation}
    \gamma r_{i}^{q - 1}\sin{(q\theta_i)} = 0,
\end{equation}
which assuming $\gamma \neq 0$ and $r_i \neq 0$ has the solutions
\begin{equation}
    \theta_{i} = \frac{k\pi}{q},\ \ k \in \mathbb{Z}.
\end{equation}
Stability requires
\begin{equation}
    \frac{\partial} {\partial \theta_i} \frac{\dd \theta_i}{\dd t}
    = -\gamma\,q\,r_{i}^{\,q-2}\cos\bigl(q\theta_i\bigr) < 0,
\end{equation}
which in $\theta_i \in [0, 2\pi]$ is fulfilled for
\begin{equation}
    \theta_{i}^{(k)} = \frac{2\pi k}{q},\ \ k = 0, \ldots, q - 1.
\end{equation}
This corresponds to the $q$ discrete phases of the planar Potts model.
\subsubsection[Stable fixed point phases with coupling (beta > 0) assuming small offset from Potts model phases]{Stable fixed point phases with coupling (\texorpdfstring{$\beta > 0$}{beta > 0}) assuming small offset from Potts model phases}\label{app:StabilityDiscretization}
We consider the case with coupling between spins ($\beta > 0$) and analyze the stability of fixed points close to the planar Potts model phases, such that each phase can be written as a small deviation from its nearest planar Potts model phase:
\begin{equation}
    \theta_{i} = \theta_{i}^{(k)} + \varepsilon_{i},\ \ |\varepsilon_{i}| \ll 1,
\end{equation}
and let
\begin{equation}
    \Delta_{ij} = \theta_{j}^{(k)} - \theta_{i}^{(k)} = \frac{2\pi}{q}(k_j - k_i).
\end{equation}
Expanding the fixed point equation to first order in $\varepsilon_i$ gives
\begin{equation}
    0 = -\gamma r_{i}^{q - 2}q\varepsilon_i + \frac{\beta}{r_i}\sum_{j}^{N}{J_{ij}r_{j}\left(\sin(\Delta_{ij}) + \cos(\Delta_{ij})(\varepsilon_j - \varepsilon_i)\right)}.
\end{equation}
Collecting this into a linear system for the offsets vector 
\(\bm{\varepsilon} = (\varepsilon_1, \ldots, \varepsilon_N)^{T}\in\mathbb{R}^N\) gives
\begin{equation} \label{eq:phase_linear_system}
    \bigl(\gamma q\,\mathbf{D} + \beta\,\mathbf{L}_{c}\bigr)\,\bm{\varepsilon}
    \;=\;\beta\,\mathbf{b},
\end{equation}
where
\begin{align}
    \mathbf{D} &= \mathrm{diag}\bigl(r_1^{q-2},\ldots,r_N^{q-2}\bigr) \in\mathbb{R}^{N\times N},\\
    \mathbf{L}_{c} &= \bigl\{(\mathbf{L}_{c}\,\mathbf{v})_i
        = \frac{1}{r_i}\sum_{j=1}^N J_{ij}\,r_j\cos(\Delta_{ij})\,(v_i - v_j)\bigr\} \in\mathbb{R}^{N\times N},\\
    \mathbf{b} &= (b_1,\ldots,b_N)^{T} \in\mathbb{R}^N,\quad
    b_i = \frac{1}{r_i}\sum_{j=1}^N J_{ij}\,r_j\sin(\Delta_{ij}).
\end{align}
The stability of a fixed point in phase space is determined by the eigenvalues of the Jacobian matrix of the phase equations, with elements
\begin{equation}\label{eq:phase_jacobian}
    \mathcal{V}_{ij} = \frac{\partial}{\partial \theta_{j}}\frac{\dd \theta_{i}}{\dd t}
    = -\gamma q\,r_{i}^{q-2}\cos(q\theta_i)\delta_{ij}
    + \beta\left(\frac{1}{r_i}J_{ij}r_j\cos(\theta_j - \theta_i)
    - \frac{1}{r_i}\delta_{ij}\sum_{k}^{N}{J_{ik}r_{k}\cos{(\theta_{k} - \theta_{i})}}\right).
\end{equation}
Assuming $\gamma > 0$, the first term is negative around the planar Potts phases as $\cos(q\theta_i^{(k)}) = 1$. The second term can be positive or negative depending on the coupling matrix and the phases.

If $\beta$ and $\gamma$ are increased simultaneously such that their ratio remains constant, the offsets will remain constant and the stability will remain unaffected, as both the linear system in Eq. \eqref{eq:phase_linear_system} and the Jacobian matrix in Eq. \eqref{eq:phase_jacobian} are simply scaled by the same positive factor.

As \(\beta \to 0\), fixed points approach the planar Potts model phases, since the linear system in Eq. \eqref{eq:phase_linear_system} reduces to \(\gamma q\,\mathbf{D}\,\bm{\varepsilon} = 0\), which has the unique solution \(\bm{\varepsilon} = \mathbf{0}\) because \(\mathbf{D}\) is positive definite (we consider only fixed points with non-zero amplitudes). In this limit, the Jacobian matrix in Eq. \eqref{eq:phase_jacobian} becomes diagonal with negative diagonal elements, confirming that these fixed points are stable. 

For $\beta > 0$ and $\beta / \gamma \ll 1$, the linear system in Eq. \eqref{eq:phase_linear_system} can be approximated as
\begin{equation}
    \bm{\varepsilon} \approx \frac{\beta}{\gamma q}\,\mathbf{D}^{-1}\, \mathbf{b},
\end{equation}
showing that the offsets grow linearly with the $\beta/\gamma$ ratio when this ratio is small. Note that $\mathbf{D}$ is positive definite and therefore invertible. From Eq. \eqref{eq:phase_jacobian}, we see that increasing $\beta/\gamma$ will increase the magnitude of the off-diagonal coupling terms relative to the diagonal self-interaction term, which may destabilize the fixed point if the off-diagonal part has nonnegative eigenvalues.

\subsubsection{Conditions for phase discretization enforcement}
The above analysis shows that stable fixed points close to the planar Potts model phases exist when the $\beta/\gamma$ ratio is sufficiently small. An annealing scheme that gradually reduces the $\beta/\gamma$ ratio during the optimization process will therefore enforce phase discretization, as fixed points will approach the planar Potts model phases and become stable if not already stable.
\section{Checks of dynamical properties}\label{app:dynamical_requirements} \label{app:dynamical_properties}
The three dynamical properties introduced in the main text can be evaluated directly from the governing equations assuming the annealing schedules of Table~\ref{tab:models}. Below we outline how each property is checked and then evaluate each model based on the stability and fixed-point analysis of section~\ref{app:stability}.

\subsection{How dynamical properties are checked}

\paragraph{Connected origin-to-solution branch}
For models with amplitude dynamics, this property requires that the trivial fixed point at the origin is initially stable and bifurcates into a nontrivial solution branch that evolves continuously from the origin under the applied annealing schedule. As shown in section~\ref{app:StabilityOrigin}, the origin stability depends on the principal eigenvalue of the Jacobian matrix, which crosses zero at a threshold determined by the annealing parameters. Section~\ref{app:FixedPointEvolution} demonstrates how fixed points evolve after the origin loses stability. Models without amplitude dynamics do not have an origin fixed point and thus this property is not applicable.

\paragraph{Enforced amplitude homogeneity}
Enforcing amplitude homogeneity requires a mechanism that drives all spin amplitudes to the same value during evolution. As analyzed in section~\ref{app:AmplitudeHomogeneity}, for general coupling matrices, homogeneous amplitude fixed points generically do not exist unless additional mechanisms are present. This can be achieved through dynamic gain control (NEC model), saturating nonlinearities (sigmoid models), or by fixing the amplitudes at all times ($q$-SHIL model).

\paragraph{Enforced phase discretization}
Phase discretization means that, through evolution of the system, the continuous spin phases converge to the discrete set of phases corresponding to the Potts model states. As shown in section~\ref{app:StabilityDiscretization}, stable fixed points approach the planar Potts phases $\theta_i^{(k)} = 2\pi k / q$ when the $\beta/\gamma$ ratio is small. Decreasing this ratio during annealing ensures that fixed points both approach the Potts phases and become (or remain) stable.

\subsection{NEC model}

\paragraph{Connected origin-to-solution branch (Fulfilled):} As shown in section~\ref{app:StabilityNEC}, the origin is stable for $\alpha < \alpha_{\mathrm{th}} = -\mu_{\max}$, where $\mu_{\max}$ is the largest eigenvalue of the coupling matrix. The NEC gain equation~\eqref{eq:nec_alpha} increases $\alpha_i$ over time while $r_i < r_{\mathrm{target}}$. As $\alpha$ passes $\alpha_{\mathrm{th}}$, the origin loses stability and a continuous branch of growing amplitudes emerges along one of $q$ equispaced discrete phases, as demonstrated in section~\ref{app:FixedPointEvolutionPolynomial}. The NEC model thus has a connected origin-to-solution branch.

\paragraph{Enforced amplitude homogeneity (Fulfilled):} As analyzed in section~\ref{app:AmplitudeHomogeneity}, the NEC model employs dynamic gain control through Eq.~\eqref{eq:nec_alpha}, where the spin-dependent gains $\alpha_i$ evolve to drive all amplitudes toward the target value $r_{\mathrm{target}}$. The negative feedback mechanism ensures that at steady state, $r_i = r_{\mathrm{target}}$ for all $i$, regardless of the coupling structure. Therefore, amplitude homogeneity is enforced.

\paragraph{Enforced phase discretization (Fulfilled):} For the NEC model, $\beta = 1$ by definition. As $\gamma$ is increased during annealing, the $\beta/\gamma$ ratio decreases. By the analysis of section~\ref{app:StabilityDiscretization}, this ensures that fixed points approach the Potts phases and remain stable. Therefore, phase discretization is enforced.

\subsection{\(q\)-PDC model}

\paragraph{Connected origin-to-solution branch (Not fulfilled):} As shown in section~\ref{app:StabilityQPDC}, with $\alpha = -1$ fixed, the origin is stable for $\beta < \beta_{\mathrm{th}} = 1/\mu_{\max}$. The annealing scheme fixes $\beta$ while increasing $\gamma$. Since $\beta$ does not cross the threshold during annealing, the origin never loses stability and no connected origin-to-solution branch exists.

\paragraph{Enforced amplitude homogeneity (Not fulfilled):} As shown in section~\ref{app:AmplitudeHomogeneity}, for general coupling matrices, the amplitude fixed point equation~\eqref{eq:amplitude_fixed_point} requires spin-dependent row sums of the effective coupling matrix to be equal for a homogeneous fixed point to exist. Since this is generically not satisfied, the $q$-PDC model does not enforce amplitude homogeneity.

\paragraph{Enforced phase discretization (Fulfilled):} With $\beta$ fixed and $\gamma$ increasing during annealing, the $\beta/\gamma$ ratio decreases. By section~\ref{app:StabilityDiscretization}, phase discretization is therefore enforced.

\subsection{\(q\)-SHIL model}

\paragraph{Connected origin-to-solution branch (Not applicable):} The $q$-SHIL model does not have amplitude dynamics (all amplitudes are fixed to unity by definition) and thus does not have an origin fixed point. Therefore, the property of a connected origin-to-solution branch is not applicable.

\paragraph{Enforced amplitude homogeneity (Fulfilled):} As noted in section~\ref{app:AmplitudeHomogeneity}, the $q$-SHIL model has all amplitudes fixed to unity ($r_i = 1\ \forall i$) by definition. Therefore, amplitude homogeneity is trivially enforced.

\paragraph{Enforced phase discretization (Fulfilled):} For the $q$-SHIL model, $\beta = 1$ and $r_i = 1\ \forall i$ by definition. As $\gamma$ is increased during annealing, the $\beta/\gamma$ ratio decreases. By section~\ref{app:StabilityDiscretization}, phase discretization is therefore enforced.

\subsection{Polynomial PM model}

\paragraph{Connected origin-to-solution branch (Fulfilled):} As shown in section~\ref{app:StabilityPolynomial}, with $\alpha = -1$ fixed, the origin is stable for $\beta < \beta_{\mathrm{th}} = 1/\mu_{\max}$. As $\beta$ is increased past $\beta_{\mathrm{th}}$ during annealing, the origin loses stability and a continuous branch of growing amplitudes emerges along one of $q$ equispaced discrete phases, as demonstrated in section~\ref{app:FixedPointEvolutionPolynomial}. The polynomial PM model thus has a connected origin-to-solution branch.

\paragraph{Enforced amplitude homogeneity (Not fulfilled):} As shown in section~\ref{app:AmplitudeHomogeneity}, for general coupling matrices, homogeneous amplitude fixed points generically do not exist because the row sums of the effective coupling matrix differ across spins. The polynomial PM model does not include any mechanism to enforce amplitude homogeneity.

\paragraph{Enforced phase discretization (Not fulfilled with current schedule):} Both $\beta$ and $\gamma$ can be controlled during annealing. Phase discretization can be enforced by increasing $\gamma$ faster than linearly with respect to $\beta$, e.g., $\gamma = \beta^p$ with $p > 1$, which decreases the $\beta/\gamma$ ratio. However, with the employed annealing scheme where $\gamma$ and $\beta$ increase linearly together ($\gamma \propto \beta$), the $\beta/\gamma$ ratio remains constant. By the analysis of section~\ref{app:StabilityDiscretization}, phase discretization is therefore not enforced with this schedule. Still, as illustrated in figure~\ref{fig:dynamics} in the main text, the phases cluster around the Potts phases for the employed annealing scheme, if not discretizing completely.

\subsection{Sigmoid PM model}

\paragraph{Connected origin-to-solution branch (Fulfilled):} As shown in section~\ref{app:StabilitySigmoid}, with $\alpha < 1$ fixed, the origin is stable for $\beta < \beta_{\mathrm{th}} = (1-\alpha)/\mu_{\max}$ (from Eq.~\eqref{eq:sigmoid_beta_threshold}). As $\beta$ is increased past $\beta_{\mathrm{th}}$ during annealing, the origin loses stability and a continuous branch of growing amplitudes emerges along one of $q$ equispaced discrete phases, as demonstrated in section~\ref{app:FixedPointEvolutionSigmoid}. The sigmoid PM model thus has a connected origin-to-solution branch.

\paragraph{Enforced amplitude homogeneity (Fulfilled):} As analyzed in section~\ref{app:AmplitudeHomogeneity}, the sigmoid model uses the saturating nonlinearity $\tanh(\cdot)$, which bounds outputs to $(-1, 1)$. As the gain parameters $\alpha$ and $\beta$ increase during annealing, the argument of the hyperbolic tangent grows large, driving $\tanh(R_i) \to 1$ for all spins. Since steady states satisfy $r_i = \tanh(R_i)$, all amplitudes converge to unity regardless of coupling structure. Therefore, amplitude homogeneity is enforced.

\paragraph{Enforced phase discretization (Not fulfilled with current schedule):} As for the polynomial PM model, both $\beta$ and $\gamma$ can be controlled during annealing. Phase discretization can be enforced by increasing $\gamma$ faster than linearly with respect to $\beta$, which decreases the $\beta/\gamma$ ratio. However, with the employed annealing scheme where $\gamma$ and $\beta$ increase linearly together, the $\beta/\gamma$ ratio remains constant. By section~\ref{app:StabilityDiscretization}, phase discretization is therefore not enforced with this schedule. Still, as illustrated in figure~\ref{fig:dynamics} in the main text, the phases cluster around the Potts phases for the employed annealing scheme, if not discretizing completely.
\section{Requirements on nonlinearity order to ensure amplitude bounding in polynomial-based Potts machines} \label{app:bounding}
For three of the five studied Potts machine (PM) models, namely the NEC, $q$-PDC, and polynomial PM models, the governing equation for the real-valued amplitude $r$ of an uncoupled spin can be written in the following form (see equation \ref{eq:polynomial_amplitude_eq}):
\begin{equation}
    \frac{\dd{r}}{\dd{t}} = \alpha r - r^n + \gamma r^{q-1}\cos{q\theta}
\end{equation}

Without coupling, the spin phase quickly converges to a discrete Potts state, such that $\cos{q\theta}=1$, as shown in section \ref{app:StabilityDiscretization}. Thus, the values of $n$ and $q$ determine the bounding of the amplitude $r$ in the following way:
\begin{center}
\begin{tabular}{r c l c l}
 $n>q-1$ & $\rightarrow$ & $r^n$ grows faster than $\gamma r^{q-1}$ & $\rightarrow$ & Amplitude always bounded \\
 $n=q-1$ & $\rightarrow$ & $\gamma r^{q-1}$ grows faster than $r^n$ for $\gamma > 1$ & $\rightarrow$ & Amplitude runaway for $\gamma > 1$ \\
 $n<q-1$ & $\rightarrow$ & $\gamma r^{q-1}$ grows faster than $r^n$ for $\gamma > 0$ & $\rightarrow$ & Amplitude runaway for $\gamma > 0$ \\
\end{tabular}
\end{center}

To ensure that amplitudes are bounded for any value of $\gamma$, we must therefore require $n\geq q$. It should be noted, however, that this analysis does not consider coupling, which can assist in bounding the amplitudes.
\section{Time-domain simulations with the Euler-Maruyama method} \label{app:euler-maruyama}
The Potts and Ising machine models can be expressed as multi-dimensional stochastic differential equations (SDEs) of the form:
\begin{equation}
    \dv{\mathbf{x}}{t} = \mathbf{F}(\mathbf{x},t) + f_\xi \bm{\upxi}(t)
\end{equation}
where $\mathbf{x}$ is the state vector representing the spin amplitudes and/or phases, $\mathbf{F}(\mathbf{x},t)$ is the drift term representing the deterministic dynamics, $\bm{\upxi}(t)$ represents Gaussian white noise, and $f_\xi$ is a noise factor scaling the intensity.

Given a time step size $\Delta t$, the Euler-Maruyama method \cite{maruyama_continuous_1955} discretizes the integration of the equation as:
\begin{equation}
    \mathbf{x}[k+1] = \mathbf{x}[k] + \mathbf{F}(\mathbf{x}[k],k)\Delta t + f_\xi \Delta \bm{\upxi}[k]
\end{equation}
where the stochastic process increment $\Delta \bm{\upxi}[k]$ is sampled from a normal distribution with mean zero and variance $\Delta t$. For models with complex-valued spins, the stochastic process is applied independently to both the real and imaginary parts, with variance $\Delta t/2$ for each part, such that the total variance is still $\Delta t$.
\section{Simulation and optimization parameters} \label{app:SimParams}

\subsection{GSet benchmark graph specifications}

\begin{table}[ht!]
  \centering
  \setlength{\extrarowheight}{3pt}
  \caption{GSet graphs used for benchmarking with optimum cut values for Max-3-Cut and Max-4-Cut.}
  \label{tab:gset_graphs}
  \begin{tabular}{lcccc}
    \toprule
    Graph
    & Vertices
    & Edges
    & \makecell{Optimum\\Max-3-Cut}
    & \makecell{Optimum\\Max-4-Cut}
    \\
    \midrule
    G1 & 800 & 19176 & 15165 & 16803 \\
    G2 & 800 & 19176 & 15172 & 16809 \\
    G3 & 800 & 19176 & 15173 & 16806 \\
    G4 & 800 & 19176 & 15184 & 16814 \\
    G5 & 800 & 19176 & 15193 & 16816 \\
    \bottomrule
  \end{tabular}
  \caption*{The optimum cut values for the GSet graphs interpreted as Max-3-Cut and Max-4-Cut problems are found in the literature \cite{ma_multiple_2017}.}
\end{table}

In the following subsections, parameter sweeps are denoted by \textit{start : step : end}, where $\mathrm{start}$ is the first value, $\mathrm{step}$ is the increment between values, and $\mathrm{end}$ is the last value. For example, $1:0.1:2$ means that the sweep is equispaced between 1 and 2 in increments of 0.1, i.e. $1, 1.1, 1.2, \ldots, 2$.

\FloatBarrier

\subsection{Non-Equilibrium Condensates (NEC) model}

\begin{table}[h!]
\setlength{\extrarowheight}{3pt}
\centering
\caption{NEC model default parameters. These parameters are used if nothing else is specified.}
\label{tab:nec_params_default}
\begin{tabularx}{\textwidth}{cXc}
\toprule
Symbol & Meaning & Value \\
\midrule
$T$ & Total simulation time & $5\times10^3$ \\
$\Delta t$ & Time step size & $10^{-2}$ \\
$n_\mathrm{steps}$ & Number of time steps & $5\times 10^5$ \\
$n_\mathrm{runs}$ & Number of runs per hyperparameter combination & 100 \\
$f_\xi$ & Noise factor & $10^{-4}$ \\
$\varepsilon_\alpha$ & Gain relaxation rate & $10^{-2}$ \\
$\alpha_\mathrm{0}$ & Initial value of $\alpha$ & $\alpha_\mathrm{th}$ \\
$\alpha_\mathrm{th}$ & Threshold value of $\alpha$ for non-zero fixed points & Eq. \eqref{eq:GeneralAlphaBetaThreshold} \\
$\gamma$ & Strength of phase discretization & $\gamma_\mathrm{0} + \varepsilon_\gamma t$ \\
$\gamma_\mathrm{0}$ & Initial value of $\gamma$ & $0$ \\
\bottomrule
\end{tabularx}
\end{table}

\begin{table}[h!]
\setlength{\extrarowheight}{3pt}
\centering
\caption{NEC model parameters for the dynamic property study, i.e. figure \ref{fig:dynamics} in the main text.}
\label{tab:nec_params_dynamics}
\begin{tabularx}{\textwidth}{cXc}
\toprule
Symbol & Meaning & Value \\
\midrule
$T$ & Total simulation time & $10^3$ \\
$n_\mathrm{steps}$ & Number of time steps & $10^5$ \\
$n$ & Nonlinearity order & 3 \\
$\varepsilon_\alpha$ & Gain relaxation rate & $4\times10^{-2}$ \\
$\varepsilon_\gamma$ & Rate of increase of $\gamma$ & $3\times 10^{-3} $ \\ 
$r_\mathrm{target}$ & Target amplitude & $1$ \\
\bottomrule
\end{tabularx}
\end{table}

\begin{table}[h!]
\setlength{\extrarowheight}{3pt}
\centering
\caption{NEC model GSet Max-3-Cut benchmark optimization sweep parameters.}
\label{tab:nec_params_benchmark_k3}
\begin{tabularx}{\textwidth}{cXc}
\toprule
Symbol & Meaning & Value \\
\midrule
$n$ & Nonlinearity order & 3 \\
$r_\mathrm{target}$ & Target amplitude & $1.5 : 0.1 : 2.5$ \\
$\varepsilon_\gamma$ & Rate of increase of $\gamma$ & $2\times 10^{-4} : 1\times 10^{-4} : 3\times 10^{-3}$ \\
\bottomrule
\end{tabularx}
\end{table}

\begin{table}[h!]
\setlength{\extrarowheight}{3pt}
\centering
\caption{NEC model GSet Max-4-Cut benchmark optimization sweep parameters.}
\label{tab:nec_params_benchmark_k4}
\begin{tabularx}{\textwidth}{cXc}
\toprule
Symbol & Meaning & Value \\
\midrule
$n$ & Nonlinearity order & 4 \\
$r_\mathrm{target}$ & Target amplitude & $1 : 0.2 : 3$ \\
$\varepsilon_\gamma$ & Rate of increase of $\gamma$ & $0 : 4\times 10^{-5} : 4\times 10^{-4}$ \\
\bottomrule
\end{tabularx}
\end{table}

\begin{table}[h!]
\setlength{\extrarowheight}{3pt}
\centering
\caption{NEC model ER50 Max-3-Cut benchmark optimization sweep parameters.}
\label{tab:nec_params_benchmark_rudy50}
\begin{tabularx}{\textwidth}{cXc}
\toprule
Symbol & Meaning & Value \\
\midrule
$n$ & Nonlinearity order & 3 \\
$r_\mathrm{target}$ & Target amplitude & $0 : 0.25 : 4$ \\
$\varepsilon_\gamma$ & Rate of increase of $\gamma$ & $1\times 10^{-3} : 1\times 10^{-3} : 2.5\times 10^{-2}$ \\
\bottomrule
\end{tabularx}
\end{table}

\FloatBarrier

\subsection{$q$-Photon Down-Conversion ($q$-PDC) model} \label{app:sim_settings_qpdc}

\begin{table}[h!]
\setlength{\extrarowheight}{3pt}
\centering
\caption{$q$-PDC model default parameters. These parameters are used if nothing else is specified.}
\label{tab:qpdc_params_default}
\begin{tabularx}{\textwidth}{cXc}
\toprule
Symbol & Meaning & Value \\
\midrule
$T$ & Total simulation time & $50$ \\
$\Delta t$ & Time step size & $10^{-3}$ \\
$n_\mathrm{steps}$ & Number of time steps & $5\times 10^4$ \\
$n_\mathrm{runs}$ & Number of runs per hyperparameter combination & 100 \\
$f_\xi$ & Noise factor & $10^{-4}$ \\
$\beta$ & Global coupling strength & Eq. \eqref{eq:qpdc_beta} \\
$\kappa_l$ & Intrinsic loss & $1$ as per \cite{honari-latifpour_combinatorial_2022} \\
$\gamma$ &  Strength of phase discretization & $\gamma_\mathrm{0} + \varepsilon_\gamma t$ \\
$\gamma_\mathrm{0}$ & Initial value of $\gamma$ & $0$ \\
$\gamma_\mathrm{th}$ & Threshold value of $\gamma$ for non-zero fixed points & $\frac{4}{3^{3/4}}$ \\
\bottomrule
\end{tabularx}
\end{table}

\begin{table}[h!]
\setlength{\extrarowheight}{3pt}
\centering
\caption{$q$-PDC model parameters for the dynamic property study, i.e. figure \ref{fig:dynamics} in the main text.}
\label{tab:qpdc_params_dynamics}
\begin{tabularx}{\textwidth}{cXc}
\toprule
Symbol & Meaning & Value \\
\midrule
$T$ & Total simulation time & $10$ \\
$n_\mathrm{steps}$ & Number of time steps & $10^4$ \\
$\varepsilon_\gamma$ & Rate of increase of $\gamma$ & $0.5 \gamma_\mathrm{th}$ \\
\bottomrule
\end{tabularx}
\end{table}

\begin{table}[h!]
\setlength{\extrarowheight}{3pt}
\centering
\caption{$q$-PDC model GSet Max-3-Cut and Max-4-Cut benchmark optimization sweep parameters.}
\label{tab:qpdc_params_benchmark}
\begin{tabularx}{\textwidth}{cXc}
\toprule
Symbol & Meaning & Value \\
\midrule
$\varepsilon_\gamma$ & Rate of increase of $\gamma$ & $0 : 0.04 \gamma_\mathrm{th} : \gamma_\mathrm{th}$  \\
\bottomrule
\end{tabularx}
\end{table}

\begin{table}[h!]
\setlength{\extrarowheight}{3pt}
\centering
\caption{$q$-PDC model ER50 Max-3-Cut benchmark optimization sweep parameters.}
\label{tab:qpdc_params_benchmark_rudy50}
\begin{tabularx}{\textwidth}{cXc}
\toprule
Symbol & Meaning & Value \\
\midrule
$\varepsilon_\gamma$ & Rate of increase of $\gamma$ & $0 : 0.1 \gamma_\mathrm{th} : 2 \gamma_\mathrm{th}$  \\
\bottomrule
\end{tabularx}
\end{table}

\FloatBarrier

\subsection{$q^\mathrm{th}$-harmonic Sub-Harmonic Injection Locking ($q$-SHIL) model}

\begin{table}[h!]
\setlength{\extrarowheight}{3pt}
\centering
\caption{$q$-SHIL model default parameters. These parameters are used if nothing else is specified.}
\label{tab:qshil_params_default}
\begin{tabularx}{\textwidth}{cXc}
\toprule
Symbol & Meaning & Value \\
\midrule
$T$ & Total simulation time & $10^3$ \\
$\Delta t$ & Time step size & $10^{-3}$ \\
$n_\mathrm{steps}$ & Number of time steps & $10^6$ \\
$n_\mathrm{runs}$ & Number of runs per hyperparameter combination & 100 \\
$f_\xi$ & Noise factor & $10^{-4}$ \\
$\gamma$ & Strength of phase discretization & $\gamma_\mathrm{0} + \varepsilon_\gamma t$ \\
$\gamma_\mathrm{0}$ & Initial value of $\gamma$ & $0$ \\
$\varepsilon_\gamma$ & Rate of increase of $\gamma$ & $10^{-2}$ \\
$\mathbf{\theta}_0$ & Initial phase vector & Chosen from uniform distribution over $[-\pi, \pi]$ \\
\bottomrule
\end{tabularx}
\end{table}

\begin{table}[h!]
\setlength{\extrarowheight}{3pt}
\centering
\caption{$q$-SHIL model parameters for the dynamic property study, i.e. figure \ref{fig:dynamics} in the main text.}
\label{tab:qshil_params_dynamics}
\begin{tabularx}{\textwidth}{cXc}
\toprule
Symbol & Meaning & Value \\
\midrule
$\Delta t$ & Time step size & $10^{-2}$ \\
$n_\mathrm{steps}$ & Number of time steps & $10^5$ \\
$\varepsilon_\gamma$ & Rate of increase of $\gamma$ & $2\times 10^{-3}$ \\
\bottomrule
\end{tabularx}
\end{table}

\FloatBarrier

\subsection{Polynomial Potts machine model}

\begin{table}[h!]
\setlength{\extrarowheight}{3pt}
\centering
\caption{Polynomial Potts machine model default parameters. These parameters are used if nothing else is specified.}
\label{tab:polynomial_potts_params_default}
\begin{tabularx}{\textwidth}{cXc}
\toprule
Symbol & Meaning & Value \\
\midrule
$T$ & Total simulation time & $10^3$ \\
$\Delta t$ & Time step size & $10^{-3}$ \\
$n_\mathrm{steps}$ & Number of time steps & $10^6$ \\
$n_\mathrm{runs}$ & Number of runs per hyperparameter combination & 100 \\
$f_\xi$ & Noise factor & $10^{-4}$ \\
$\beta$ & Global coupling strength & $\beta_0 + \varepsilon_\beta t$ \\
$\beta_0$ & Initial value of $\beta$ & $\beta_\mathrm{th}$ \\
$\beta_\mathrm{th}$ & Threshold value of $\beta$ for non-zero fixed points & Eq. \eqref{eq:GeneralAlphaBetaThreshold} \\
$\varepsilon_\beta$ & Rate of increase of $\beta$ & $2\times 10^{-3}$ \\
$\gamma$ & Strength of phase discretization & $f_\gamma \beta$ \\   
\bottomrule
\end{tabularx}
\end{table}

\begin{table}[h!]
\setlength{\extrarowheight}{3pt}
\centering
\caption{Polynomial Potts machine model parameters for the dynamic property study, i.e. figure \ref{fig:dynamics} in the main text.}
\label{tab:polynomial_potts_params_dynamics}
\begin{tabularx}{\textwidth}{cXc}
\toprule
Symbol & Meaning & Value \\
\midrule
$\Delta t$ & Time step size & $10^{-2}$ \\
$n_\mathrm{steps}$ & Number of time steps & $10^5$ \\
$n$ & Nonlinearity order & 3 \\
$\varepsilon_\beta$ & Rate of increase of $\beta$ & $2\times 10^{-4}$ \\
$f_\gamma$ & Phase discretization vs. coupling factor & $1$ \\
\bottomrule
\end{tabularx}
\end{table}

\begin{table}[h!]
\setlength{\extrarowheight}{3pt}
\centering
\caption{Polynomial Potts machine model GSet Max-3-Cut benchmark optimization sweep parameters.}
\label{tab:polynomial_potts_params_benchmark_k3}
\begin{tabularx}{\textwidth}{cXc}
\toprule
Symbol & Meaning & Value \\
\midrule
$n$ & Nonlinearity order & $3:1:11$ \\
$f_\gamma$ & Phase discretization vs. coupling factor & $0 : 0.2 : 3$ \\
\bottomrule
\end{tabularx}
\end{table}

\begin{table}[h!]
\setlength{\extrarowheight}{3pt}
\centering
\caption{Polynomial Potts machine model GSet Max-4-Cut benchmark optimization sweep parameters.}
\label{tab:polynomial_potts_params_benchmark_k4}
\begin{tabularx}{\textwidth}{cXc}
\toprule
Symbol & Meaning & Value \\
\midrule
$n$ & Nonlinearity order & $4:1:11$ \\
$f_\gamma$ & Phase discretization versus coupling factor & $0 : 5\times 10^{-2} : 1$ \\
\bottomrule
\end{tabularx}
\end{table}

\begin{table}[h!]
\setlength{\extrarowheight}{3pt}
\centering
\caption{Polynomial Potts machine model ER50 Max-3-Cut benchmark optimization sweep parameters.}
\label{tab:polynomial_potts_params_benchmark_rudy50}
\begin{tabularx}{\textwidth}{cXc}
\toprule
Symbol & Meaning & Value \\
\midrule
$n$ & Nonlinearity order & $3:1:10$ \\
$f_\gamma$ & Phase discretization versus coupling factor & $0 : 0.5 : 10$ \\
\bottomrule
\end{tabularx}
\end{table}

\FloatBarrier

\subsection{Sigmoid Potts machine model}

\begin{table}[h!]
\setlength{\extrarowheight}{3pt}
\centering
\caption{Sigmoid Potts machine model default parameters. These parameters are used if nothing else is specified.}
\label{tab:sigmoid_potts_params_default}
\begin{tabularx}{\textwidth}{cXc}
\toprule
Symbol & Meaning & Value \\
\midrule
$T$ & Total simulation time & $2\times10^3$ \\
$\Delta t$ & Time step size & $10^{-3}$ \\
$n_\mathrm{steps}$ & Number of time steps & $10^6$ \\
$n_\mathrm{runs}$ & Number of runs per hyperparameter combination & 100 \\
$f_\xi$ & Noise factor & $10^{-4}$ \\
$\beta$ & Global coupling strength & $\beta_0 + \varepsilon_\beta t$ \\
$\beta_0$ & Initial value of $\beta$ & $\beta_\mathrm{th}$ \\
$\beta_\mathrm{th}$ & Threshold value of $\beta$ for non-zero fixed points & Eq. \eqref{eq:sigmoid_beta_threshold} \\
$\varepsilon_\beta$ & Rate of increase of $\beta$ & $10^{-2}$ \\
$\gamma$ & Strength of phase discretization & $f_\gamma \beta$ \\
\bottomrule
\end{tabularx}
\end{table}

\begin{table}[h!]
\setlength{\extrarowheight}{3pt}
\centering
\caption{Sigmoid Potts machine model parameters for the dynamic property study, i.e. figure \ref{fig:dynamics} in the main text.}
\label{tab:sigmoid_potts_params_dynamics}
\begin{tabularx}{\textwidth}{cXc}
\toprule
Symbol & Meaning & Value \\
\midrule
$T$ & Total simulation time & $10^3$ \\
$\Delta t$ & Time step size & $10^{-2}$ \\
$n_\mathrm{steps}$ & Number of time steps & $10^5$ \\
$\varepsilon_\beta$ & Rate of increase of $\beta$ & $2\times 10^{-3}$ \\
$f_\gamma$ & Phase discretization versus coupling factor & $2$ \\
\bottomrule
\end{tabularx}
\end{table}

\begin{table}[h!]
\setlength{\extrarowheight}{3pt}
\centering
\caption{Sigmoid Potts machine model GSet Max-3-Cut benchmark optimization sweep parameters.}
\label{tab:sigmoid_potts_params_benchmark_k3}
\begin{tabularx}{\textwidth}{cXc}
\toprule
Symbol & Meaning & Value \\
\midrule
$\alpha$ & Amplitude gain & $-50 : 10 : 0$ \\
$f_\gamma$ & Phase discretization versus coupling factor & $0 : 0.5 : 8$ \\
\bottomrule
\end{tabularx}
\caption*{The sweep of $\alpha$ is limited to a minimum value of $-50$ to limit computational time, as lower values slow spin dynamics as mentioned in the main text section \ref{sec:sigmoid}.}
\end{table}

\begin{table}[h!]
\setlength{\extrarowheight}{3pt}
\centering
\caption{Sigmoid Potts machine model GSet Max-4-Cut benchmark optimization sweep parameters.}
\label{tab:sigmoid_potts_params_benchmark_k4}
\begin{tabularx}{\textwidth}{cXc}
\toprule
Symbol & Meaning & Value \\
\midrule
$\alpha$ & Amplitude gain & $-50 : 20 : 10$ \\
$f_\gamma$ & Phase discretization versus coupling factor & $0 : 0.1 : 2$ \\
\bottomrule
\end{tabularx}
\caption*{The sweep of $\alpha$ is limited to a minimum value of $-50$ to limit computational time, as lower values slow spin dynamics as mentioned in the main text section \ref{sec:sigmoid}.}
\end{table}

\begin{table}[h!]
\setlength{\extrarowheight}{3pt}
\centering
\caption{Sigmoid Potts machine model ER50 Max-3-Cut benchmark optimization sweep parameters.}
\label{tab:sigmoid_potts_params_benchmark_rudy50}
\begin{tabularx}{\textwidth}{cXc}
\toprule
Symbol & Meaning & Value \\
\midrule
$\alpha$ & Amplitude gain & $-50 : 10 : 0$ \\
$f_\gamma$ & Phase discretization versus coupling factor & $0 : 0.5 : 8$ \\
\bottomrule
\end{tabularx}
\caption*{The sweep of $\alpha$ is limited to a minimum value of $-50$ to limit computational time, as lower values slow spin dynamics as mentioned in the main text section \ref{sec:sigmoid}.}
\end{table}

\FloatBarrier

\subsection{Reference Ising machine model}

\begin{table}[h!]
\setlength{\extrarowheight}{3pt}
\centering
\caption{Reference Ising machine model default parameters. These parameters are used if nothing else is specified.}
\label{tab:sigmoid_ising_params_default}
\begin{tabularx}{\textwidth}{cXc}
\toprule
Symbol & Meaning & Value \\
\midrule
$T$ & Total simulation time & $2\times10^3$ \\
$\Delta t$ & Time step size & $10^{-3}$ \\
$n_\mathrm{steps}$ & Number of time steps & $10^6$ \\
$n_\mathrm{runs}$ & Number of runs per hyperparameter combination & 100 \\
$f_\xi$ & Noise factor & $10^{-4}$ \\
$\alpha$ & Amplitude gain & $-50$ \\
$\beta$ & Global coupling strength & $\beta_0 + \varepsilon_\beta t$ \\
$\beta_0$ & Initial value of $\beta$ & 0 \\
$\varepsilon_\beta$ & Rate of increase of $\beta$ & $10^{-5}$ \\
\bottomrule
\end{tabularx}
\caption*{The value of $\alpha$ is set to $-50$, and not lower, in order to limit computational time, as lower values slow spin dynamics as mentioned in the main text section \ref{sec:sigmoid_im}.}
\end{table}

\begin{table}[h!]
\setlength{\extrarowheight}{3pt}
\centering
\caption{Reference Ising machine model parameters for the dynamic property study, i.e. figure \ref{fig:dynamics} in the main text.}
\label{tab:sigmoid_ising_params_dynamics}
\begin{tabularx}{\textwidth}{cXc}
\toprule
Symbol & Meaning & Value \\
\midrule
$T$ & Total simulation time & $10^3$ \\
$\Delta t$ & Time step size & $10^{-2}$ \\
$n_\mathrm{steps}$ & Number of time steps & $10^5$ \\
$\alpha$ & Amplitude gain & $0.9$ \\
$\varepsilon_\beta$ & Rate of increase of $\beta$ & $4\times10^{-4}$ \\
$B/A$ & Edge- vs. one-hot constraint factor (see \cite{prins_how_2025}) & 1 \\
$\zeta$ & External field scaling factor (see \cite{prins_how_2025}) & 0.5 \\
\bottomrule
\end{tabularx}
\end{table}

\begin{table}[h!]
\setlength{\extrarowheight}{3pt}
\centering
\caption{Reference Ising machine model GSet Max-3-Cut benchmark optimization sweep parameters.}
\label{tab:sigmoid_ising_params_benchmark_k3}
\begin{tabularx}{\textwidth}{cXc}
\toprule
Symbol & Meaning & Value \\
\midrule
$B/A$ & Edge- vs. one-hot constraint factor (see \cite{prins_how_2025}) & $100 n_\mathrm{vertices}^{-1} : 10 n_\mathrm{vertices}^{-1} : 200 n_\mathrm{vertices}^{-1}$ \\
$\zeta$ & External field scaling factor (see \cite{prins_how_2025}) & $0.4 : 0.05 : 0.7$ \\
\bottomrule
\end{tabularx}
\end{table}

\begin{table}[h!]
\setlength{\extrarowheight}{3pt}
\centering
\caption{Reference Ising machine model GSet Max-4-Cut benchmark optimization sweep parameters.}
\label{tab:sigmoid_ising_params_benchmark_k4}
\begin{tabularx}{\textwidth}{cXc}
\toprule
Symbol & Meaning & Value \\
\midrule
$B/A$ & Edge- vs. one-hot constraint factor (see \cite{prins_how_2025}) & $150 n_\mathrm{vertices}^{-1} : 10 n_\mathrm{vertices}^{-1} : 250 n_\mathrm{vertices}^{-1}$ \\
$\zeta$ & External field scaling factor (see \cite{prins_how_2025}) & $0.5 : 0.05 : 0.8$ \\
\bottomrule
\end{tabularx}
\end{table}

\begin{table}[h!]
\setlength{\extrarowheight}{3pt}
\centering
\caption{Reference Ising machine model ER50 Max-3-Cut benchmark optimization sweep parameters.}
\label{tab:sigmoid_ising_params_benchmark_rudy50}
\begin{tabularx}{\textwidth}{cXc}
\toprule
Symbol & Meaning & Value \\
\midrule
$B/A$ & Edge- vs. one-hot constraint factor (see \cite{prins_how_2025}) & $5 n_\mathrm{vertices}^{-1} : 5 n_\mathrm{vertices}^{-1} : 100 n_\mathrm{vertices}^{-1}$ \\
$\zeta$ & External field scaling factor (see \cite{prins_how_2025}) & $0.1 : 0.05 : 1.0$ \\
\bottomrule
\end{tabularx}
\end{table}

\FloatBarrier
\section{Benchmark results} \label{app:benchmark_results}

\subsection{Optimality gap distribution for best hyperparameter combinations}

\begin{sidewaystable}[h!]
\setlength{\extrarowheight}{2pt}
\centering
\caption{Benchmark results for studied Potts machine and Ising machine models on Max-3-Cut problems with best hyperparameter combination.}
\label{tab:benchmark_results_k3}
\begin{tabularx}{\textwidth}{cccccccc}
\toprule
Model & Graph & Hyperparameter 1 & Hyperparameter 2 & \makecell{Mean relative\\optimality gap (\%)} & \makecell{Median relative\\optimality gap (\%)} & \makecell{Minimum relative\\optimality gap (\%)} & \makecell{SD on relative\\optimality gap (\%)} \\
\midrule
\multirow{5}{*}{NEC}
  & G1 & $f_\gamma=8.00$ & $r_\mathrm{target}=2.10$ & 0.25 & 0.24 & 0.16 & 0.05 \\
  & G2 & $f_\gamma=14.50$ & $r_\mathrm{target}=1.90$ & 0.25 & 0.23 & 0.13 & 0.11 \\
  & G3 & $f_\gamma=12.50$ & $r_\mathrm{target}=2.20$ & 0.21 & 0.17 & 0.13 & 0.08 \\
  & G4 & $f_\gamma=10.50$ & $r_\mathrm{target}=2.20$ & 0.26 & 0.20 & 0.14 & 0.09 \\
  & G5 & $f_\gamma=10.00$ & $r_\mathrm{target}=2.10$ & 0.26 & 0.25 & 0.15 & 0.05 \\
\midrule
\multirow{5}{*}{$q$-PDC}
  & G1 & $\varepsilon_\gamma / \gamma_\mathrm{th}=0.48$ & --- & 4.01 & 4.04 & 3.37 & 0.24 \\
  & G2 & $\varepsilon_\gamma / \gamma_\mathrm{th}=0.48$ & --- & 4.02 & 4.01 & 3.34 & 0.25 \\
  & G3 & $\varepsilon_\gamma / \gamma_\mathrm{th}=0.48$ & --- & 4.05 & 4.06 & 3.38 & 0.24 \\
  & G4 & $\varepsilon_\gamma / \gamma_\mathrm{th}=0.48$ & --- & 4.07 & 4.10 & 3.54 & 0.24 \\
  & G5 & $\varepsilon_\gamma / \gamma_\mathrm{th}=0.48$ & --- & 4.15 & 4.15 & 3.69 & 0.25 \\
\midrule
\multirow{5}{*}{$q$-SHIL}
  & G1 & --- & --- & 1.45 & 1.45 & 0.99 & 0.16 \\
  & G2 & --- & --- & 1.42 & 1.42 & 0.97 & 0.17 \\
  & G3 & --- & --- & 1.44 & 1.44 & 1.01 & 0.19 \\
  & G4 & --- & --- & 1.40 & 1.40 & 1.07 & 0.15 \\
  & G5 & --- & --- & 1.47 & 1.48 & 1.02 & 0.16 \\
\midrule
\multirow{5}{*}{Polynomial PM}
  & G1 & $n=3.00$ & $f_\gamma=2.80$ & 0.85 & 0.84 & 0.42 & 0.15 \\
  & G2 & $n=3.00$ & $f_\gamma=2.80$ & 0.84 & 0.85 & 0.57 & 0.11 \\
  & G3 & $n=3.00$ & $f_\gamma=2.80$ & 0.78 & 0.76 & 0.47 & 0.16 \\
  & G4 & $n=3.00$ & $f_\gamma=3.80$ & 0.85 & 0.86 & 0.58 & 0.12 \\
  & G5 & $n=3.00$ & $f_\gamma=3.00$ & 0.81 & 0.80 & 0.45 & 0.16 \\
\midrule
\multirow{5}{*}{Sigmoid PM}
  & G1 & $f_\gamma=5.00$ & $\alpha=-50.00$ & 0.67 & 0.68 & 0.35 & 0.12 \\
  & G2 & $f_\gamma=5.50$ & $\alpha=-50.00$ & 0.61 & 0.60 & 0.44 & 0.07 \\
  & G3 & $f_\gamma=4.50$ & $\alpha=-50.00$ & 0.49 & 0.36 & 0.34 & 0.17 \\
  & G4 & $f_\gamma=5.00$ & $\alpha=-40.00$ & 0.53 & 0.57 & 0.40 & 0.09 \\
  & G5 & $f_\gamma=3.50$ & $\alpha=-50.00$ & 0.59 & 0.56 & 0.56 & 0.08 \\
\midrule
\multirow{5}{*}{Reference IM}
  & G1 & $(B/A)n_\mathrm{vertices}=200.00$ & $\zeta=0.55$ & 0.09 & 0.09 & 0.00 & 0.05 \\
  & G2 & $(B/A)n_\mathrm{vertices}=200.00$ & $\zeta=0.55$ & 0.12 & 0.11 & 0.05 & 0.04 \\
  & G3 & $(B/A)n_\mathrm{vertices}=190.00$ & $\zeta=0.55$ & 0.08 & 0.08 & 0.01 & 0.04 \\
  & G4 & $(B/A)n_\mathrm{vertices}=200.00$ & $\zeta=0.50$ & 0.14 & 0.14 & 0.00 & 0.05 \\
  & G5 & $(B/A)n_\mathrm{vertices}=200.00$ & $\zeta=0.50$ & 0.06 & 0.03 & 0.00 & 0.07 \\
\bottomrule
\end{tabularx}
\end{sidewaystable}

\begin{sidewaystable}[h!]
\setlength{\extrarowheight}{2pt}

\centering
\caption{Benchmark results for studied Potts machine and Ising machine models on Max-4-Cut problems with best hyperparameter combination.}
\label{tab:benchmark_results_k4}
\begin{tabularx}{\textwidth}{cccccccc}
\toprule
Model & Graph & Hyperparameter 1 & Hyperparameter 2 & \makecell{Mean relative\\optimality gap (\%)} & \makecell{Median relative\\optimality gap (\%)} & \makecell{Minimum relative\\optimality gap (\%)} & \makecell{SD on relative\\optimality gap (\%)} \\
\midrule
\multirow{5}{*}{NEC}
  & G1 & $f_\gamma=1.80$ & $r_\mathrm{target}=2.60$ & 3.32 & 3.29 & 3.12 & 0.11 \\
  & G2 & $f_\gamma=0.40$ & $r_\mathrm{target}=2.60$ & 3.32 & 3.28 & 3.28 & 0.12 \\
  & G3 & $f_\gamma=1.20$ & $r_\mathrm{target}=2.00$ & 3.24 & 3.21 & 3.17 & 0.08 \\
  & G4 & $f_\gamma=0.60$ & $r_\mathrm{target}=2.20$ & 3.15 & 3.15 & 3.15 & 0.04 \\
  & G5 & $f_\gamma=0.80$ & $r_\mathrm{target}=1.80$ & 3.35 & 3.32 & 3.21 & 0.12 \\
\midrule
\multirow{5}{*}{$q$-PDC}
  & G1 & $\varepsilon_\gamma / \gamma_\mathrm{th}=0.20$ & --- & 6.03 & 6.00 & 5.35 & 0.29 \\
  & G2 & $\varepsilon_\gamma / \gamma_\mathrm{th}=0.20$ & --- & 6.03 & 6.03 & 5.23 & 0.28 \\
  & G3 & $\varepsilon_\gamma / \gamma_\mathrm{th}=0.24$ & --- & 6.06 & 5.99 & 5.33 & 0.36 \\
  & G4 & $\varepsilon_\gamma / \gamma_\mathrm{th}=0.20$ & --- & 6.01 & 5.94 & 5.17 & 0.40 \\
  & G5 & $\varepsilon_\gamma / \gamma_\mathrm{th}=0.44$ & --- & 6.08 & 6.05 & 5.29 & 0.35 \\
\midrule
\multirow{5}{*}{$q$-SHIL}
  & G1 & --- & --- & 3.69 & 3.69 & 3.29 & 0.18 \\
  & G2 & --- & --- & 3.68 & 3.68 & 3.22 & 0.15 \\
  & G3 & --- & --- & 3.65 & 3.66 & 3.25 & 0.17 \\
  & G4 & --- & --- & 3.66 & 3.66 & 3.23 & 0.19 \\
  & G5 & --- & --- & 3.67 & 3.68 & 3.07 & 0.17 \\
\midrule
\multirow{5}{*}{Polynomial PM}
  & G1 & $n=9.00$ & $f_\gamma=0.55$ & 3.66 & 3.66 & 3.22 & 0.18 \\
  & G2 & $n=10.00$ & $f_\gamma=0.55$ & 3.64 & 3.64 & 3.18 & 0.18 \\
  & G3 & $n=8.00$ & $f_\gamma=0.75$ & 3.55 & 3.52 & 3.12 & 0.20 \\
  & G4 & $n=10.00$ & $f_\gamma=0.20$ & 3.57 & 3.56 & 3.21 & 0.17 \\
  & G5 & $n=9.00$ & $f_\gamma=0.55$ & 3.63 & 3.62 & 3.16 & 0.19 \\
\midrule
\multirow{5}{*}{Sigmoid PM}
  & G1 & $f_\gamma=0.30$ & $\alpha=-50.00$ & 3.49 & 3.49 & 3.49 & 0.00 \\
  & G2 & $f_\gamma=1.10$ & $\alpha=-50.00$ & 3.43 & 3.43 & 3.43 & 0.00 \\
  & G3 & $f_\gamma=1.00$ & $\alpha=-30.00$ & 3.28 & 3.28 & 3.28 & 0.03 \\
  & G4 & $f_\gamma=1.20$ & $\alpha=-30.00$ & 3.32 & 3.32 & 3.32 & 0.00 \\
  & G5 & $f_\gamma=0.30$ & $\alpha=-30.00$ & 3.29 & 3.29 & 3.29 & 0.00 \\
\midrule
\multirow{5}{*}{Reference IM}
  & G1 & $(B/A)n_\mathrm{vertices}=220.00$ & $\zeta=0.75$ & 0.18 & 0.19 & 0.02 & 0.06 \\
  & G2 & $(B/A)n_\mathrm{vertices}=220.00$ & $\zeta=0.75$ & 0.19 & 0.20 & 0.04 & 0.06 \\
  & G3 & $(B/A)n_\mathrm{vertices}=220.00$ & $\zeta=0.75$ & 0.17 & 0.17 & 0.05 & 0.06 \\
  & G4 & $(B/A)n_\mathrm{vertices}=220.00$ & $\zeta=0.75$ & 0.18 & 0.18 & 0.05 & 0.05 \\
  & G5 & $(B/A)n_\mathrm{vertices}=220.00$ & $\zeta=0.75$ & 0.18 & 0.18 & 0.04 & 0.06 \\
\bottomrule
\end{tabularx}
\end{sidewaystable}

\FloatBarrier

\subsection{Performance sensitivity to hyperparameter settings} \label{app:hyperparam_sensitivity}

\subsubsection{Max-3-Cut problems}

\begin{figure}[htbp]
  \centering
  \includegraphics[scale=.8]{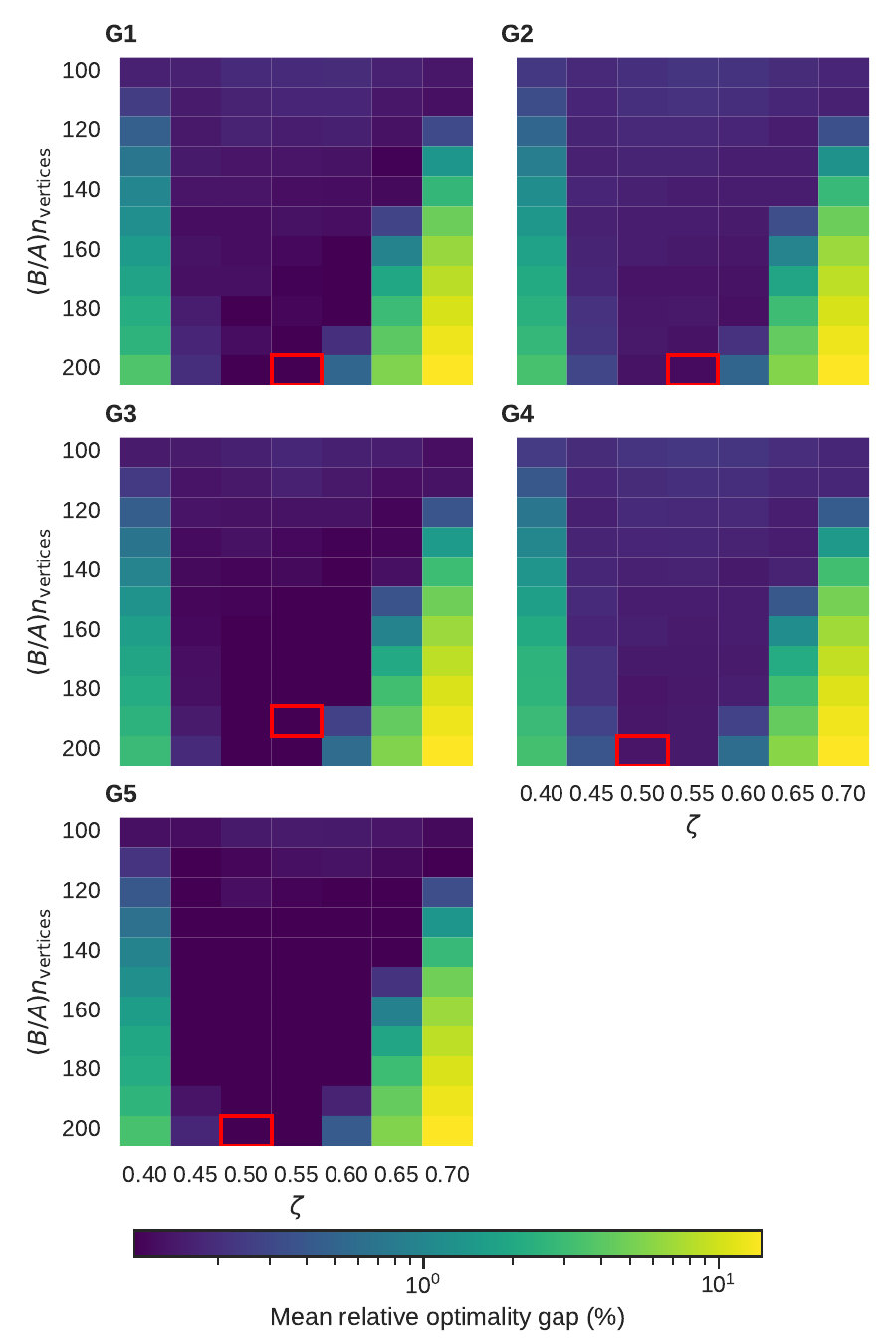}
  \caption{Heatmap showing the mean relative optimality gap (\%) for the reference IM model over the G1-G5 Max-3-Cut problems as a function of the hyperparameters $\alpha$ and $(B/A)n_\mathrm{vertices}$. The values of the mean relative optimality gap are shown in each cell of the heatmap. The hyperparameter configuration with the minimum mean relative optimality gap is marked by a red border.}
  \label{fig:cim_3cut_hyperparam}
\end{figure}

\begin{figure}[htbp]
  \centering
  \includegraphics[scale=.8]{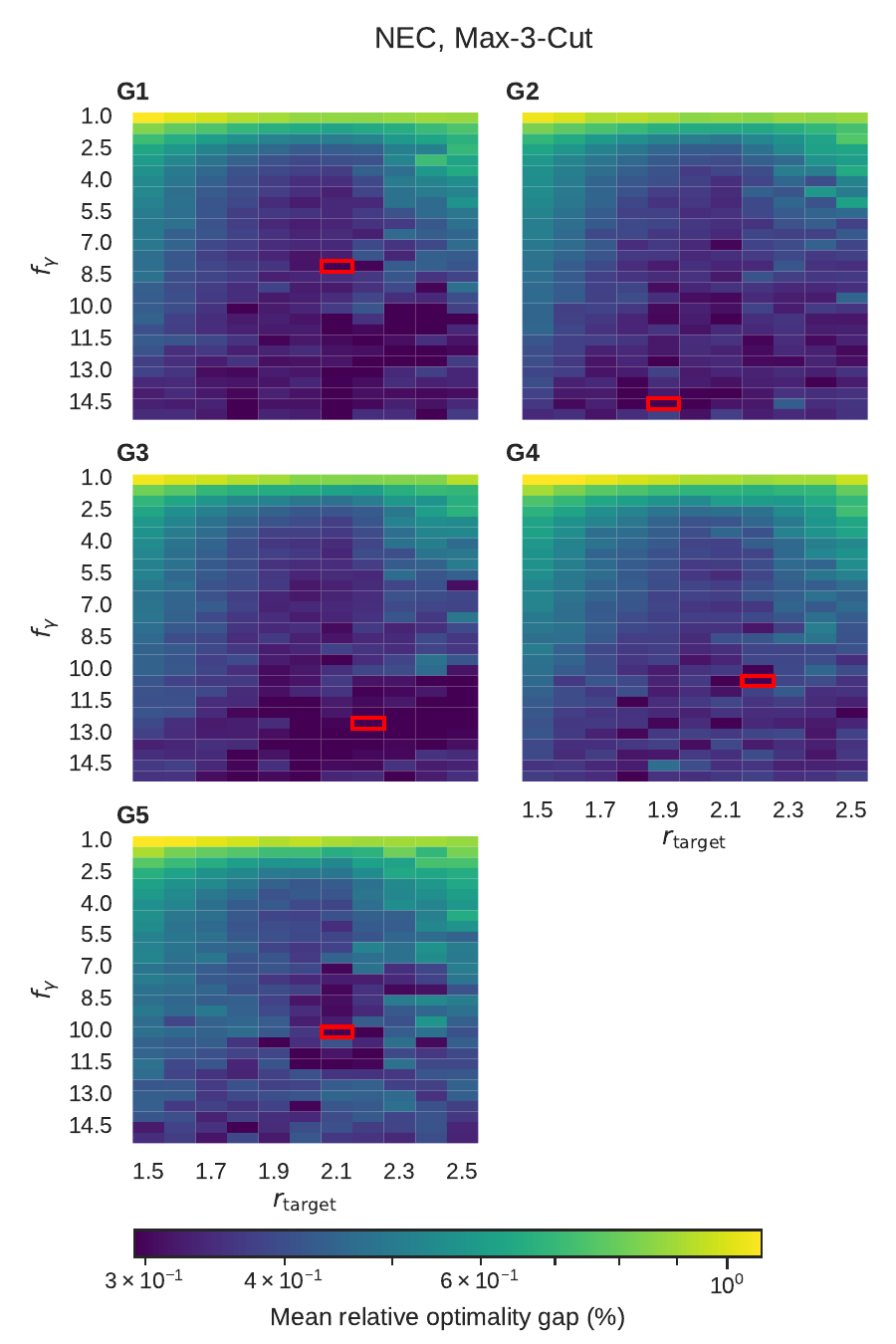}
  \caption{Heatmap showing the mean relative optimality gap (\%) for the NEC model over the G1-G5 Max-3-Cut problems as a function of the hyperparameters $f_\gamma$ and $r_\mathrm{target}$. The values of the mean relative optimality gap are shown in each cell of the heatmap. The hyperparameter configuration with the minimum mean relative optimality gap is marked by a red border.}
  \label{fig:nec_3cut_hyperparam}
\end{figure}

\begin{figure}[htbp]
  \centering
  \includegraphics[scale=.8]{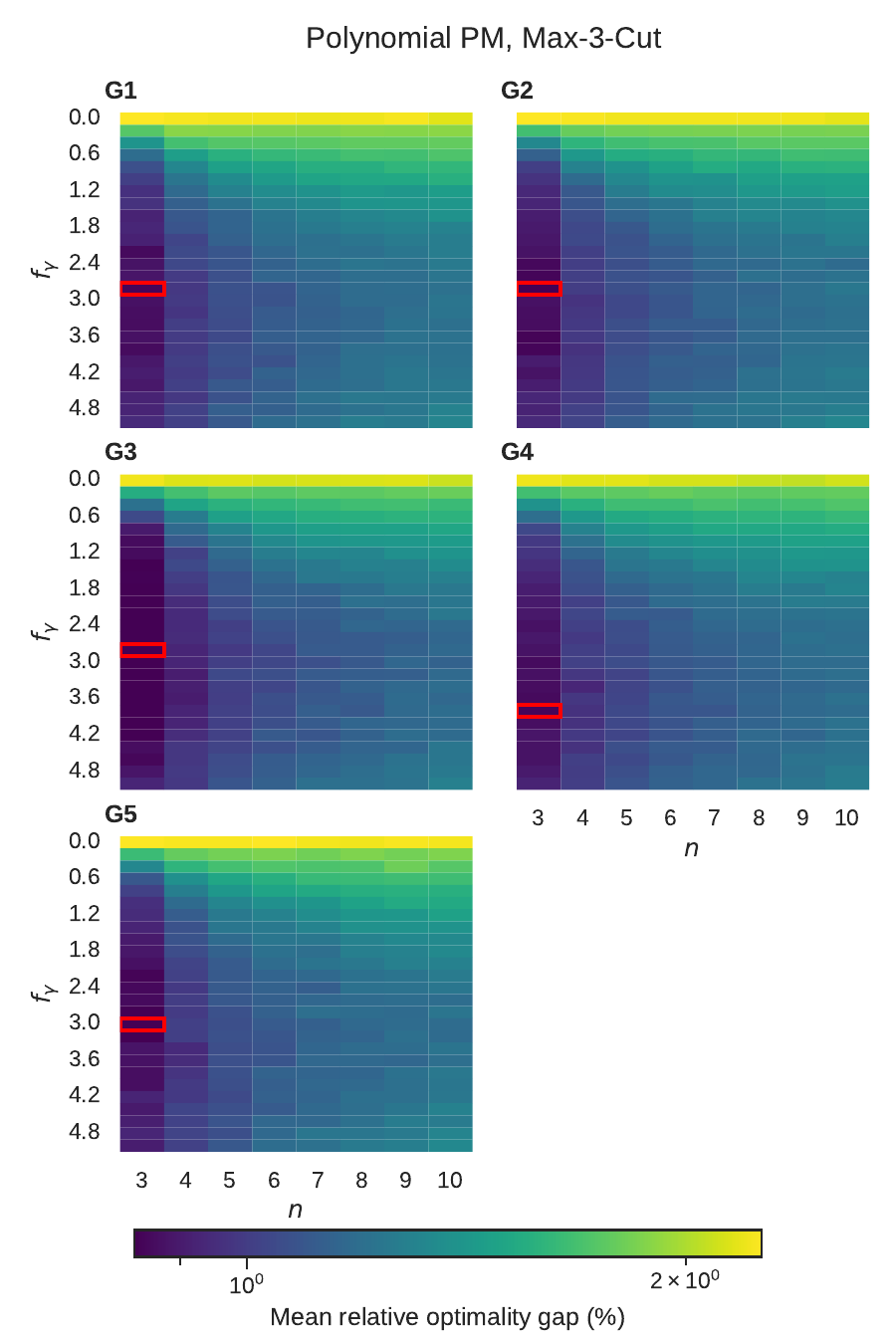}
  \caption{Heatmap showing the mean relative optimality gap (\%) for the polynomial PM model over the G1-G5 Max-3-Cut problems as a function of the hyperparameters $n$ and $f_\gamma$. The values of the mean relative optimality gap are shown in each cell of the heatmap. The hyperparameter configuration with the minimum mean relative optimality gap is marked by a red border.}
  \label{fig:polynomial_3cut_hyperparam}
\end{figure}

\begin{figure}[htbp]
  \centering
  \includegraphics[scale=.8]{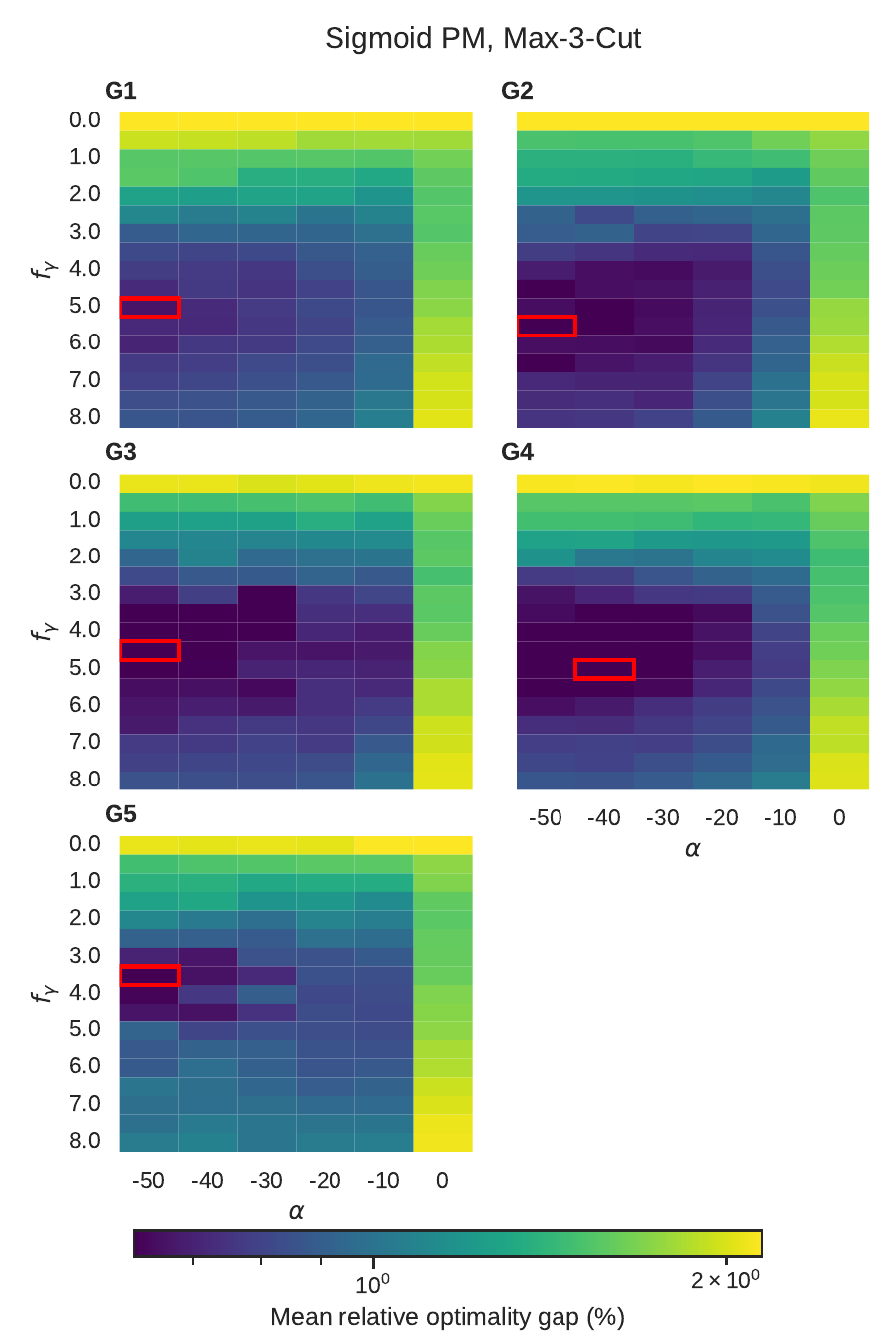}
  \caption{Heatmap showing the mean relative optimality gap (\%) for the sigmoid PM model over the G1-G5 Max-3-Cut problems as a function of the hyperparameters $f_\gamma$ and $\alpha$. The values of the mean relative optimality gap are shown in each cell of the heatmap. The hyperparameter configuration with the minimum mean relative optimality gap is marked by a red border. The sweep of $\alpha$ is limited to a minimum value of $-50$ to limit computational time, as lower values slow spin dynamics as mentioned in the main text section \ref{sec:sigmoid}.}
  \label{fig:sigmoid_3cut_hyperparam}
\end{figure}

\begin{figure}[htbp]
  \centering
  \includegraphics[scale=.8]{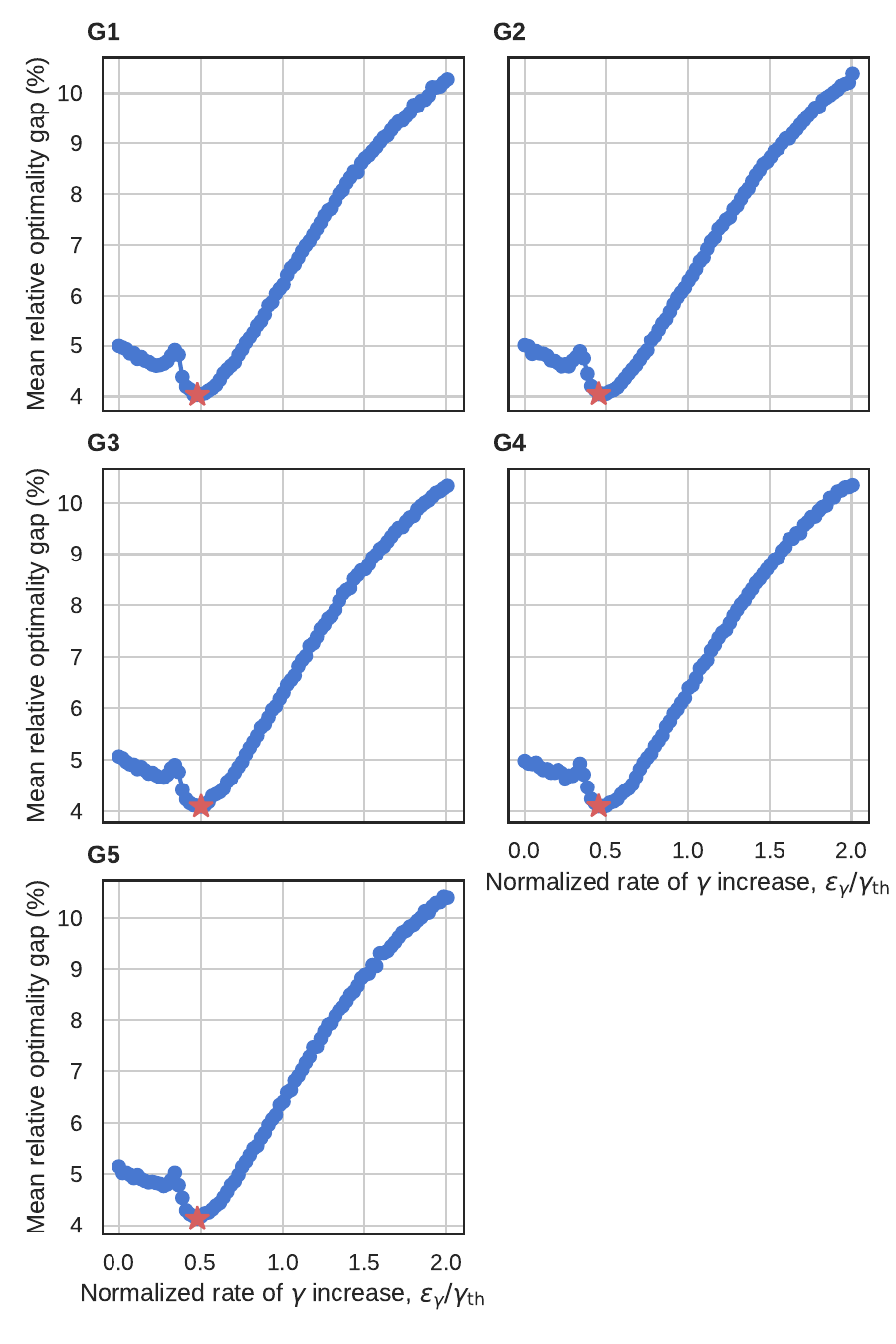}
  \caption{Mean relative optimality gap (\%) for the $q$-PDC model over the G1-G5 Max-3-Cut problems as a function of the hyperparameter $f_\gamma$. The minimum mean relative optimality gap is marked by a red star.}
  \label{fig:q_pdc_3cut_hyperparam}
\end{figure}

\FloatBarrier

\subsubsection{Max-4-Cut problems}

\begin{figure}[htbp]
  \centering
  \includegraphics[scale=.8]{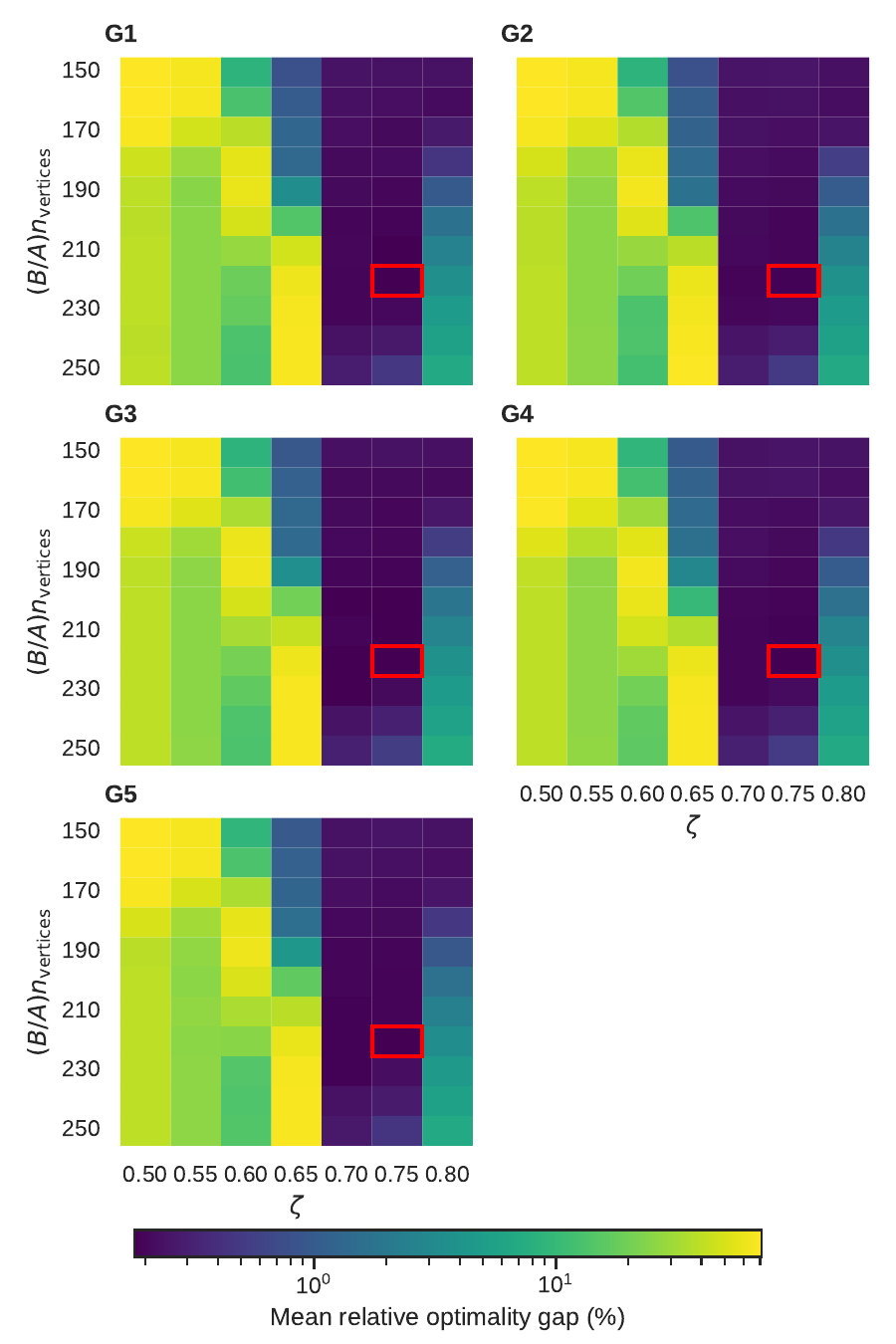}
  \caption{Heatmap showing the mean relative optimality gap (\%) for the reference IM model over the G1-G5 Max-4-Cut problems as a function of the hyperparameters $(B/A)n_\mathrm{vertices}$ and $\zeta$. The values of the mean relative optimality gap are shown in each cell of the heatmap. The hyperparameter configuration with the minimum mean relative optimality gap is marked by a red border.}
  \label{fig:cim_4cut_hyperparam}
\end{figure}

\begin{figure}[htbp]
  \centering
  \includegraphics[scale=.8]{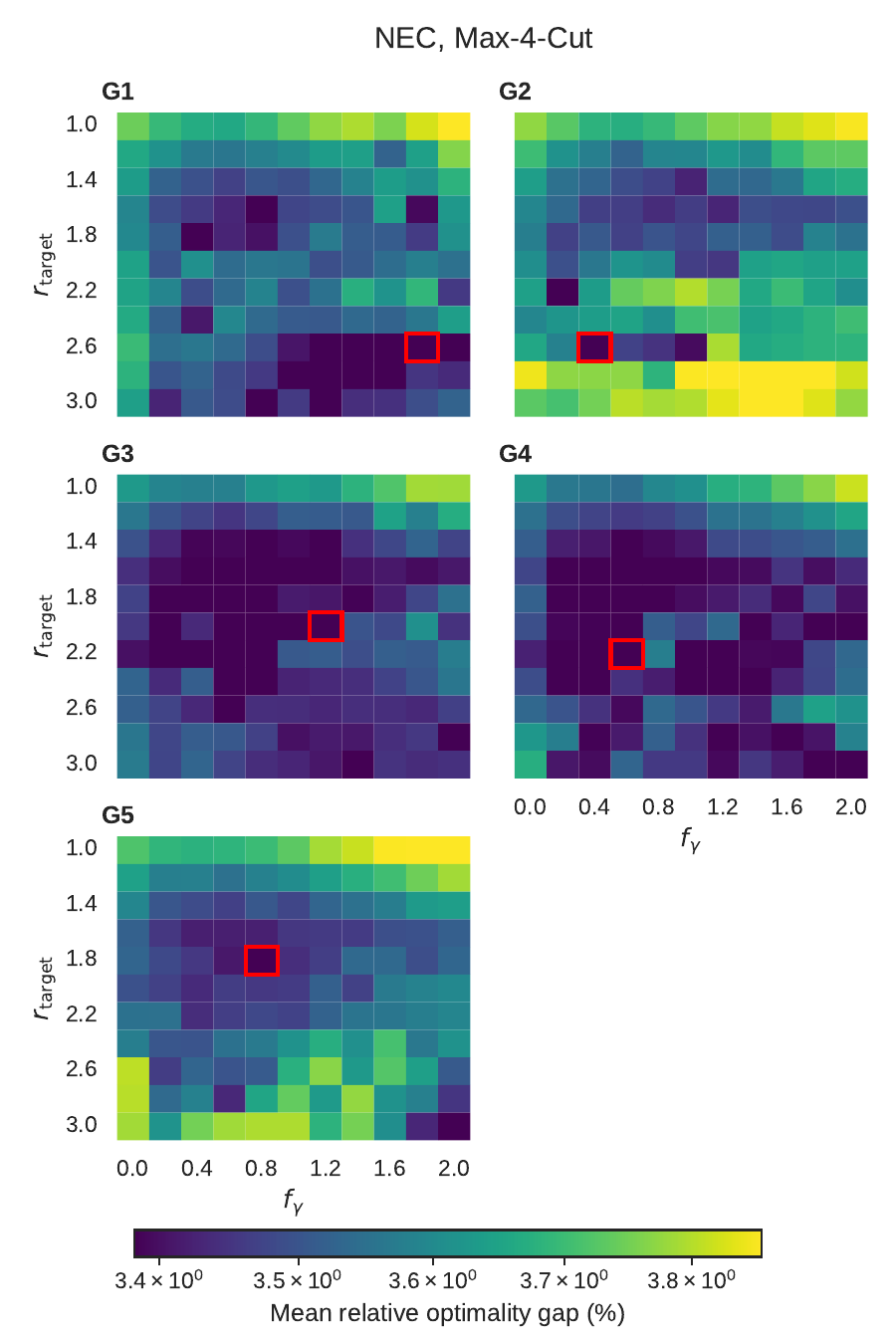}
  \caption{Heatmap showing the mean relative optimality gap (\%) for the NEC model over the G1-G5 Max-4-Cut problems as a function of the hyperparameters $f_\gamma$ and $r_\mathrm{target}$. The values of the mean relative optimality gap are shown in each cell of the heatmap. The hyperparameter configuration with the minimum mean relative optimality gap is marked by a red border.}
  \label{fig:nec_4cut_hyperparam}
\end{figure}

\begin{figure}[htbp]
  \centering
  \includegraphics[scale=.8]{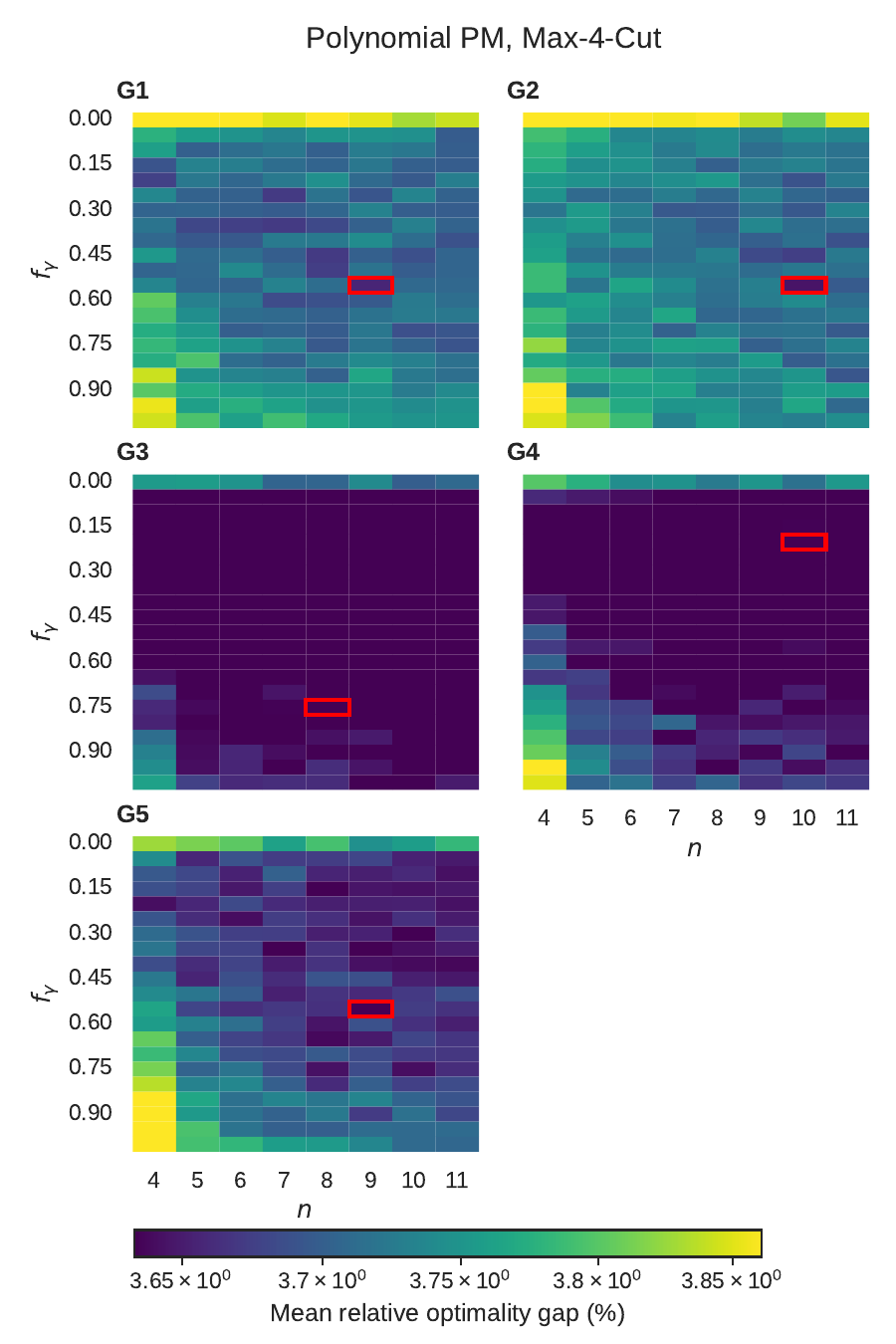}
  \caption{Heatmap showing the mean relative optimality gap (\%) for the polynomial PM model over the G1-G5 Max-4-Cut problems as a function of the hyperparameters $n$ and $f_\gamma$. The values of the mean relative optimality gap are shown in each cell of the heatmap. The hyperparameter configuration with the minimum mean relative optimality gap is marked by a red border.}
  \label{fig:polynomial_4cut_hyperparam}
\end{figure}

\begin{figure}[htbp]
  \centering
  \includegraphics[scale=.8]{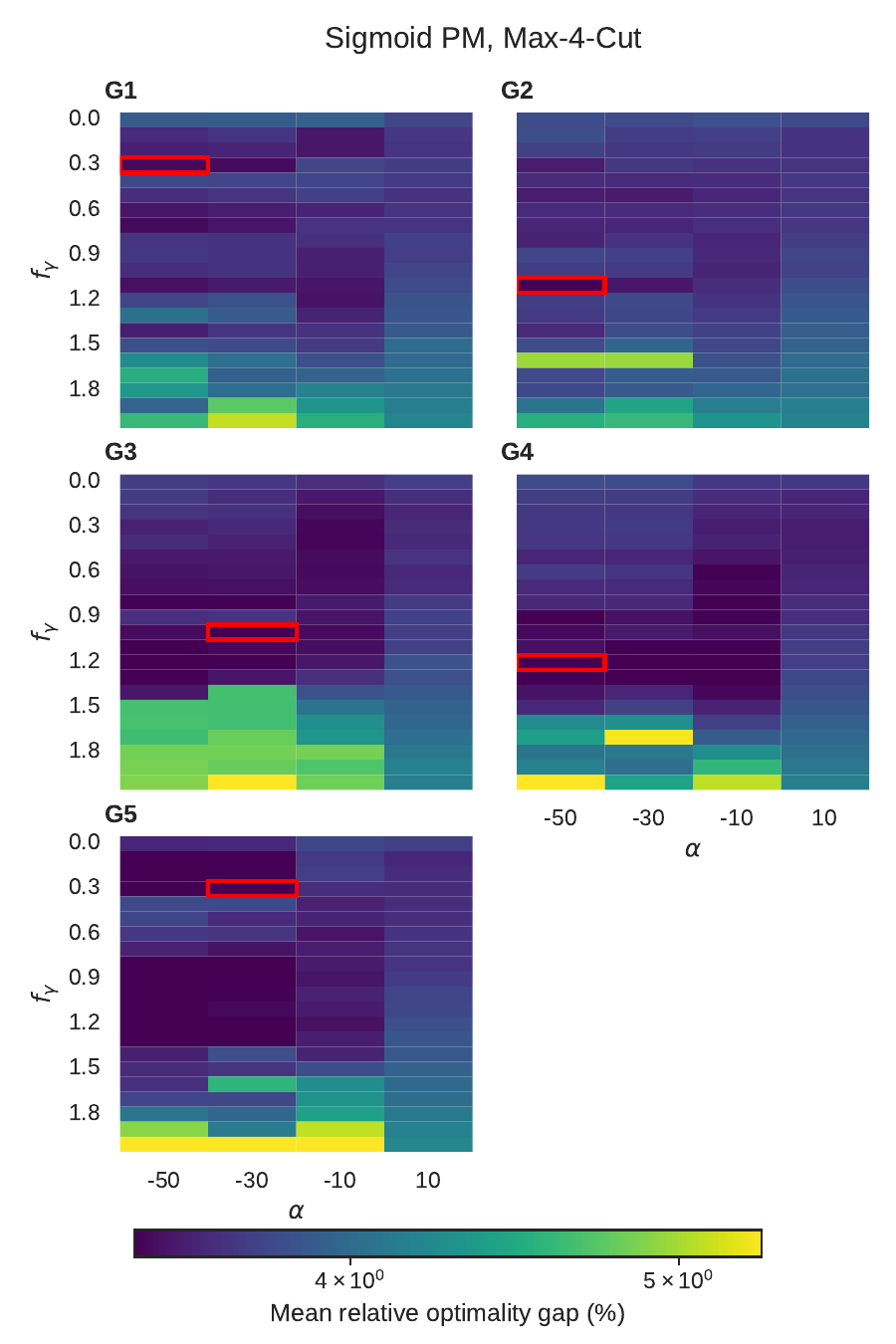}
  \caption{Heatmap showing the mean relative optimality gap (\%) for the sigmoid PM model over the G1-G5 Max-4-Cut problems as a function of the hyperparameters $f_\gamma$ and $\alpha$. The values of the mean relative optimality gap are shown in each cell of the heatmap. The hyperparameter configuration with the minimum mean relative optimality gap is marked by a red border. The sweep of $\alpha$ is limited to a minimum value of $-50$ to limit computational time, as lower values slow spin dynamics as mentioned in the main text section \ref{sec:sigmoid}.}
  \label{fig:sigmoid_4cut_hyperparam}
\end{figure}

\begin{figure}[htbp]
  \centering
  \includegraphics[scale=.8]{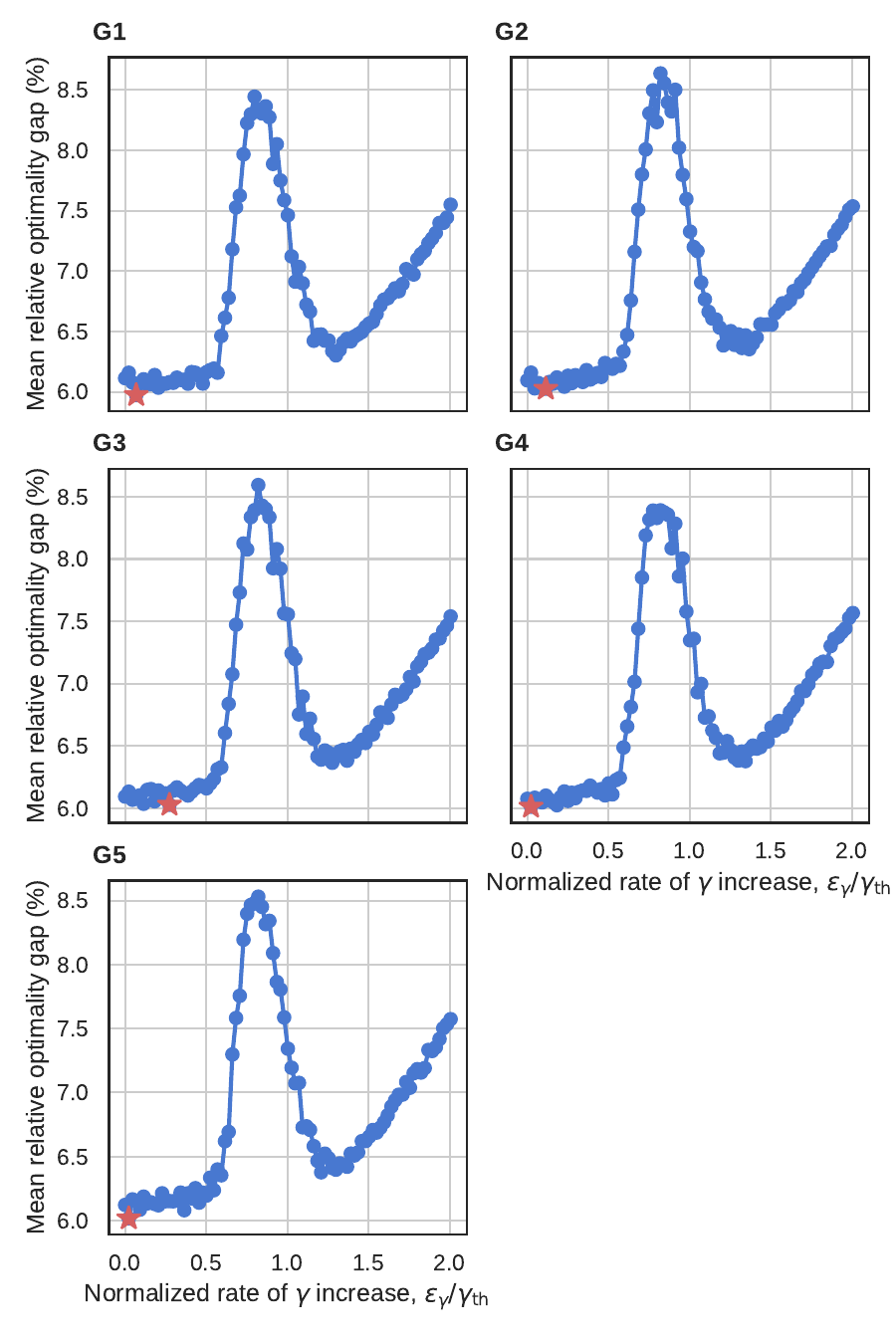}
  \caption{Mean relative optimality gap (\%) for the $q$-PDC model over the G1-G5 Max-4-Cut problems as a function of the hyperparameter $f_\gamma$. The minimum mean relative optimality gap is marked by a red star.}
  \label{fig:q_pdc_4cut_hyperparam}
\end{figure}

\FloatBarrier
\section{Convergence tests} \label{app:convergence_tests}

This section presents convergence tests for the studied Potts machine and Ising machine models. We examine three types of convergence: (1) simulation time convergence, which tests how performance changes with total simulation time $T$; (2) time step convergence, which tests the numerical stability of the Euler-Maruyama integration scheme with respect to the time step $\mathrm{d}t$; and (3) annealing parameter convergence, which tests how performance depends on the rate of parameter annealing. Each convergence test varies a single parameter while keeping the remaining hyperparameters fixed at the values used in the main benchmarking study.

Convergence tests are performed on the three benchmark problem sets defined in section~\ref{sec:problem_selection}: GSet Max-3-Cut, GSet Max-4-Cut, and ER50 Max-3-Cut. For the GSet problems, we report the mean relative optimality gap over 100 runs per graph, while for the ER50 problems, we report the success rate, also over 100 runs per graph. Error bars indicate standard deviation across runs.

\newpage
\subsection{Simulation time convergence}

These tests vary the total simulation time $T$ while keeping other parameters fixed. Longer simulation times allow the system to explore more of the solution space and potentially find better solutions, but at increased computational cost.

\subsubsection{GSet Max-3-Cut}

\begin{figure}[htbp]
  \centering
  \includegraphics[width=\textwidth]{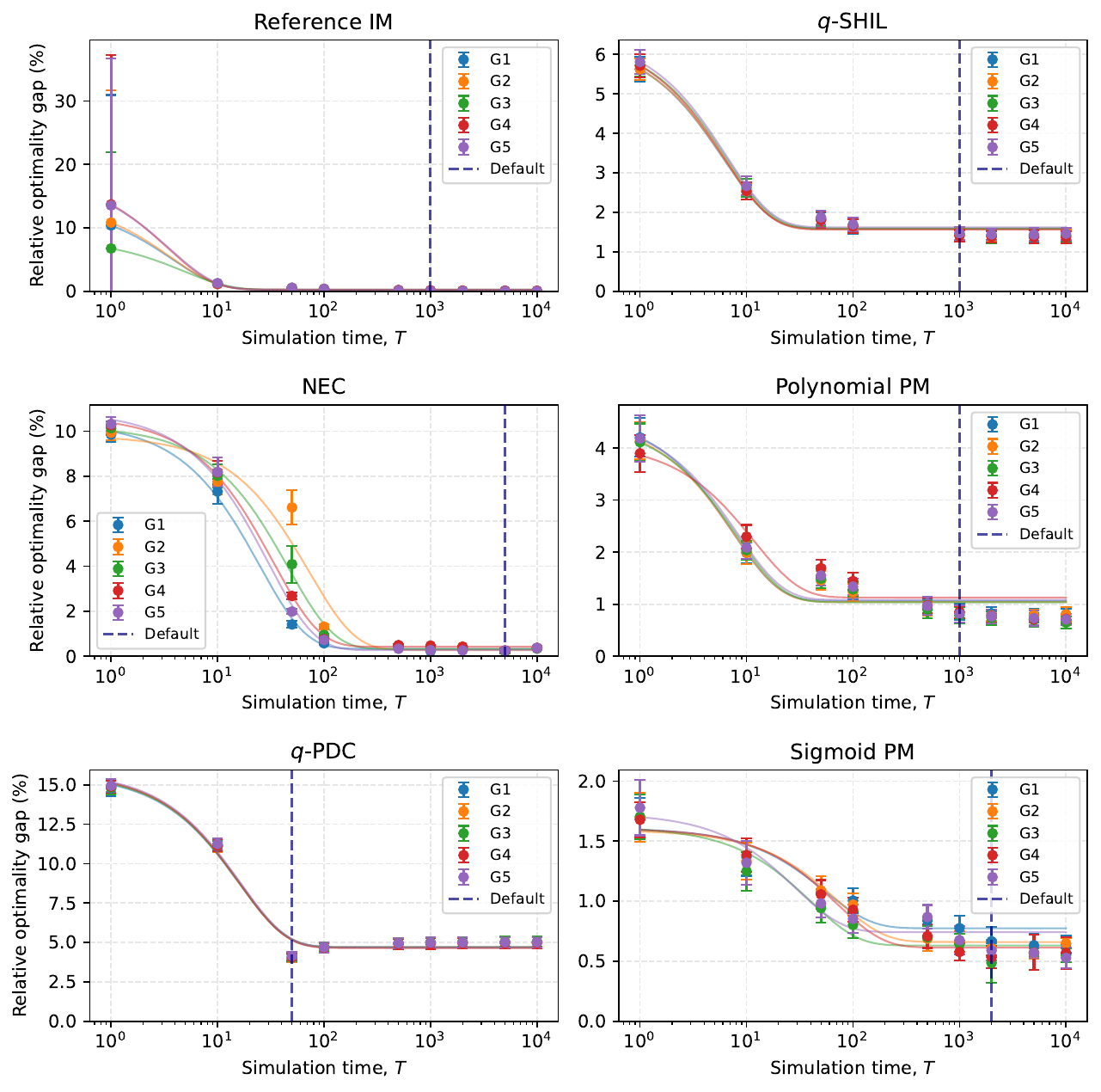}
  \caption{Simulation time convergence on GSet Max-3-Cut problems (G1--G5, 800 vertices each). Each subplot shows a different model, with colored series representing individual graphs (G1--G5). The mean relative optimality gap is shown as a function of total simulation time $T$, with error bars indicating standard deviation across 100 runs. Dashed lines show exponential saturation fits, and vertical dashed lines mark the parameter value selected for the main benchmarking study.}
  \label{fig:conv_sim_time_gset_3cut}
\end{figure}
\FloatBarrier

\newpage
\subsubsection{GSet Max-4-Cut}

\begin{figure}[htbp]
  \centering
  \includegraphics[width=\textwidth]{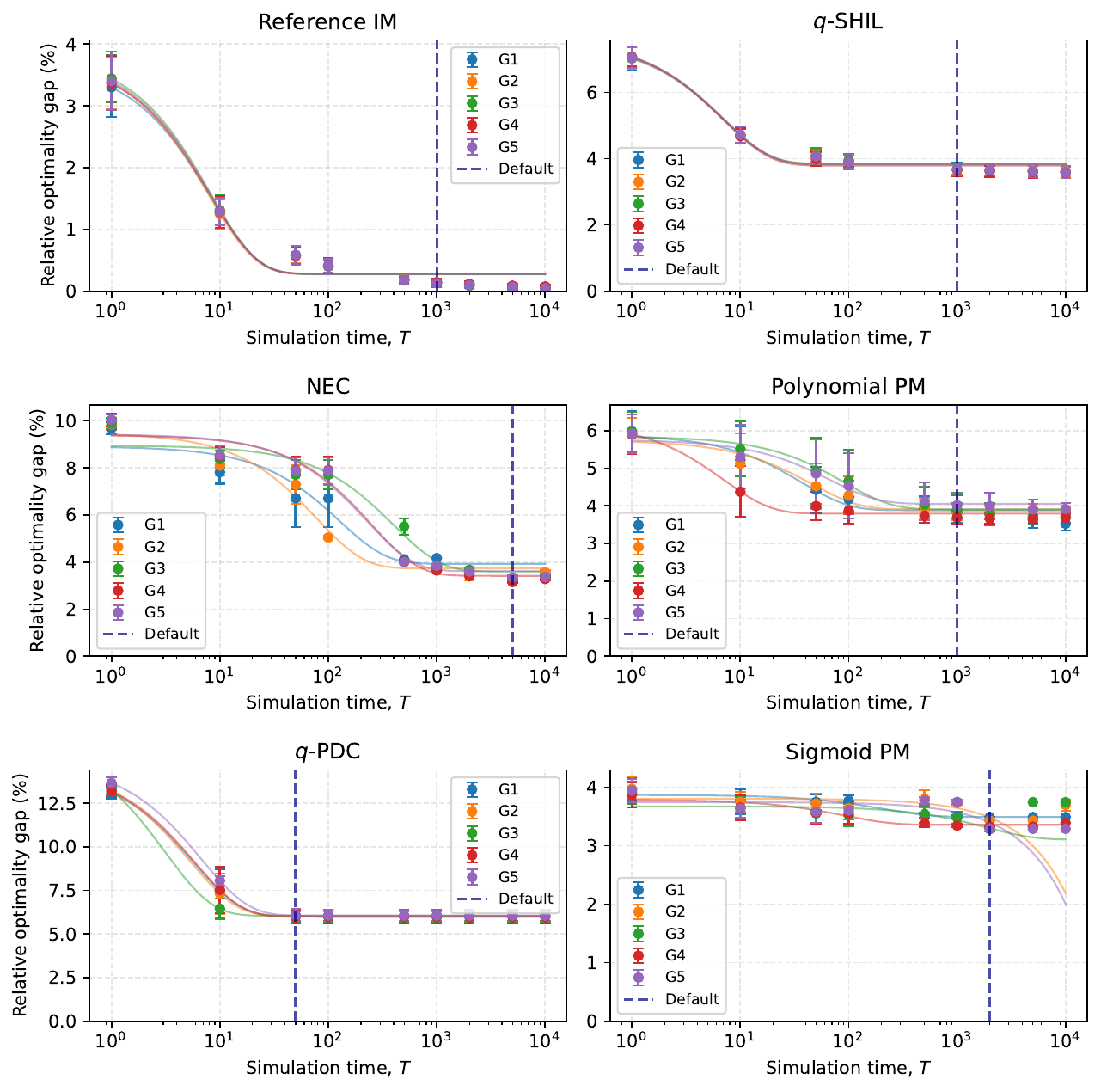}
  \caption{Simulation time convergence on GSet Max-4-Cut problems (G1--G5, 800 vertices each). Each subplot shows a different model, with colored series representing individual graphs (G1--G5). The mean relative optimality gap is shown as a function of total simulation time $T$, with error bars indicating standard deviation across 100 runs. Dashed lines show exponential saturation fits, and vertical dashed lines mark the selected parameter value.}
  \label{fig:conv_sim_time_gset_4cut}
\end{figure}
\FloatBarrier

\newpage
\subsubsection{ER50 Max-3-Cut}

\begin{figure}[htbp]
  \centering
  \includegraphics[width=\textwidth]{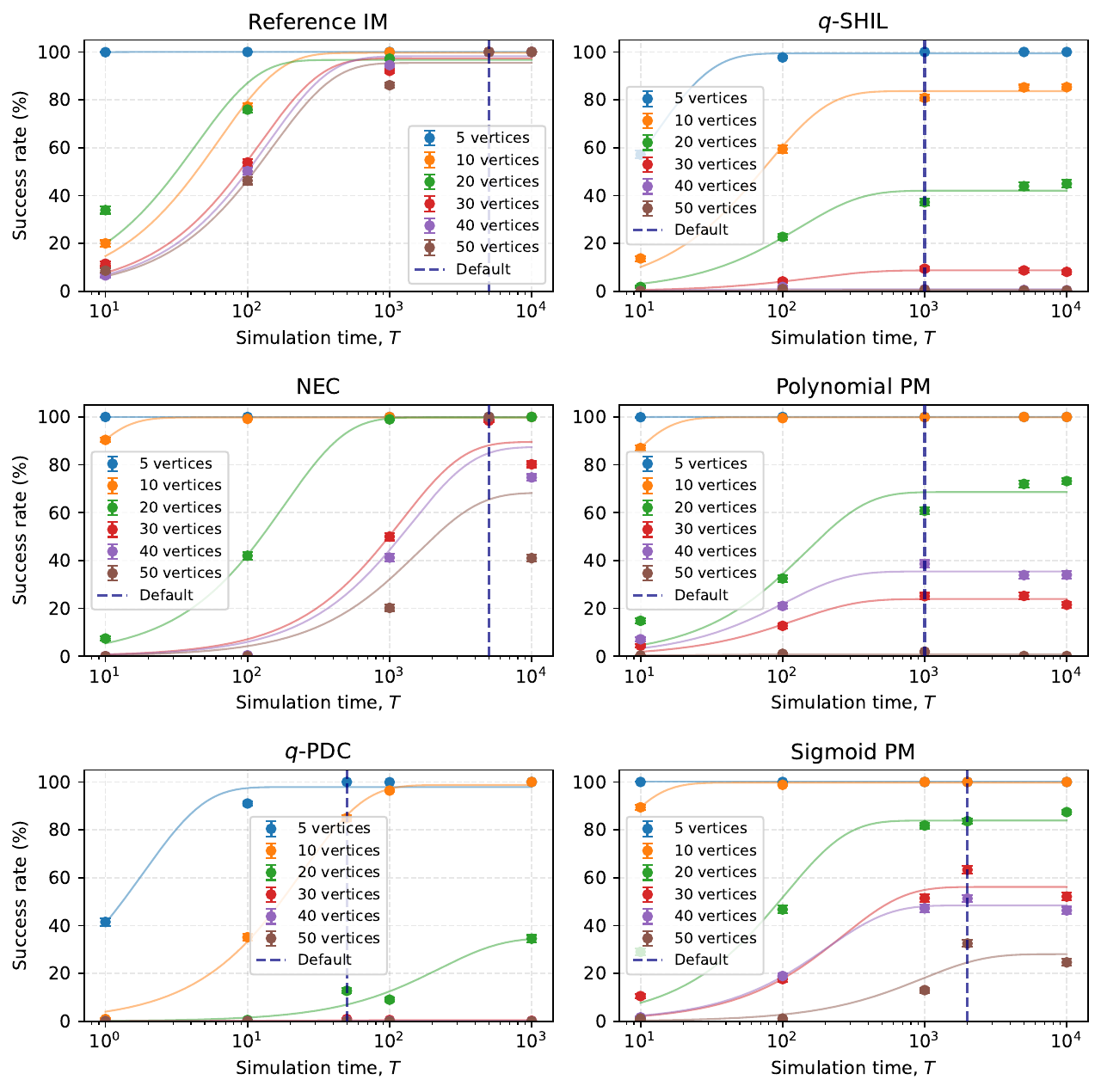}
  \caption{Simulation time convergence on ER50 Max-3-Cut problems. Each subplot shows a different model, with colored series representing different graph sizes (5--50 vertices). Success rate is shown as a function of total simulation time $T$, with error bars indicating standard deviation across runs. Dashed lines show exponential saturation fits, and vertical dashed lines mark the selected parameter value.}
  \label{fig:conv_sim_time_g05_3cut}
\end{figure}
\FloatBarrier

\newpage
\subsection{Time step convergence}

These tests vary the integration time step $\mathrm{d}t$ while keeping the total simulation time fixed. Smaller time steps provide more accurate numerical integration but require more computational steps. The goal is to identify the largest time step that maintains numerical stability and solution quality.

\subsubsection{GSet Max-3-Cut}

\begin{figure}[htbp]
  \centering
  \includegraphics[width=\textwidth]{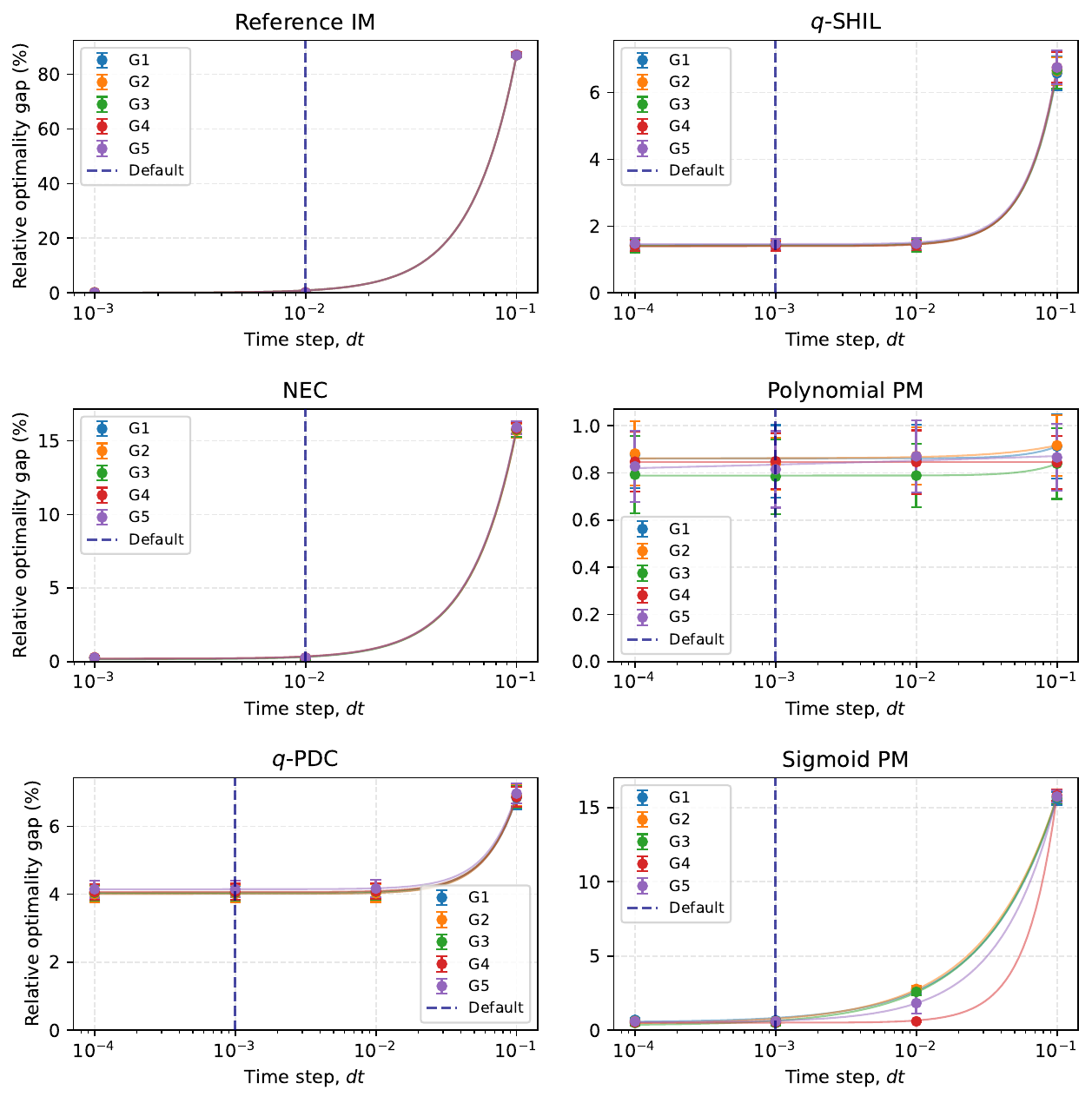}
  \caption{Time step convergence on GSet Max-3-Cut problems (G1--G5, 800 vertices each). Each subplot shows a different model, with colored series representing individual graphs (G1--G5). The mean relative optimality gap is shown as a function of integration time step $\mathrm{d}t$, with error bars indicating standard deviation across 100 runs. Dashed lines show power-law saturation fits, and vertical dashed lines mark the parameter value selected for the main benchmarking study.}
  \label{fig:conv_time_step_gset_3cut}
\end{figure}
\FloatBarrier

\newpage
\subsubsection{GSet Max-4-Cut}

\begin{figure}[htbp]
  \centering
  \includegraphics[width=\textwidth]{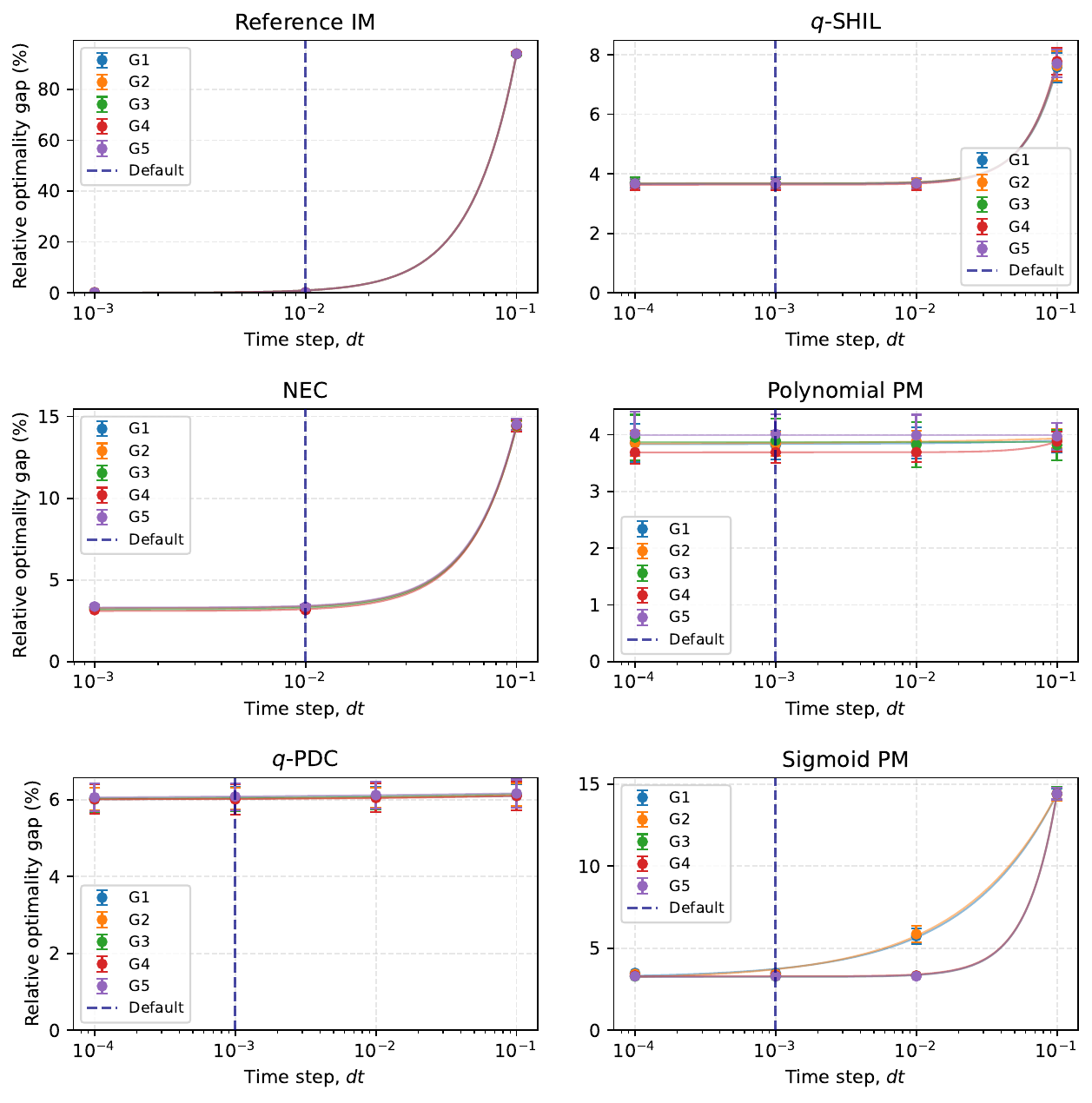}
  \caption{Time step convergence on GSet Max-4-Cut problems (G1--G5, 800 vertices each). Each subplot shows a different model, with colored series representing individual graphs (G1--G5). The mean relative optimality gap is shown as a function of integration time step $\mathrm{d}t$, with error bars indicating standard deviation across 100 runs. Dashed lines show power-law saturation fits, and vertical dashed lines mark the selected parameter value.}
  \label{fig:conv_time_step_gset_4cut}
\end{figure}
\FloatBarrier

\newpage
\subsubsection{ER50 Max-3-Cut}

\begin{figure}[htbp]
  \centering
  \includegraphics[width=\textwidth]{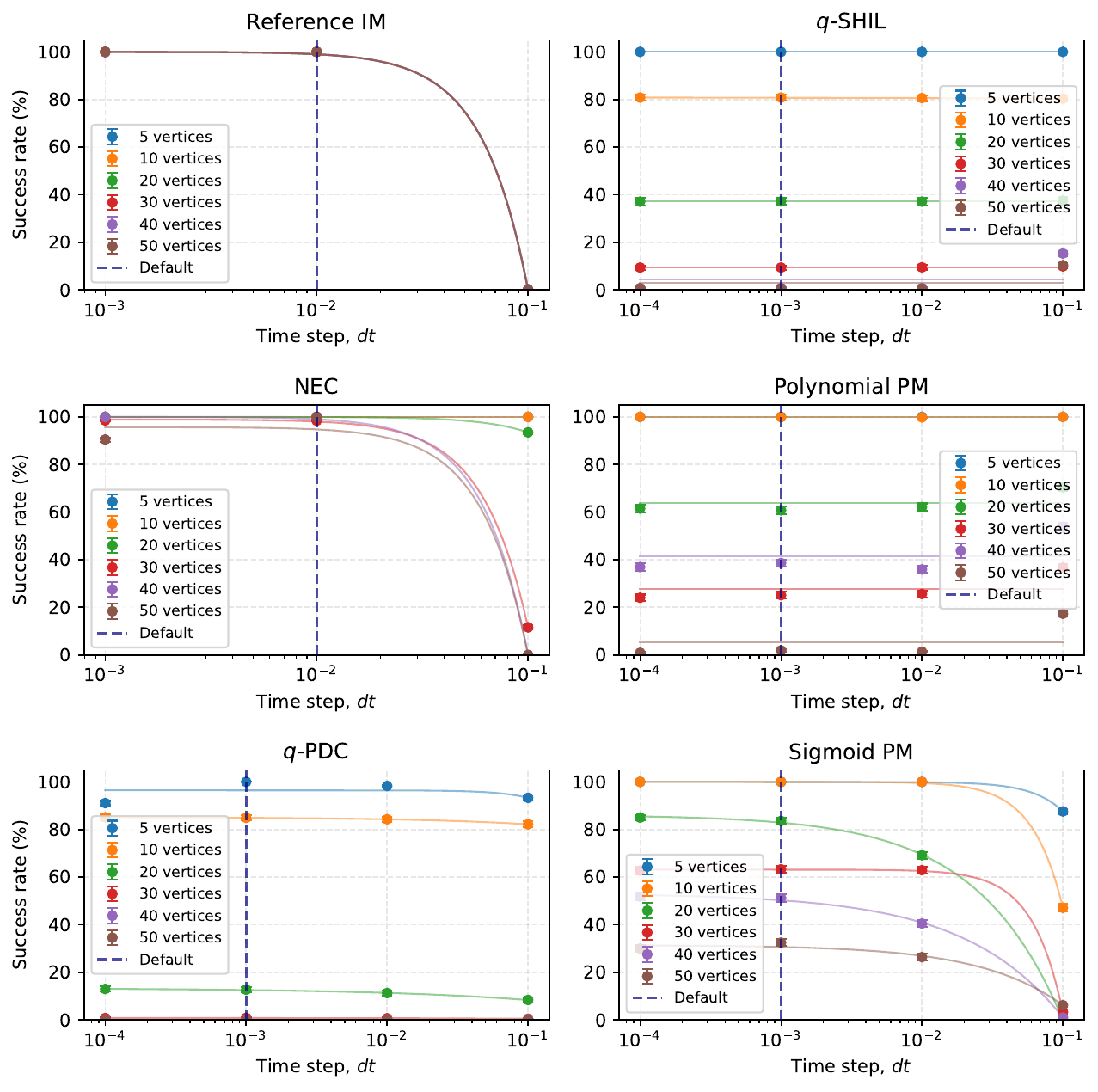}
  \caption{Time step convergence on ER50 Max-3-Cut problems. Each subplot shows a different model, with colored series representing different graph sizes (5--50 vertices). Success rate is shown as a function of integration time step $\mathrm{d}t$, with error bars indicating standard deviation across runs. Dashed lines show power-law saturation fits, and vertical dashed lines mark the selected parameter value.}
  \label{fig:conv_time_step_g05_3cut}
\end{figure}
\FloatBarrier

\newpage
\subsection{Annealing parameter convergence}

These tests vary model-specific annealing parameters to assess how the rate of parameter change affects solution quality. Slower annealing generally allows more thorough exploration of the solution space. The specific parameters varied are: $\varepsilon_\alpha$ (gain relaxation rate) for NEC, $\gamma$ span for $q$-SHIL, $\beta$ span for polynomial PM and sigmoid PM, and $\beta$ span for reference IM. The $q$-PDC model is excluded from these tests as its annealing behavior is already captured in the hyperparameter sensitivity analysis (Section~\ref{app:hyperparam_sensitivity}).

\subsubsection{GSet Max-3-Cut}

\begin{figure}[htbp]
  \centering
  \includegraphics[width=\textwidth]{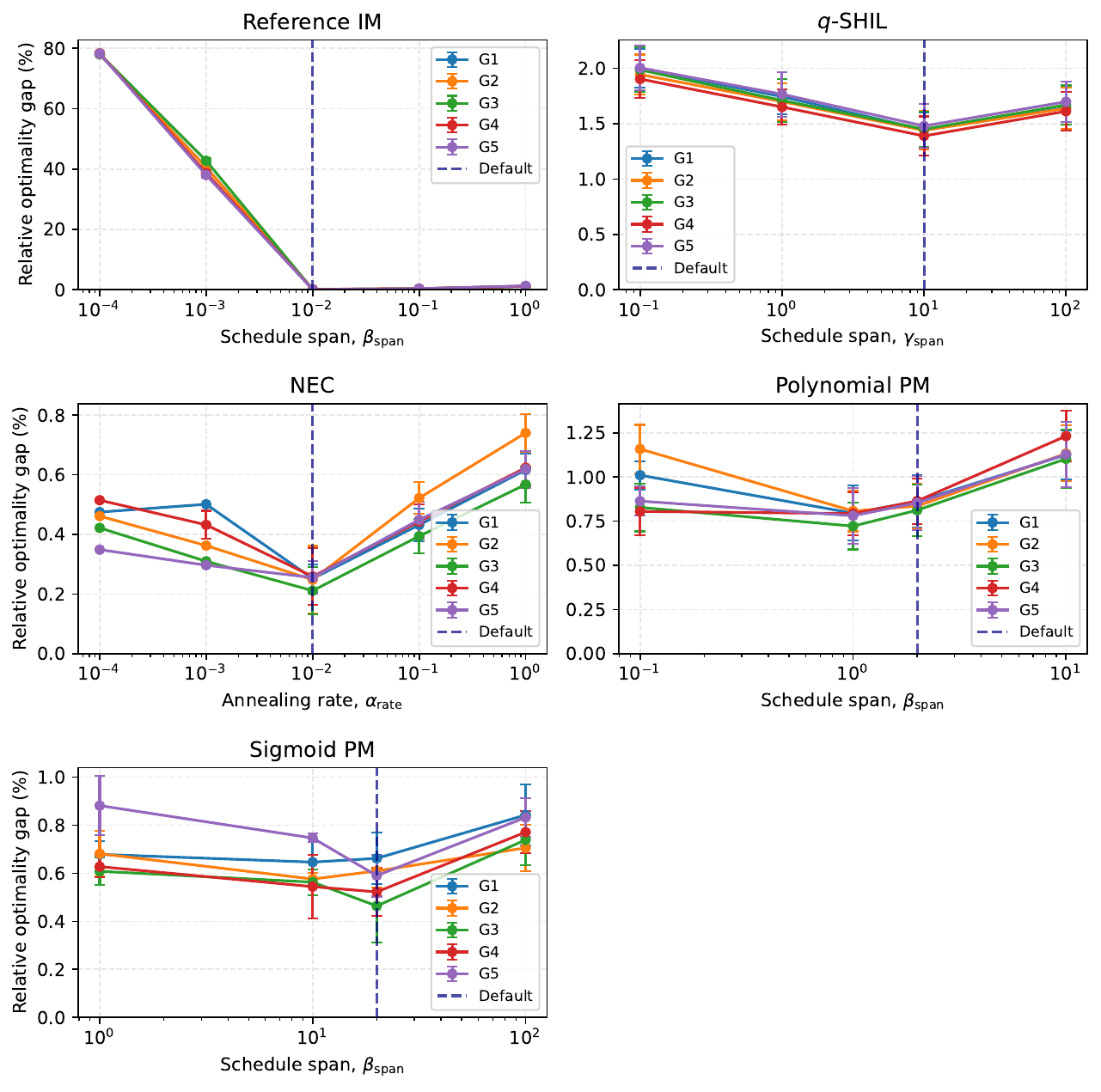}
  \caption{Annealing parameter convergence on GSet Max-3-Cut problems (G1--G5, 800 vertices each). Each subplot shows a different model (excluding $q$-PDC), with colored series representing individual graphs (G1--G5). The mean relative optimality gap is shown as a function of the model-specific annealing parameter, with error bars indicating standard deviation across 100 runs. Vertical dashed lines mark the parameter value selected for the main benchmarking study.}
  \label{fig:conv_annealing_gset_3cut}
\end{figure}
\FloatBarrier

\newpage
\subsubsection{GSet Max-4-Cut}

\begin{figure}[htbp]
  \centering
  \includegraphics[width=\textwidth]{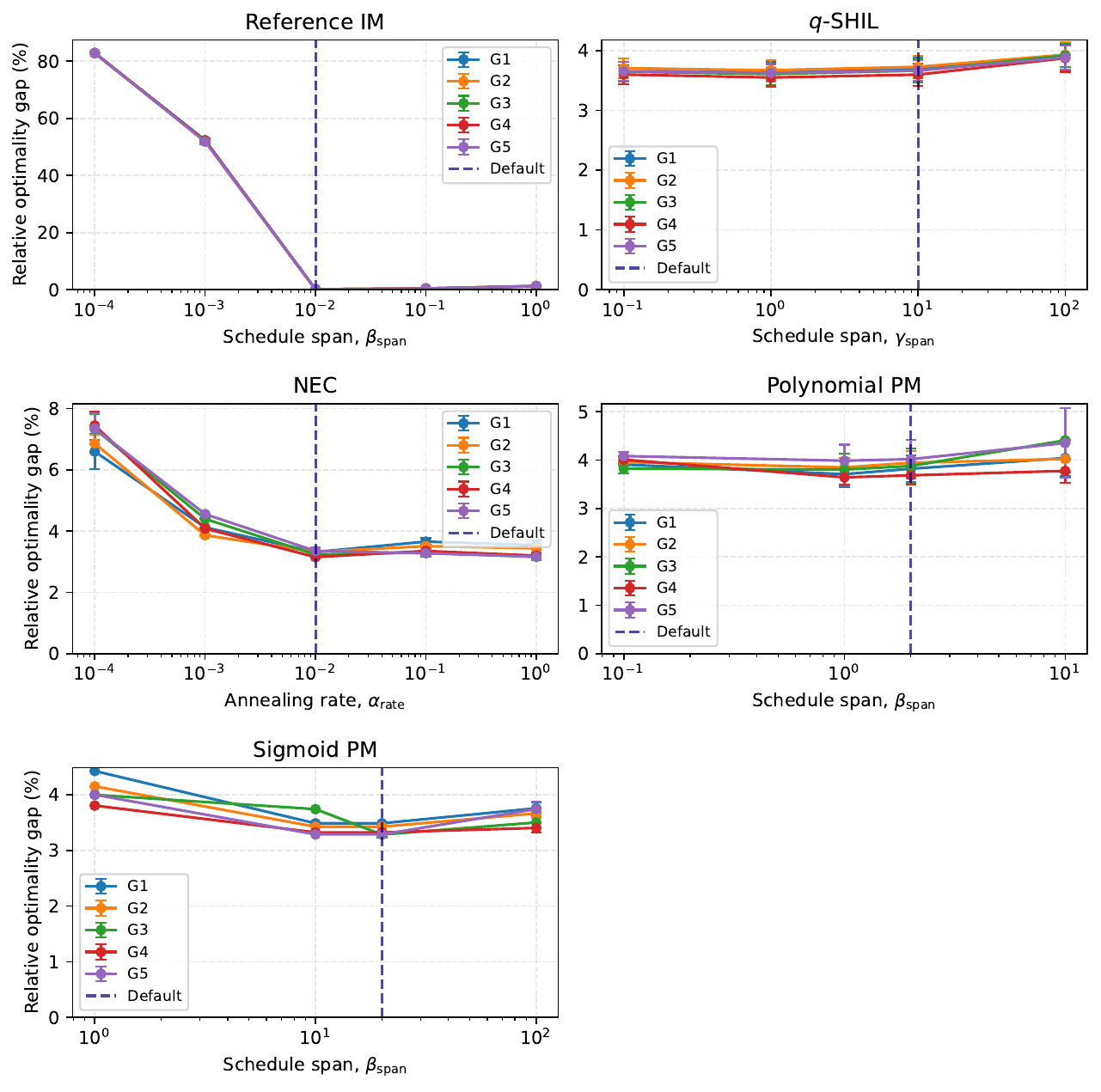}
  \caption{Annealing parameter convergence on GSet Max-4-Cut problems (G1--G5, 800 vertices each). Each subplot shows a different model (excluding $q$-PDC), with colored series representing individual graphs (G1--G5). The mean relative optimality gap is shown as a function of the model-specific annealing parameter, with error bars indicating standard deviation across 100 runs. Vertical dashed lines mark the selected parameter value.}
  \label{fig:conv_annealing_gset_4cut}
\end{figure}
\FloatBarrier

\newpage
\subsubsection{ER50 Max-3-Cut}

\begin{figure}[htbp]
  \centering
  \includegraphics[width=\textwidth]{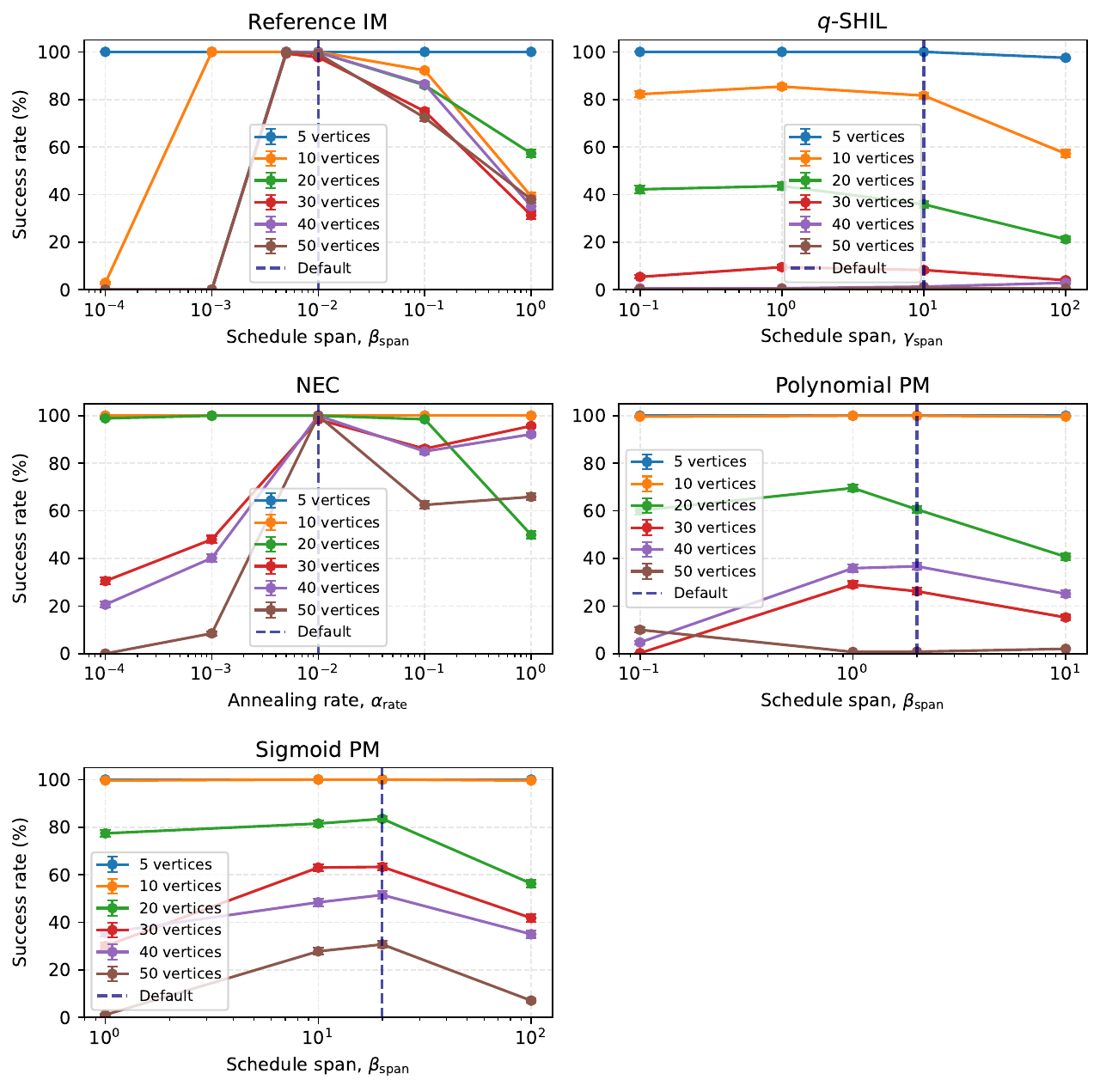}
  \caption{Annealing parameter convergence on ER50 Max-3-Cut problems. Each subplot shows a different model (excluding $q$-PDC), with colored series representing different graph sizes (5--50 vertices). Success rate is shown as a function of the model-specific annealing parameter, with error bars indicating standard deviation across runs. Vertical dashed lines mark the selected parameter value.}
  \label{fig:conv_annealing_g05_3cut}
\end{figure}
\FloatBarrier

\end{document}